\documentclass[11pt]{article}

\usepackage{amsfonts,latexsym,amsthm,amssymb,amsmath,amscd,euscript,float}
\usepackage{fullpage}
\usepackage[margin=1in]{geometry}
\usepackage{hyperref}
\usepackage{mathtools}
\usepackage{charter}
\usepackage{natbib}
\usepackage{rotating}
\usepackage{multicol}
\usepackage[dvipsnames]{xcolor}
\usepackage[usenames,dvipsnames]{pstricks}

\usepackage{colortbl}

\usepackage{authblk}

\usepackage{hyperref}
\hypersetup{
    pdffitwindow=false,            
    pdfstartview={Fit},            
    pdftitle={Evolution of social norms for moral judgment},     
    pdfauthor={Taylor Kessinger},         
    pdfsubject={},                 
    pdfnewwindow=true,             
    colorlinks=true,               
    linkcolor=OrangeRed,      
    citecolor=ForestGreen,       
    filecolor=Orchid,            
    urlcolor=Cerulean           
}

\mathtoolsset{showonlyrefs=true}

\definecolor{royalpurple}{rgb}{0.47, 0.32, 0.66}

\newcommand{\ass}{u_a}
\newcommand{\act}{u_x}

\newcommand{\p}{p}
\newcommand{\q}{q}

\newcommand{\g}{\gamma}
\newcommand{\G}{\Gamma}
\newcommand{\M}{\mathcal{M}}
\newcommand{\Gij}[2]{G^{#1,#2}}
\newcommand{\out}{\omega}

\newcommand{\pgc}{P_{GC}}
\newcommand{\pgd}{P_{GD}}
\newcommand{\pbc}{P_{BC}}
\newcommand{\pbd}{P_{BD}}

\newcommand{\bin}{b^\text{in}}

\newcommand{\bout}{b^\text{out}}

\title{Evolution of social norms for moral judgment}

\author[1]{Taylor A. Kessinger}
\author[2]{Corina E. Tarnita}
\author[1,3]{Joshua B. Plotkin}

\affil[1]{\footnotesize {Department of Biology, University of Pennsylvania, Philadelphia, PA 19104, USA}}
\affil[2]{\footnotesize {Department of Ecology and Evolutionary Biology, Princeton University, Princeton, NJ 08544, USA}}
\affil[3]{Center for Mathematical Biology, University of Pennsylvania, Philadelphia, PA 19104, USA}

\begin{document}

\maketitle

\begin{abstract}
\noindent Reputations provide a powerful mechanism to sustain cooperation, as individuals cooperate with those of good social standing. But how should moral reputations be updated as we observe social behavior, and when will a population converge on a common norm of moral assessment? Here we develop a mathematical model of cooperation conditioned on reputations,
for a population that is stratified into groups. Each group may subscribe to a different social norm for assessing reputations, and so norms compete as individuals choose to move from one group to another. 
We show that a group initially comprising a minority of the population may nonetheless overtake the entire population--especially if it adopts the \emph{Stern Judging} norm, which assigns a bad reputation to individuals who cooperate with those of bad standing. When individuals do not change group membership, stratifying reputation information into groups tends to destabilize cooperation, unless individuals are strongly insular and favor in-group social interactions.  
We discuss the implications of our results for the structure of information flow in a population and the evolution of social norms of moral judgment.

\end{abstract}

\section{Introduction}

Societies depend on cooperation.
This is generally considered paradoxical, because cooperation is often costly and may be selected against.
But a wide range of well-understood mechanisms help to explain cooperation in human and non-human populations, such as kin selection \citep{hamilton1964, smith1964}, population structure \citep{ohtsuki2006rule,ohtsuki2006replicator}, and direct reciprocity \citep{trivers1971, wilkinson1984}.
There is ample evidence for these mechanisms acting across diverse taxa \citep{bourke2011, vanvliet2022microbes}, including controlled experiments on repeated interactions among human subjects \citep{rand2013evolution}.
Nonetheless, large human societies often require cooperation between unrelated strangers who have little prospect for future interactions. 

Reputations and social norms provide a compelling explanation for cooperation in large, human societies, as people tend to cooperate with others of good social standing \citep{alexander1985biological,elster1989social,cialdini1991focus,kandori1992social,fehr2004social,bicchieri2005grammar, tomasello2013origins, fehr2018normative,hechter2018norms, curry2019good}.
In disciplines ranging from psychology to economics, there is broad recognition that social norms of moral assessment, and institutions that support these norms, are critical for shaping cooperative behavior and solving problems of collective action \citep{zucker1986production,ostrom1990governing,gintis2005moral}.
But how exactly a population comes to mutual agreement about a social norm remains an active area of research in these fields \citep{van2022neurons}.
Here, we adapt the theory of indirect reciprocity to provide some theoretical insights into this outstanding question.

Under the theory of indirect reciprocity, individuals who cooperate with others of good social standing will maintain their own good reputation and thereby increase the chance of reciprocal cooperation from strangers \citep{nowak1998evolution,milinski2001cooperation}.
This simple idea of cooperation conditioned on moral reputations has substantial empirical support, as reputations are known to be related to cooperation, and cooperation can in turn lead to higher social status \citep{rege2004impact,bereczkei2007public,von2019dynamics,fehr2004social,gurerk2006competitive}.
Moreover, studies that include functional neuroimaging have shown that people can reliably detect different social norms about cooperation \citep{hackel2020shifting}.
People tend to join the dominant institution of moral assessment, which then facilitates subsequent cooperation, in behavioral economic experiments \citep{gurerk2006competitive}.
Even disinterested third party observers punish those who do not play by the dominant social norm for cooperation \citep{fehr2004third}. 

Given this empirical basis, evolutionary game theory provides a modeling framework to study the spread and maintenance of cooperative behavior conditioned on reputations \citep{nowak1998evolution,milinski2001cooperation,milinski2002reputation,nowak2005evolution,pacheco2006stern,ohtsuki2006leading,sigmund2010calculus,santos2016evolution}.
Mathematical models of indirect reciprocity keep track of reputations for all individuals in a population, as well as of their behavioral strategies, which are both updated over time.
This theoretical framework has been largely successful in delineating what conditions can sustain cooperation in terms of the nature of pairwise social interactions, 
the availability of information about reputations, and the social norm for awarding a good or bad reputation to an individual based on their observed behavior. 

Two factors have emerged as critical for cooperation under indirect reciprocity: the extent to which individuals share a consensus view of each others' reputations, and the social norm by which individuals are assessed and reputations assigned.
First, it is known that cooperation can flourish when there is consensus about reputations in a population, which can be achieved through rapid gossip \citep{sommerfeld2007gossip} or by a centralized institution that broadcasts reputation information \citep{radzvilavicius2021nature}.
But cooperation tends to collapse when individuals each have their own idiosyncratic views of each others' reputations \citep{uchida2013private}.
Second, stable cooperation depends strongly on what norm the society adopts for moral judgment.
For example, a norm that judges an individual as good for cooperating with a bad actor (i.e., a norm that values forgiveness) is less likely to foster cooperation than a norm that judges such an individual as bad (i.e., a norm that values punishment)\footnote{Note that the term ``social norm'' has a specific technical meaning in the literature on indirect reciprocity, which may differ from the broader notions of descriptive and injunctive norms used in psychology \citep{cialdini1991focus}.} \citep{ohtsuki2006leading}. 

Crucially, however, all these theoretical insights have been derived for homogeneous societies---where individuals all embrace the same, exogenously imposed social norm---and assuming either fully public \citep{radzvilavicius2021nature} or entirely private information \citep{uchida2013private}.
But, in reality, most societies are multicultural and structured into groups that may hold different views about the reputations of individuals \citep{bouwmeester2017registered, henrich2001search}. 
Information about reputations may flow freely within a group of individuals who share a common language, ethnic or religious identity, political affiliation, etc., but flow more slowly between groups.
Members of a group may therefore share a common view on the reputations of others in the population, but different groups may disagree.
Different groups may even subscribe to different social norms for evaluating reputations in the first place. This reality raises two fundamental questions: in a heterogeneous society with imperfect information flow, under what conditions will everyone eventually converge on a common norm of judgment, and when will the prevailing norm be socially optimal?

These questions have proven difficult to answer, despite a few notable attempts \citep{pacheco2006stern,chalub2006norms,nakamura2012favoritism}.
Here we approach the problem by developing a general theory of indirect reciprocity in group-structured populations.
We allow for multiple co-occurring norms and explore a continuum of scenarios between fully public and fully private information, by partitioning the population into distinct and disjoint ``gossip groups''.
Members of a gossip group share the same views about the reputations of all individuals in the population, but different groups may hold different views.
Individuals learn behavioral strategies from each other, or they may change their group membership, by biased imitation---so that the strategic composition and group sizes evolve over time.
When different groups subscribe to different norms of judgment, the model describes competition between social norms arising from individual-level decisions.
This approach allows us to predict whether, and by what dynamics, a population will come to adhere to a single social norm. 

Our analysis identifies \emph{Stern Judging}---a social norm that values defection against individuals of bad standing---as the best competitor against 
other norms in the literature on indirect reciprocity.
And \emph{Stern Judging} is an even stronger competitor when individuals preferentially interact with their in-group. More generally, 
we obtain a simple condition that determines whether a social norm that is initially used by a minority of the population will spread via social contagion.

We also analyze how group structure affects the prospects for cooperation, when group memberships are fixed.
Even when all groups use the same social norm, a population may fail to secure a high level of cooperation when information about reputations is fractured into independent groups. 
The destabilizing effect on cooperation grows rapidly with the number of groups, but it can be partly mitigated by a preference for in-group social interactions. We compare our results to the theoretical literature on indirect reciprocity and to empirical work on reputations and norms for cooperation in human societies.
We conclude by discussing implications for the evolution of social norms, and for the number of groups with independent judgments that a well-functioning society can sustain.

\section{Model}

Our analysis extends a well-established body of mathematical models for how cooperation emerges from indirect reciprocity \citep{nowak1998evolution,milinski2001cooperation,milinski2002reputation,nowak2005evolution,pacheco2006stern,ohtsuki2006leading,sigmund2010calculus,santos2016evolution}.
These models describe how individuals in a large population behave in pairwise interactions. Individuals choose actions according to strategies that may account for each others' reputations; reputations themselves are assessed and updated according to a social norm of moral judgment.
Strategies that provide larger payoffs tend to spread by biased imitation. 

In particular, we consider a population of $N$ individuals who play a series of one-shot donation games with each other.
Each round, every individual plays the game twice with everyone else, acting once as a donor and once as a recipient.
Donors may either cooperate, paying a cost $c$ to convey a benefit $b > c$ to their interaction partner, or defect, paying no cost and conveying no benefit.
This game constitutes a minimal example of a social dilemma \citep{rapoport1965prisoner,nowak2005evolution}.

Each individual has a view of the current social standing, or reputation, of every other individual in the population.
A donor chooses whether to cooperate or not according to their behavioral strategy and the reputation of the recipient: cooperators (denoted ALLC or $X$) always cooperate, defectors (denoted ALLD or $Y$) always defect, and discriminators (DISC, $Z$) cooperate with those they consider to have a good reputation and defect with those they consider to have a bad reputation.

We extend the standard model by partitioning the population into $K$ distinct and disjoint ``gossip groups'', which comprise fractions $\nu_1, \nu_2, \ldots \nu_K$ of the total population.
Individuals within a gossip group share the same view about the reputations in the population, but different gossip groups may disagree about reputations.
Different groups may even employ different social norms for assigning reputations.
This model describes a situation in which individuals transmit information about reputations to other members of their group via rapid gossip \citep{sommerfeld2007gossip} or, alternatively, in which each group has its own centralized ``institution'' \citep{radzvilavicius2021nature} that broadcasts reputation judgments to everyone in the group. (Mathematically, the reputational views of the $K$ groups can be described by $K$ vectors $\{ 0, 1\}^N$. An entry of $0$ or $1$ denotes that an individual has a bad or good reputation according that group, respectively.) In the case of a single group, $K=1$, our analysis reduces to the standard model of indirect reciprocity with public information \citep{milinski2001cooperation,milinski2002reputation,nowak2005evolution,pacheco2006stern,ohtsuki2006leading,sigmund2010calculus,santos2016evolution}.

After a round of pairwise game interactions, each group updates its views of everyone's reputations. An individual's reputation is updated as follows.
A random interaction in which they acted as a donor is sampled from the most recent round, and their reputation is assessed based on what action they took towards the recipient.
The rule for assessing the reputation of a donor, called the \emph{social norm}
considers the donor's action and the group's view of the recipient's current reputation (i.e., the social norm is second-order \citep{pacheco2006stern}).
We focus most of our analysis on the four norms that are the most common in the literature \citep{radzvilavicius2021nature}. All four norms regard it as good to cooperate with an individual who has a good reputation, and bad to defect with an individual who has a good reputation; but the norms differ in how they assess interactions with bad individuals:
\begin{enumerate}
    \item under \emph{Scoring}, cooperating with a bad recipient will result in the donor being assigned a good reputation, whereas defecting with a bad recipient results in a bad reputation.
    \item under \emph{Shunning}, any interaction with a bad recipient yields a bad reputation.
    \item under \emph{Simple Standing}, any interaction with a bad recipient yields a good reputation.
    \item under \emph{Stern Judging}, cooperating with bad results in a bad reputation, but defecting with a bad recipient results in a good reputation.
\end{enumerate}

Our model also allows for two types of error: errors in strategy execution and errors in reputation assessment \citep{sasaki2017evolution}. Whenever an donor intends to cooperate with a recipient, 
there is a chance $\act$ that the donor will accidentally defect, which we call an execution error. (Individuals who intend to defect never accidentally cooperate.) In addition, there is a chance $\ass$ that a group will erroneously assign the wrong reputation to the donor, which we call an assessment error.

\subsubsection*{Strategy dynamics under payoff-biased imitation}

We describe our model in two parts: first, by specifying how strategy frequencies evolve under payoff-biased imitation, when group membership is fixed; second, by specifying how group sizes themselves evolve, when individuals switch groups by payoff-biased imitation.

To keep track of strategy frequencies and reputations in the population--and the resulting actions and payoffs that arise--we must account for the fact that different groups may hold different views of an individual's reputation. We let $f_s^I$ denote the fraction of individuals in group $I$ who follow strategy $s \in \{X, Y, Z\}$. Further, we let $g_s^{I,J}$ denote the fraction of individuals following strategy $s$ in group $I$ whom group $J$ sees as good (the first superscript index denotes ``who'', and the second index denotes ``in whose eyes''). Finally, we define $g^{I,J}$ as the average fraction of individuals in group $I$ whom group $J$ sees as good, and $g^{\bullet,J}$ as the average fraction of individuals in the entire population whom group $J$ sees as good.
These average reputations are given by
\begin{equation}
    \begin{split}
        g^{I,J} & = \sum_s f_s^I g_s^{I,J},
        \\
        g^{\bullet,J} & = \sum_I g^{I,J}.
    \end{split}
\end{equation}
(Throughout our presentation, sums over $s$ are always interpreted as sums over strategic types, $s \in \{X, Y, Z\}$; and sums over capital letters are interpreted as sums over groups, e.g.~$J \in \{ 1 \ldots K \}$.)

Each individual accrues a payoff $b$ from their interactions with cooperators, and also from discriminators who view them as good. In addition, cooperators pay the cost of cooperation, $c$, and discriminators pay this cost when they interact with recipients their group sees as good.
Thus, in the limit of large $N$, the net payoff of each strategic type in group $I$, averaged over all pairwise interactions they engage in and accounting for execution errors, is given by
\begin{equation}
    \begin{split}
        \Pi_X^I & = (1 - \act) \Big[ b \sum_J \nu_J (f_X^J + f_Z^J g_X^{I,J}) - c \Big]
        \\
        \Pi_Y^I & = (1 - \act) \Big[ b \sum_J \nu_J (f_X^J + f_Z^J g_Y^{I,J}) \Big]
        \\
        \Pi_Z^I & = (1 - \act) \Big[ b \sum_J \nu_J (f_X^J + f_Z^J g_Z^{I,J}) - c g^{\bullet,I} \Big].
    \end{split}
    \label{eq:fit_multi_groups}
\end{equation}

After all groups have updated their views of everyone's reputations, 
a randomly chosen individual updates their behavioral strategy according to payoff-biased imitation \citep{sigmund2010calculus, radzvilavicius2019evolution}.
That is, a randomly chosen individual, following strategy $s$, chooses a random other individual in the population, following strategy $s^\prime$, and compares their payoffs. If the focal individual is in group $I$, and their comparison partner is in group $J$, the focal individual switches their \emph{behavioral strategy} to $s^\prime$ with a probability given by the Fermi function
\begin{equation}
    \phi(\Pi_s^I,\Pi_{s^\prime}^J) = \frac{1}{1 + \exp \big[ \beta (\Pi_s^I - \Pi_{s^\prime}^J) \big] }.
    \label{eq:fermi}
\end{equation}
The parameter $\beta$ here is called the strength of selection \citep{traulsen2007pairwise,traulsen2010human}, which we assume to be weak ($\beta \ll 1$) for the entirety of our analysis.

The resulting strategy dynamics can be described by a replicator equation \citep{taylor1978evolutionary,sigmund2010calculus} in the limit of a large population size. We derive the replicator equation for our model (SI Section 8) under the standard assumption that reputations equilibrate quickly before individuals update strategies \citep{okada2018solution}. The form of the resulting replicator equation depends on how exactly comparison partners are chosen for strategy imitation. If comparison partners are chosen irrespective of group identity, the strategic frequencies $f_s^I$ quickly equalize across all groups $I$ and converge to a common set of values $f_s$, whose dynamics then satisfy
\begin{equation}
    \begin{split}
        \dot{f}_s & = f_s \big( \Pi_s - \Bar{\Pi} \big), \text{with}
        \\
        \Pi_s & = \frac{\sum_J \nu_J f_s^J \Pi_s^J}{\sum_J \nu_J f_s^J} = \sum_J \nu_J \Pi_s^J,
        \\
        \Bar{\Pi} & = \sum_J \sum_s \nu_J f_s^J \Pi_s^J = \sum_J \sum_s \nu_J f_s \Pi_s^J.
    \end{split}
    \label{eq:mixed_replicator}
\end{equation}

Strategic evolution (Eq.~\eqref{eq:mixed_replicator}) and the resulting levels of cooperation have been the primary focus of research on indirect reciprocity. Depending upon the social norm, it is known that strategic evolution can lead to a stable equilibrium of discriminators ($f_Z=1$) who generate high levels of cooperation sustained by indirect reciprocity \citep{milinski2001cooperation,milinski2002reputation,nowak2005evolution,pacheco2006stern,ohtsuki2006leading,sigmund2010calculus,santos2016evolution}.
Aside from strategic evolution, however, we also study the evolution of competing gossip groups themselves, which may subscribe to different social norms for judging reputations. 

\subsubsection*{Dynamics of competing gossip groups}
To study competing gossip groups, we develop a model in which, instead of strategic imitation, individuals engage in imitation of group membership, so that group sizes may change over time.
In particular, after a round of population-wide pairwise gameplay and reputation assessment, a randomly chosen individual, with group membership $I$, chooses a random other individual in the population, with group membership $J$, and compares their payoffs.
The focal individual switches their \emph{group membership} to $J$ with a probability given by Eq.~\eqref{eq:fermi}. When analyzing dynamic group membership, we make the simplifying assumption (which can be relaxed; see Supplementary Information Section 9.6) that all groups are fixed for the same behavioral strategy, so all $s = s^\prime$ in Eq.~\eqref{eq:fermi} and each gossip group $I$ is characterized by a single average payoff $\Pi^I$. We obtain a replicator equation for the relative sizes of gossip groups:
\begin{equation}
    \begin{split}
        \dot{\nu}_I & = \nu_I \big( \Pi^I - \Bar{\Pi} \big), \text{with}
        \\
        \Bar{\Pi} & = \sum_J \nu_J \Pi^J.
    \end{split}
    \label{eq:replicatornu}
\end{equation}
We analyze this model in the case of $K = 2$ competing groups and when all individuals use the discriminator strategy, meaning that they attend to a co-player's reputation when choosing whether or not to donate to them.

\section{Results}

\subsection{Dynamics of group membership}

We first analyze the dynamics of $K=2$ gossip groups that form independent assessments of all reputations.
In most cases, we find that group sizes are bistable: above a critical frequency, $\nu_1^*$, the size of group $1$ will increase towards one, and below the critical frequency the size of group $1$ will decrease towards zero (see Supplementary Information Section 4).
And so eventually the population will be dominated by one group or the other.

The precise value of $\nu_1^*$, which determines when group $1$ will eventually out-compete group $2$, depends on the social norms the two groups follow.
Although there is no simple expression for $\nu_1^*$ in general, we can exploit the model's bistability to gain some analytic insight.
If $\dot{\nu}_1$ is positive when $\nu_1 = 1/2$, then because the system is at most bistable, we have $\nu_1^* < 1/2$ (because $\dot{\nu}_1$ must cross the $\nu_1$-axis at a value lower than $1/2$). Likewise, if $\dot{\nu}_1$ is negative when $\nu_1 = 1/2$, then $\nu_1^* > 1/2$.
By rewriting Eq.~\eqref{eq:fit_multi_groups} and setting $\nu_1 = \nu_2 = 1/2$, we find that $\dot{\nu}_1|_{\nu_1 = 1/2} > 0$ only if
\begin{equation}
    \big[ (b-c) (g^{1,1} - g^{2,2}) + (b+c) (g^{1,2} - g^{2,1}) \big] \big|_{\nu_1 = 1/2} > 0.
    \label{eq:nu_condition}
\end{equation} 
When this condition is satisfied, we have $\nu_1^*<1/2$, which means that group $1$ can grow and eventually overtake the entire population even when it starts as the smaller group.

There is a simple intuition associated with the terms in Eq.~\eqref{eq:nu_condition}, which governs whether group $1$ or group $2$ has the advantage.
The first term represents the difference in the payoff to group $1$ versus group $2$ due to within-group interactions.
The second term represents the difference in the payoff to group $1$ versus group $2$ due to out-group interactions;  
it can be thought of as the payoff difference between groups due to cooperation that is not reciprocated by the opposing group.
If the net advantage to group $1$ is positive, then group $1$ will grow.

We can understand the terms in Eq.~\eqref{eq:nu_condition} more explicitly by considering in-group and out-group rates of donation. Individuals in group $1$ cooperate with each other (barring execution errors) at a rate $g^{1,1}$---each paying a cost $c$, as donor, and earning a benefit $b$, as recipient.
Individuals in group 1 thereby accrue average payoff $(b-c) g^{1,1}$ from their in-group interactions, and likewise group $2$ individuals accrue $(b-c) g^{2,2}$ for their in-group interactions.
And so the first term, $(b-c) (g^{1,1}-g^{2,2})$, represents the fitness difference between groups $1$ and $2$ arising from in-group interactions.
In addition, an individual in group $2$ donates to an individual in group $1$ with probability $g^{1,2}$, paying a cost $c$ and providing benefit $b$ to the member of group $1$. This produces a fitness difference of $(b + c) g^{1,2}$ arising when a (potential) donor in group $2$ interacts with a recipient in group $1$. Likewise, there is a fitness difference of $-(b + c) g^{2,1}$ arising from donors in group $1$ interacting with recipients in group $2$. And so the second term in Eq.~\eqref{eq:nu_condition} represents the difference between the payoffs to groups $1$ and $2$ due to between-group interactions.

When the two groups follow the same social norm, we have $g^{1,1} = g^{2,2}$ and $g^{1,2} = g^{2,1}$, and the inequality of Eq.~\eqref{eq:nu_condition} becomes an equality. In other words, when both groups follow the same norm then $\nu_1^* = 1/2$, and so whichever gossip group is initially larger will grow and eventually dominate the entire population. 
(The sole exception to this is \emph{Scoring}: when \emph{Simple Standing} and \emph{Stern Judging} compete against it, they can  invade starting from frequency zero, i.e., $\nu_1^* = 0$, and $\nu_1 = 1$ is the only stable equilibrium.) 
There is a simple intuition for this result: the larger gossip group has an advantage, all else being equal, because members of that group share their reputational views with a larger portion of the population, which reduces the rate of unreciprocated cooperation \citep{radzvilavicius2021nature}.

\subsubsection*{Competing social norms}

We can use our model of dynamic group membership (Eq.~\eqref{eq:replicatornu}) to study competition between social norms -- that is, the dynamics of groups that subscribe to \textit{different} norms of moral assessment. We again find bistability: group $1$ will grow and overtake the entire population when its initial frequency exceeds a critical value, $\nu_1^*$; otherwise it will shrink to size zero. When the groups follow different norms, the critical frequency $\nu_1^*$ above which group $1$ will eventually fix need not equal $1/2$. And so one social norm may start in the minority and yet eventually outcompete the other norm.

\begin{figure}[ht!]
    \begin{center}
    \includegraphics[width=6.2in]{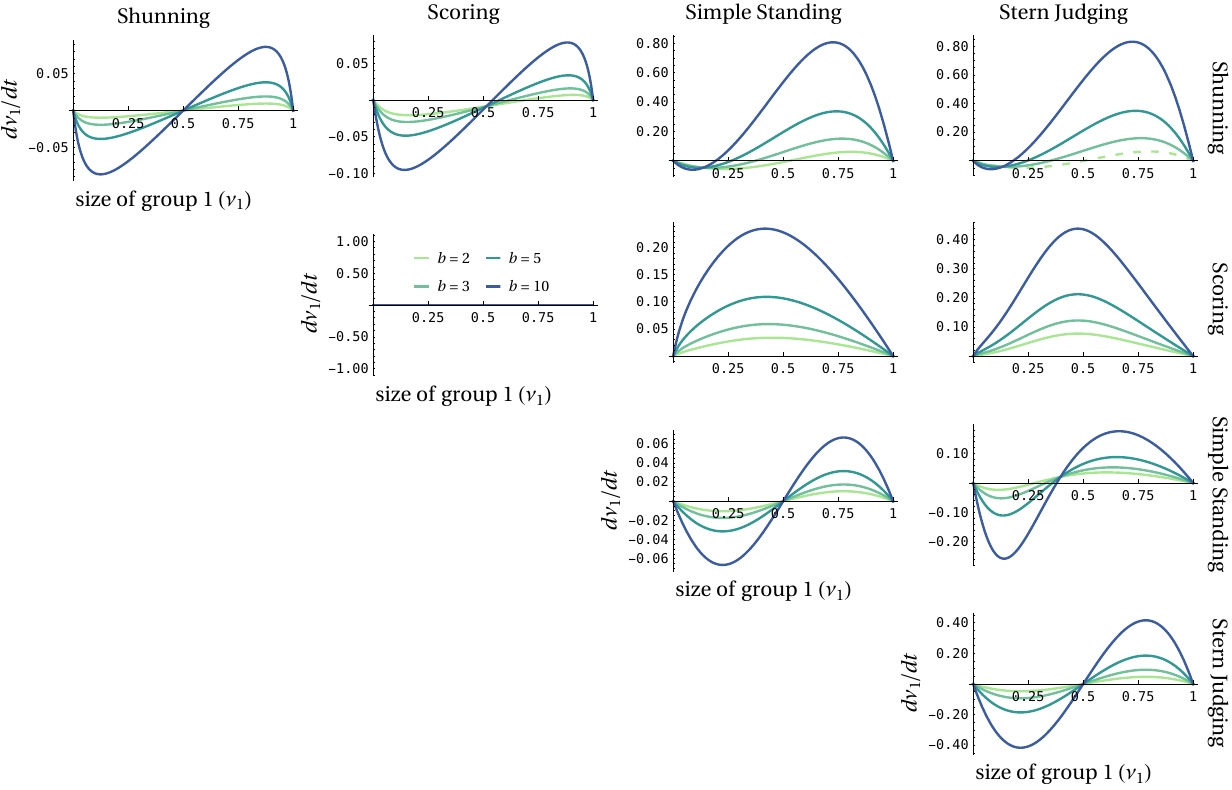}
    \end{center}
    \caption{\small{The dynamics of competing social norms for $K = 2$ groups 
    are typically bistable: group $1$ will grow and overtake group $2$ when its size exceeds a threshold, $\nu_1>\nu_1^*$, or else it will shrink to zero. The social norm used in group $1$ is indicated at the top of each column, and the norm used in group $2$ is indicated at the right of each row.  Each panel shows the rate of change of group $1$'s size, $\dot{\nu_1}$, as a function of its current size, $\nu_1$, with different colors corresponding to different values of the benefit $b$ in the donation game. The threshold $\nu_1^*$ is determined by where the curve crosses the $x$-axis. Note that \emph{Stern Judging} is usually the strongest norm, with $\nu_1^*<1/2$ against all other norms. In all plots, $c  = 1$, $\ass = \act = 0.02$, and the value of $b$ is colored according to the inset shown in the \emph{Scoring}-\emph{Scoring} figure.
    The dotted line for $b = 2$, when \emph{Stern Judging} competes against \emph{Shunning}, indicates the special situation that, as Shunning grows, the population passes through a regime where it is vulnerable to invasion by unconditional defectors.}}
    \label{fig:group_switch_b}
\end{figure}

The outcome of competing norms is captured by Eq.~\eqref{eq:nu_condition}, whose terms need not be zero, so that $\nu_1^*$ may differ from $1/2$. In particular, if the inequality is satisfied, then $\nu_1^* < 1/2$.
This means that group $1$ may initially comprise a minority of the population (but not too small a minority), and yet eventually out-compete group $2$ that subscribes to a different social norm. When this happens, we say that group $1$ follows a ``stronger'' norm, meaning that individuals who follow that norm can enjoy a fitness advantage over others, even when their own group is smaller, thus enticing others to adopt their norm.
We find that \emph{Stern Judging} is the ``strongest'' norm among the four we study, enjoying a value $\nu_1^* < 1/2$ against any other norm (Fig.~\ref{fig:group_switch_b}).

In general, the rate at which group $1$ grows, $\dot{\nu}_1$, depends upon its current frequency, the social norm of group $1$, and the social norm of group $2$.
Fig.~\ref{fig:group_switch_b} illustrates that \emph{Stern Judging} generally outcompetes any other norm, even when starting from a minority 
of the population, regardless of the game payoff $b$. The sole exception is when \emph{Stern Judging} is pitted against \emph{Shunning} for sufficiently small $b$, such as $b = 2$, when \emph{Stern Judgers} engage in more unreciprocated cooperation than do \emph{Shunners}.
However, this case is exceptional, because as the \emph{Shunning} group grows in size, the population passes through a regime where they can be invaded by defectors (see Supplementary Information Section 4.2). This means that, in a model with both strategic copying and group switching (Eq.~\eqref{eq:copying_both}), defectors could invade the population.
And so in summary, \emph{Stern Judging} is a stronger norm than all others, whenever the population is robust enough to prevent unconditional defectors from invading as individuals update their group membership (Fig.~\ref{fig:group_switch_b}).

The strength of the \emph{Stern Judging} norm can also be understood in terms of Eq.~\eqref{eq:nu_condition}.
\emph{Stern Judging} affords a high level of within-group consensus, but it is less tolerant of the opposing group insofar as they do not share reputational views, so both terms in Eq.~\eqref{eq:nu_condition} can be positive. As a result, a group following \emph{Stern Judging} may grow and outcompete another norm even when it starts in the minority (i.e.~$\nu_1^*<1/2$).
By contrast, \emph{Shunning} cannot guarantee a high level of within-group consensus, whereas \emph{Simple Standing} is too accommodating of differences across group opinions to compete vigorously against \emph{Stern Judging}.
\emph{Scoring} is unique in that its members may have higher views of their \emph{out-group} than of their in-group, and so it typically loses in competition against any other norm.
The results on intra- and inter-group reputational views are summarized in Table \ref{tab:group_views}, which helps to explain why \emph{Stern Judging} is the strongest norm.

\begin{table}[ht!]
\centering
    \begin{tabular}{|c|c|c|c|c|}
        \multicolumn{5}{c}{within-group opinions ($g^{1,1}$)}
        \\
        \hline
        & SJ & SS & SC & SH
        \\
        \hline
        SJ &
            \cellcolor{blue!48.5}$0.97$
        &
            \cellcolor{blue!48}$0.96$
        &
            \cellcolor{blue!48}$0.96$
        &
            \cellcolor{blue!48.5}$0.97$
        \\
        \hline
        SS &
            \cellcolor{blue!48}$0.96$
        &
            \cellcolor{blue!48}$0.96$
        &
            \cellcolor{blue!48}$0.96$
        &
            \cellcolor{blue!48}$0.96$
        \\
        \hline
        SC &
            \cellcolor{blue!36.5}$0.73$
        &
            \cellcolor{blue!40}$0.80$
        &
            \cellcolor{blue!17}$0.34$
        &
            \cellcolor{blue!4.5}$0.09$
        \\
        \hline
        SH &
            \cellcolor{blue!5}$0.10$
        &
            \cellcolor{blue!5}$0.10$
        & 
            \cellcolor{blue!3}$0.06$
        &
            \cellcolor{blue!3}$0.06$
        \\
        \hline
    \end{tabular}   \begin{tabular}{|c|c|c|c|c|}
        \multicolumn{5}{c}{between-group opinions ($g^{1,2}$)}
        \\
        \hline
        & SJ & SS & SC & SH
        \\
        \hline
        SJ &
            \cellcolor{red!23.5}$0.47$
        &
            \cellcolor{red!37.5}$0.75$
        &
            \cellcolor{red!32.5}$0.65$
        &
            \cellcolor{red!18}$0.36$
        \\
        \hline
        SS &
            \cellcolor{red!41.5}$0.83$
        &
            \cellcolor{red!45}$0.90$
        &
            \cellcolor{red!41}$0.82$
        &
            \cellcolor{red!19}$0.38$
        \\
        \hline
        SC &
            \cellcolor{red!39}$0.78$
        &
            \cellcolor{red!43}$0.86$
        &
            \cellcolor{red!17}$0.34$
        &
            \cellcolor{red!3}$0.06$
        \\
        \hline
        SH &
            \cellcolor{red!3.5}$0.07$
        &
            \cellcolor{red!3.5}$0.07$
        & 
            \cellcolor{red!1}$0.02$
        &
            \cellcolor{red!1}$0.02$
        \\
        \hline
    \end{tabular}
    \caption{\small{In-group and out-group reputations for $K=2$ equally sized groups. The social norm used in group $1$ is indicated at the top of each column; the norm used in group $2$ is indicated at the left of each row. Darker colors denote more strongly positive opinions. When group $1$ follows \emph{Stern Judging}, it typically has a high view of itself but a somewhat low view of group $2$, so that its members will often cooperate with each other and are less likely to engage in un-reciprocated cooperation with the opposing group. No other norm satisfies both of these conditions, which is why \emph{Stern Judging} tends to outcompete other social norms across a wide range of costs and benefits $c$ and $b$ (Fig.~\ref{fig:group_switch_b}). Error rates are $\ass = \act = 0.02$.}}
    \label{tab:group_views}
\end{table}

Aside from identifying \emph{Stern Judging} as the strongest norm, our analysis identifies several other key features of norm competition. For example, \emph{Stern Judging} and \emph{Simple Standing} always out-compete \emph{Scoring}, regardless of how small their initial frequencies are (Fig.~\ref{fig:group_switch_b}).
\emph{Stern Judging} or \emph{Simple Standing} are also strong in competition against \emph{Shunning}, which will be displaced even if its initial frequency is has high as $~80\%$ (for $b=10$).
Finally, considering competition between \emph{Stern Judging} and \emph{Simple Standing}, \emph{Stern Judging} is always stronger ($v_1^*<1/2$), and the basin of attraction towards \emph{Stern Judging} is larger when the benefit of cooperation $b$ is smaller.

We have focused on competition among the four social norms that are most common in the literature.
But some prior work has considered a larger set of ``leading eight'' social norms \citep{ohtsuki2006leading}, including six third-order norms that consider the current reputation of the \emph{donor} when deciding what their new reputation should be. In the Supplementary Information (Section 6), we develop equations for reputation dynamics with third-order norms in multiple gossip groups, and we analyze the dynamics of competition between two groups that follow different norms among the ``leading eight'' (SI Figure $2$).
We find that \emph{Stern Judging} generally outcompetes all other ``leading eight'' norms, in the sense that the critical frequency $\nu_1^*$, above which 
it is guaranteed to grow, is always less than $1/2$.
The sole exception is competition with norm $s_8$, a third-order norm that differs only slightly from \emph{Stern Judging}, and which is stronger than \emph{Stern Judging} only for low values of the benefit $b$.

Finally, we have developed and solved equations for a model where two groups use different norms, but individuals rely on private assessment (Supplementary Information Section 4).
In such a scheme, switching group membership is tantamount to switching which norm an individual uses, but it does not guarantee that an individual will hold the same reputational assessments as other members of their group. \emph{Stern Judging} performs poorly in this context of norm competition with private assessment, whereas \emph{Simple Standing} out-competes the other second-order norms across a range of values of $b$ (SI Figure $1$). 

\subsection*{Competing norms with insular interactions}

So far we have assumed that individuals interact with all others in the population, accumulating payoffs from both within-group and between-group interactions without any bias. However, if group membership determines not only how an individual views the reputations in the population but also whom they tend to interact with in game play, then differential rates of within- versus between-group interactions could influence which groups perform best during competition.

We model insularity by stipulating that any given pair of individuals will not assuredly interact, but rather interact with probability $\out^{I,J} = \out^{J,I}$ for individuals in groups $I$ and $J$. We focus on the case when individuals favor in-group interactions, so that a randomly chosen pair of individuals will always interact if they happen to be members of the same group, but the interaction will occur with probability $0 < \out < 1$ if they are from different groups.
Note that this notion of insularity is weaker than that of in-group favoritism in social psychology \cite{tajfel1971social,tajfel1982social,tajfel2004social} and game theory \cite{fu2012evolution}. Insular individuals in our model simply prefer to interact with in-group members, but they have no inherent bias towards cooperating with in-group members.

The parameter $\out \leq 1$ determines the per-capita rate at which out-group social interactions are allowed, relative to in-group interactions. Note the reducing the probability of an out-group interaction has two consequences: it changes an individual's fitness, which is averaged over interactions that actually occur, and it also changes an individual's reputation, because they are more likely to be observed interacting with an in-group partner when $\out<1$. We recover the case of well-mixed interactions when $\out=1$. 

\begin{figure}[!ht]
    \begin{center}
    \includegraphics[width=6.2in]{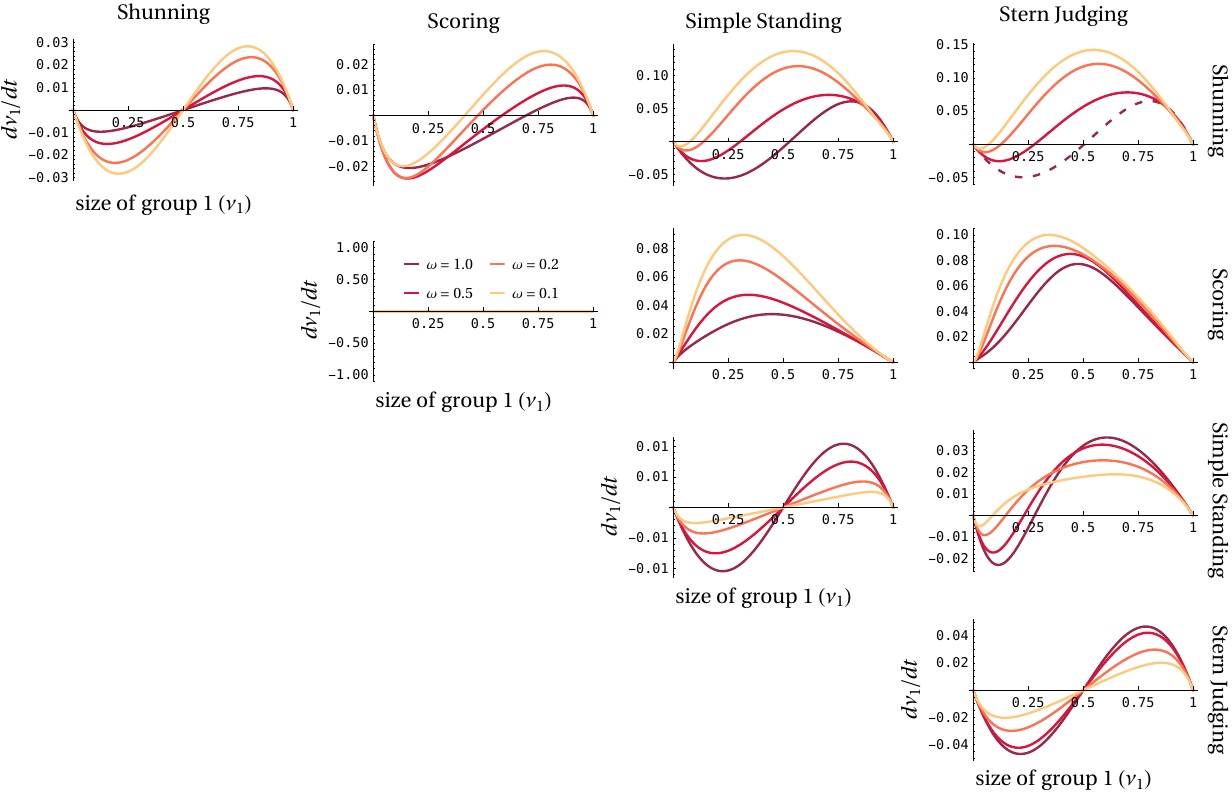}
    \end{center}
    \caption{
    \small{
    The dynamics of competing social norms in $K = 2$ groups, for different levels of insularity (out-group interaction parameter $\out$). The social norm used in group $1$ is indicated at the top of each column; the norm used in group $2$ is indicated at the right of each row.  Each panel shows the rate of change of group $1$'s size, $\dot{\nu_1}$, as a function of its current size, $\nu_1$, with different colors corresponding to different values of the outgroup interaction rate, $\out$. In all cases, insularity ($\out<1$) tends to make \emph{Stern Judging} even stronger in competition with every other norm.
    In all plots, $b = 2$, $c  = 1$, $\ass = \act = 0.02$.
    Values of $\out$ are as inset in the \emph{Scoring}-\emph{Scoring} figure.
    The dotted line for $\out = 1.0$, when \emph{Stern Judging} competes against \emph{Shunning}, indicates the special situation that, as Shunning grows, the population passes through a regime where it is vulnerable to invasion by unconditional defectors.}}
    \label{fig:group_switch_o}
\end{figure}

Even when interactions are insular, we recover the result that the larger group is guaranteed to grow in size, when two groups follow the same social norm.
However, when the groups follow different norms, a high degree of insularity (i.e., small $\out$) means that within-group interactions contribute more strongly to fitness than between-group interactions, and so in-group interactions have a greater effect on which norm will dominate.
In particular, when social interactions are insular, Eq.~\eqref{eq:nu_condition} can be generalized to the condition
\begin{equation}
    \big[ (b-c) (g^{1,1} - g^{2,2}) + \out (b+c) (g^{1,2} - g^{2,1}) \big] \big|_{\nu_1 = 1/2} > 0.
    \label{eq:nu_condition_ins}
\end{equation}
This implies that norms that might otherwise perform poorly due to a high rate of unreciprocated between-group cooperation can fare better when interactions are insular ($\out < 1$). And so insularity shifts the balance of norm competition, as reflected in Fig.~\ref{fig:group_switch_o}: \emph{Stern Judging}, \emph{Simple Standing}, and even \emph{Scoring} compete better against \emph{Shunning} at lower values of $\out$. In general, \emph{Stern Judging}'s position against every other norm is strengthened for lower rates of out-group interaction, as \emph{Stern Judging} enjoys a smaller initial frequency $\nu_1^*$ to guarantee its growth.

\subsection{Dynamics of strategy evolution}

Although our primary focus has been on the dynamics of competing gossip groups, with fixed behavioral strategies, we can alternatively analyze the effect of gossip groups on strategic evolution. In this analysis we fix group membership, so that groups do not change in size, and we assume all groups adopt the same social norm. We then analyze the dynamics of three competing strategies: cooperate, defect, and discriminate. We seek to understand how a population structure with $K>1$ distinct groups, each with independent information about reputations, alters the stability of long-term cooperative behavior.

When a population is partitioned into multiple groups that form distinct reputational judgments, we expect that it will generally be more difficult to achieve a high level of cooperation than in a single group. To demonstrate this, we first review strategy evolution in a population with a single group, which has been studied extensively in prior work \citep{hilbe2018indirect, radzvilavicius2019evolution, radzvilavicius2021nature, santos2016evolution,santos2018social,pacheco2006stern,ohtsuki2004should,ohtsuki2006leading,ohtsuki2009indirect,schmid2021nature,schmid2021scirep}. Then we compare those results to a population with $K>1$ distinct gossip groups, each with independent reputation information.

\subsubsection*{Cooperation in a well-mixed population}

In a population consisting of a single gossip group ($K=1$), which is equivalent to fully public reputations \citep{nowak2005evolution,sommerfeld2007gossip,nakamura2011indirect,ohtsuki2009indirect}, there are two stable strategic equilibria: a population composed entirely of defectors ($f_Y = 1$) or a population composed entirely of discriminators ($f_Z = 1$).
The population of discriminators can support a high level of cooperation. There is also an unstable equilibrium consisting of a mixture of defectors and discriminators at $f_Z = c/[b(\epsilon - \ass)], f_Y = 1 - f_Z, f_X = 0$ (see Supplementary Information Section 2).

An all-defector population can never be invaded, whereas an all-discriminator population can resist invasion by defectors provided
\begin{equation}
    \frac{b}{c} > \frac{1}{\pgc - \pgd} = \frac{1}{\epsilon - \ass}.
\end{equation}
Here $\pgc$ is the chance that a donor who intends to cooperate with a good recipient is assigned a good reputation, and likewise for $\pgd$, $\pbc$, and $\pbd$ (see Methods and Supplementary Information Section 1). For small error rates, this critical benefit-to-cost ratio, which guarantees stability of the all-discriminator population and produces substantial cooperation, is a little larger than $1$, which is just barely stronger than the condition required for the game to be a prisoner's dilemma to begin with. 

In summary, when only discriminators and defectors are present, and the discriminator frequency is higher than $f_Z^*$, they are then destined to fix in the population.
Because discriminators intend to cooperate with everyone they consider good, the cooperation rate at this stable equilibrium is given by $(1 - \act) g$, where $g$ is the proportion of the population considered good, which satisfies $g_Z|_{f_Z = 1}  = g \pgc + (1 - g) \pbd$.
Solving for $g$ yields
\begin{equation}
    \begin{split}
        g & = \frac{\pbd}{1 - \pgc + \pbd}
        \\
        & = \frac{\q(1 - 2\ass) + \ass}{1 - \epsilon + \q(1 - 2\ass) + \ass} =
        \begin{dcases}
            \frac{\ass}{1 - \epsilon + \ass} = \frac{\ass}{2 \ass + \act - 2 \act \ass} & \text{\emph{Shunning}, \emph{Scoring}},
            \\
            \frac{1 - \ass}{2 - \epsilon + \ass} = \frac{1 - \ass}{1 + \act - 2 \act \ass} & \text{\emph{Stern Judging}, \emph{Simple Standing}},
        \end{dcases}
    \end{split}
\end{equation}
which provides an analytical expression for the equilibrium cooperation rate.

Under the social norms \emph{Stern Judging} and \emph{Simple Standing}, this value of $g$ is close to $1$ for small error rates, meaning that most of the population is considered good in a population of discriminators, and so the rate of cooperation is very high. For example, with $\ass = \act = .02$, the value of $g$ is roughly $0.93$, and so $(1 - \act) g \approx 91\%$ of the population will be cooperating at the all-discriminator stable equilibrium. 

As these calculations demonstrate, discriminators enjoy a substantial fitness advantage when information about reputations is fully public ($K=1$).
Public information generates a high level of agreement about reputations, which means that discriminators are likely to reward each other's good behavior by cooperating.
Thus, indirect reciprocity with public information provides a powerful mechanism not only to produce a high level of cooperation but also to protect cooperative individuals from the temptation to become defectors. 

\subsubsection*{Cooperation in a group-structured population}

Even when social interactions occur across an entire well-mixed population, the free flow of reputation information can be disrupted if the population is stratified into gossip groups with potentially different views about reputations.
Different views may be held by different groups even when all groups subscribe to the same social norm of judgment, because of independent observations and independent observational errors. And so partitioning a population into $K>1$ groups is expected to temper or even destabilize the advantage of discriminators, who may no longer agree about the reputations of their interaction partners and thus might engage in unreciprocated cooperation.

\begin{figure}[ht!]
    \begin{center}
    \includegraphics[width=0.8\textwidth]{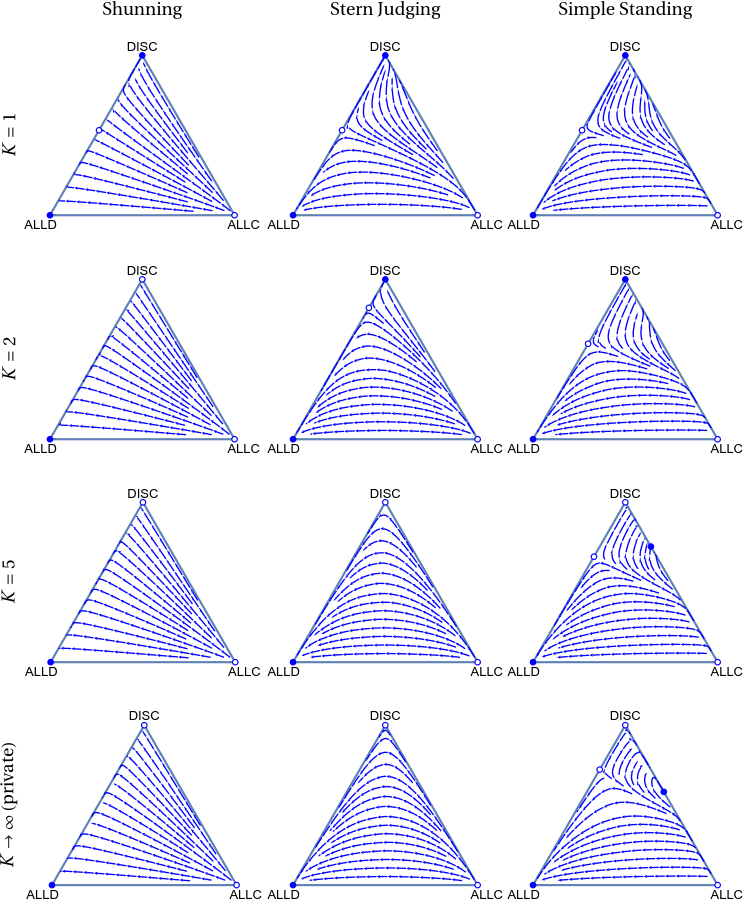}
    \end{center}
    \caption{\small{The dynamics of three competing strategies (cooperate, defect, and discriminate) under three different social norms (columns) and for different numbers $K$ of equally-sized gossip groups (rows). Arrows depict the gradient of selection within the simplex of these three strategies. Open circles indicate unstable equilibria; filled circles indicate stable equilibria.
    With a single gossip group ($K=1$), which is equivalent to public information about reputations \citep{nowak2005evolution,sommerfeld2007gossip,nakamura2011indirect,ohtsuki2009indirect}, there are large basins of attraction to the all-discriminator stable equilibrium, so that stable cooperation occurs under all three social norms.
    As the number of gossip groups $K$ increases, the dynamics rapidly approach those of a model with private assessment \citep{radzvilavicius2021nature} (fourth row), which does not support cooperation in equilibrium under \emph{Shunning} or \emph{Stern Judging}. Note that at $K$ increases, several equilibria change from stable to unstable, reducing the basin of attraction to the discriminator equilibrium; and in the case of \emph{Simple Standing}, a new stable equilibrium is born. In all panels, $b = 2$, $c  = 1$, $\ass = \act = 0.02$.}}
    \label{fig:multi_ternary}
\end{figure}

We study the effects of multiple gossip groups by solving Eq.~\eqref{eq:replicator_mixed} for various numbers of groups $K$ of equal size.
The resulting strategy dynamics under well-mixed copying are shown in the the upper panels of Fig.~\ref{fig:multi_ternary}, for a representative set of typical parameters ($b=2$, $c=1$, $\act=\ass=0.02$).

Under the \emph{Stern Judging} norm, as the number of gossip groups $K$ increases, the location of the unstable equilibrium along the DISC-ALLD edge moves towards the discriminator vertex, which reduces the basin of attraction towards the discriminator equilibrium.
Thus, a smaller portion of initial conditions yields stable cooperation. Moreover, the rate of cooperation at the all-discriminator equilibrium is also reduced (eg, for the example parameters shown in Fig.~\ref{fig:multi_ternary}, it changes from $0.93$ with one group to $0.71$ with two gossip groups).
When $K \geq 3$, the rate of cooperation drops even further, and the all-discriminator equilibrium will eventually cease to be stable altogether. This instability arises because, when there are many gossip groups, it is less likely that discriminators will interact with others who share their views of the rest of the population.

Similar results hold for the \emph{Shunning} social norm. Multiple gossip groups $K>1$ rapidly destabilize cooperative behavior in a population, in fact even more rapidly than under \emph{Stern Judging}. Under \emph{Simple Standing}, the effect of multiple groups is more subtle than for other norms. Increasing the number of gossip groups $K$ still reduces the basin of attraction to a stable equilibrium supporting cooperation. But in this case $K$ also influences the ability of \emph{cooperators} to invade the all-discriminator equilibrium (Supplementary Information Section 2): for sufficiently large $K$, the all-discriminator equilibrium is stable against invasion by defectors but \emph{not} by cooperators, producing a stable equilibrium with a mix of cooperators and discriminators that does not exist for $K=1$ (Fig.~\ref{fig:multi_ternary} for $K\geq5$). In summary, the number of gossip groups has a weak effect on the rate of stable cooperation under \emph{Simple Standing}.

Strategy dynamics under the \emph{Scoring} norm do not depend on the number or relative size of groups (Supplementary Information Section 1), and so we do not present results for \emph{Scoring} here.

\clearpage
\clearpage

\begin{figure}[ht!]
    \begin{center}
    \includegraphics[width=5.8in]{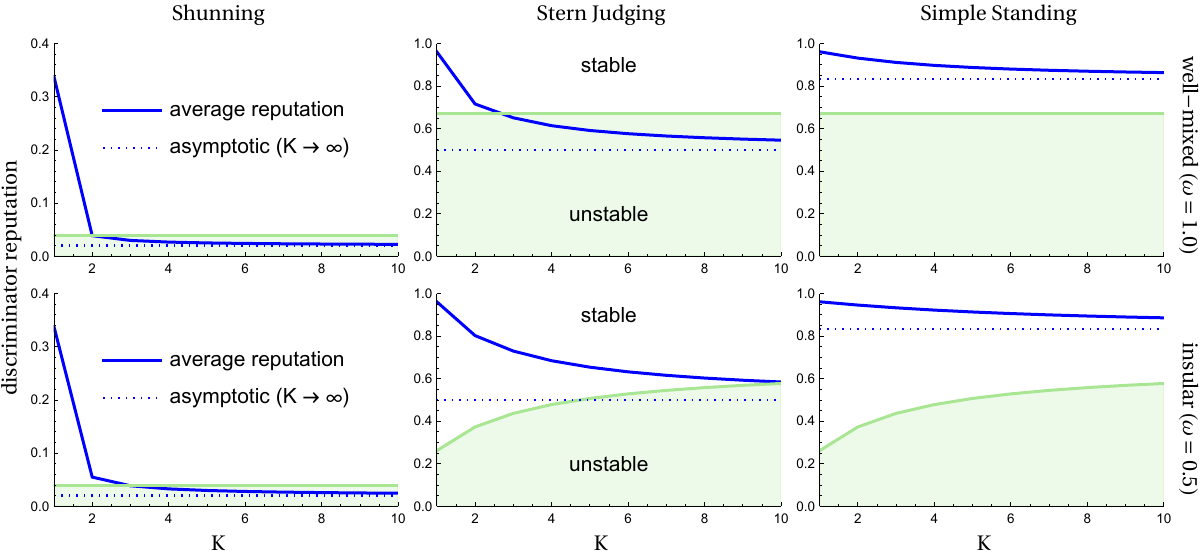}
    \end{center}
    \caption{
    \small{The average reputation $g_Z$
    in a population of discriminators depends on the number $K$ of equally sized gossip groups (solid blue lines). Social interactions are either well-mixed ($\out=1$, top row), or they are biased towards in-group partners ($\out=0.5$, bottom row). The shaded orange region indicates the regime in which discriminators are susceptible to invasion by rare defectors; above this area, discriminators are stable against invasion, so that cooperative behavior is maintained.
    Increasing the number of gossip groups, $K$, rapidly reduces the average reputation in the population, to levels that can destabilize cooperation under \emph{Shunning} or \emph{Stern Judging}; whereas \emph{Simple Standing} supports stable cooperation for arbitrarily many groups $K$.
    Insular social interactions (eg, $\out=0.5$ shown in bottom row) tend to increase the average reputation while also reducing the threshold reputation required to stabilize discriminators against defectors. As a result, in the example shown for \emph{Stern Judging}, the maximal number of gossip groups that support stable cooperation is greater when interactions are partly insular compared to well-mixed.
    The dashed blue line indicates the asymptotic value of
    $g_Z$
    in the limit of many groups, $K \rightarrow \infty$, which is equivalent to a model with private assessment \citep{uchida2013private}.
    In all panels, $b = 2$, $c  = 1$, $\ass = \act = 0.02$.}}
    \label{fig:critical_values}
\end{figure}

We can summarize the effects of multiple gossip groups $K>1$ by analyzing the stability and rate of cooperation at the all-discriminator vertex, $f_Z=1$. Discriminators can resist invasion by defectors only when their fitness exceeds the fitness of a rare defector mutant near the $f_Z=1$ vertex, i.e., when $(b - c) \bar{g}|_{f_Z = 1} > b \bar{g}_Y|_{f_Z = 1}$. This condition can be rewritten as follows:
\begin{equation}
    \begin{split}
    \bar{g}|_{f_Z = 1} & > \frac{\pbd}{1 - \pgd + \pbd - c/b}
    \\
    \therefore \bar{g}|_{f_Z = 1} & > \frac{\q(1 - 2\ass) + \ass}{1 - \ass + \q(1 - 2\ass) + \ass - c/b} =
        \begin{dcases}
            \frac{\ass}{1 - c/b} & \text{\emph{Shunning}, \emph{Scoring}},
            \\
            \frac{1-\ass}{2(1 - \ass) - c/b} & \text{\emph{Stern Judging}, \emph{Simple Standing}}.
        \end{dcases}
    \end{split}
\end{equation}
In Fig.~\ref{fig:critical_values} we plot the average reputation in an all-discriminator population, $\bar{g}_{f_Z = 1}$.
We show that, as $K$ increases, the cooperation rate in such a population decreases below the threshold for discriminator stability under \emph{Shunning} and \emph{Stern Judging}, whereas it remains above this threshold for \emph{Simple Standing}.
And so a sufficiently large number of gossip groups entirely destabilizes cooperation under two of the norms we consider, but it does not destabilize cooperation under \emph{Simple Standing}.

\subsubsection*{The limit of many groups, $K \rightarrow \infty$}
As the number of groups $K$ approaches infinity, we recover the reputation dynamics for discriminators in a population of private assessors (see Supplementary Information Section $3$):
\begin{equation}
    \begin{split}
        g_Z & = g_2 \pgc + (g - g_2) (\pgd + \pbc) + (1 - 2g + g_2) \pbd, 
        \\
        g_2 & = \sum_s f_s g_s^2.
    \end{split}
\end{equation}
These expressions are identical to those derived in \citet{radzvilavicius2019evolution} in the case of no empathy: the three terms of $g_Z$ correspond respectively to the donor and observer agreeing that the recipient is good, disagreeing about the recipient's reputation, and agreeing that the recipient is bad. This result makes intuitive sense because, in the limit of infinitely many information groups, each individual in the population effectively has an independent view from all other individuals--which is equivalent to individuals with private information about reputations.

Fig.~\ref{fig:multi_ternary} also reflects these results. We see that the average reputation of the all-discriminator population, $\bar{g}_{f_Z = 1}$, rapidly approaches the private-assessment limit as the number of groups $K$ increases. Under \emph{Simple Standing}, the asymptotic private-assessment limit still exceeds the reputation value required for discriminators to resist invasion by defectors. This is why, even under private assessment, \emph{Simple Standing} allows discriminators to persist in a sizable region of parameter space: there is a stable equilibrium that consists of a mixture of discriminators and cooperators.

\subsubsection*{Gossip groups of different size}
We also consider a scenario in which a fraction $\nu$ of the population belongs to one large group and the remaining $K-1$ groups each comprise fractions $(1 - \nu)/(K-1)$ of the population. In Supplementary Information Section $3$ we show that, as $K$ approaches infinity, this case reduces to a model with a mixture of individuals who adhere to a public institution (those in the group of size $\nu$) and individuals who act as private assessors (in the remaining groups), which has been previously studied \citep{radzvilavicius2021nature}.

\subsubsection*{Cooperation in insular gossip groups}

If group membership determines not only how an individual views the reputations in the population but also whom they tend to interact with in game play, then differential rates of within- versus between-group interactions could influence the evolution of strategies.

We find that insular social interactions ($\out<1$) tend to mitigate the otherwise destabilizing effects of gossip groups on cooperative behavior. The basic intuition for this phenomenon is simple: gossip groups destabilize cooperation because individuals from different groups may diverge in their reputational views, which lead to unreciprocated cooperation between out-group pairs. But insularity reduces the rate of out-group interactions, so that more interactions occur between individuals holding the same reputational viewpoints, which tends to restore the stability of cooperation in an all-discriminator population.

Insularity facilitates cooperation in two distinct ways. First, insularity increases the average reputations of an all-discriminator population, and, second, it also reduces the threshold reputation required for discriminators to be stable against invasion by defectors (Fig.~\ref{fig:critical_values}). Both of these effects stabilize cooperative behavior, compared to a well-mixed population. The ameliorating effects of insularity are most pronounced for the \emph{Stern Judging} social norm, where strong insularity can preserve stable cooperation with as many as $K=5$ gossip groups, for example. Insularity has a much smaller impact under the \emph{Shunning} norm. Regardless of the social norm, we have derived analytical expressions for the average reputations of discriminators, in equilibrium, and for the threshold reputation required for stability against defectors as function of the insularity parameter $\out$, the error rates, number of gossip groups, and benefits and costs of cooperation (Supplementary Information Section $5$).

\subsubsection*{The evolution of insularity}

We have seen that insularity can defray the destabilizing effects of group structure on cooperation. This raises the question of how insularity itself will evolve, in this setting.
To study this, we first analyzed the effects of a fixed level of insularity on fitness in a group-structured population, finding that more insular populations that prefer in-group social interactions generally enjoy greater mean fitness within each group.
This happens for the simple reason that high insularity increases the rate of interactions with in-group members who are most likely to share reputational views. We have also performed an invasibility analysis to determine whether a  mutant with a higher level of insularity can spread when introduced into a resident population that is less insular. For all norms in which fitness has any dependence on group identity, we find that a more insular mutant can always invade a less insular resident, so that a population will always evolve towards greater insularity.
However, if out-group social interactions are potentially more rewarding than in-group interactions (e.g., $\bout>\bin$), then a population may be be able to resist invasion by insular types, or it may evolve to stable intermediate levels of insularity (Supplementary Information Section $5$).
In general, if insularity is a transmissible phenotype, we expect groups to evolve towards increasing insularity, in the absence of countervailing mechanisms such as highly rewarding out-group interactions.

\subsubsection*{Strategic imitation restricted by population structure}

Just as groups that restrict the flow of reputation information may also restrict the choice of interaction partners for game play (insularity), groups may also restrict the choice of comparison partners for strategic imitation. We present an analysis of this situation in the Supplement (SI Section 7). We show that strategic imitation restricted by group structure has the potential to stabilize cooperation in one group, provided it is already fixed (perhaps exogenously, by policy) in another group.

\section{Discussion}
We have developed a game-theoretic model for cooperation conditioned on reputations that accounts for a population stratified into groups. Each group holds its own viewpoints about reputations, and each group subscribes to a potentially different norm of moral assessment. This model allows us to analyze competition between norms of judgment, as individuals decide to change group membership in an attempt to increase payoff. We find bistability: the population will inevitably converge on a single social norm, and which norm prevails depends on their initial frequencies. Importantly, while norms cannot generally invade when they are vanishingly rare, some norms -- especially those that value defection against bad actors -- will win out even when they initially comprise a minority of the population. In particular, \emph{Stern Judging}, which is socially optimal, emerges as the strongest competitor among norms.

Our results on norm competition help resolve an outstanding gap in the theory of cooperation mediated by reputations. Most research on indirect reciprocity assumes by fiat that everyone shares the same social norm for judging reputations, and that reputations are common knowledge. Allowing for incomplete information and competing norms is complicated, because it requires keeping track of which norm each individual (or each group) adheres to. One prior approach in the literature side-steps this difficulty by stipulating a multi-scale model of competing groups, with individual-level selection on behavior and group-level selection on norms \citep{pacheco2006stern, chalub2006norms}. In a multi-level analysis, groups adhering to different social norms accumulate payoffs and then compete with each other at the group level by playing a hawk-dove game, with victorious groups more likely to ``reproduce'' and replace other groups. Consistent with our own findings, simulations of such multi-level competition have revealed \emph{Stern Judging} to be the winning norm \citet{pacheco2006stern}. But the multi-level formulation of norm competition differs fundamentally from ours, in that no individual can unilaterally decide to change their norm for assessing reputations; rather, an entire group is instantaneously replaced by a different group that holds a different norm. Such approaches based on group-level competition provide limited intuition for evolutionary dynamics, because a trait that is beneficial for an entire group, such as group-wide adoption of a new social norm, may nonetheless be unable to proliferate through individual-level imitation and learning. 

Two prior studies have modeled competition among individuals who adhere to different norms of reputation assessment, without appealing to instantaneous group-level adoption of a new norm. \citet{uchida2010competition} analyzed a model in which individuals could choose between \emph{Simple Standing} or \emph{Stern Judging}. They found that coexistence between these norms was possible, in sharp contrast to our bistability result that eventually one norm will prevail. The reason for this discrepancy is that \citet{uchida2010competition} neglect the possibility of assessment errors ($\ass = 0$). But population dynamics without assessment errors are both unrealistic and structurally unstable when information is partly private \citep{uchida2013private}. For example, when $\ass = 0$ in a population of discriminators, everyone chooses to cooperate with everyone else, nobody's reputation depends on which norm they follow, and nobody can improve their fitness by switching norms. And so a model without any errors is a pathological boundary case, because fitness differences only arise in the presence of cooperators (ALLC).
A related study by \citet{uchida2018ecosystem} likewise finds long-term coexistence between multiple norms, but it, too, neglects errors of assessment, as well as errors of execution. Evolutionary dynamics are qualitatively different when there is some chance of committing an error while assessing reputations. In particular, when $\ass>0$, \emph{Stern Judgers} are intolerant of disagreement with the out-group and thus engage in less unreciprocated cooperation; this raises their fitness substantially, and it allows them to dominate in a population of discriminators and drive other norms to extinction (Fig.~\ref{fig:group_switch_b}).
True ``competition'' between norms only makes sense when there is differential behavior between adherents of different norms -- which only arises in the presence of errors. Consequently, studying norm competition requires that we account for errors during assessment, which is a more realistic assumption in any case.

Our model of population stratification produces an orthogonal, but complementary set of results when group membership is kept fixed and individuals attempt to increase payoff by imitating strategies rather than group membership. Even when all groups adopt the same norm of judgment, we find that population stratification decreases the prospects for cooperation overall, due to the potential for reputation disagreement as groups make independent moral judgments. These destabilizing effects on cooperation grow rapidly with the number of groups. Cooperation can be restored, however, when individuals are strongly insular and favor in-group social interactions. Thus, insularity -- even without any intrinsic bias towards in-group favoritism -- can preserve some degree of cooperation, but this comes at the cost of social isolation or tribalism.

Our analysis, therefore, predicts that cooperative societies are less likely to flourish when many distinct groups form independent moral judgments; such societies are destined to become either wholly uncooperative or tribal (unless out-group interactions are more beneficial). This account may help explain the rise of powerful, centralized entities for reputation assessment -- namely, states -- as the result of individuals maximizing their own fitness. Historically, societies have gradually shifted from competing and often overlapping territorial claims over land, such as between feudal rulers, towards recognizing singular entities as having the sole right to govern a given territory. Competing legal systems, such as religious authorities, have likewise gradually yielded to secular powers. \footnote{It is also interesting to note, in the context of political parties, that $K=2$ groups can still maintain some level of cooperation in our analysis, whereas $K \geq 3$ independent groups are highly deleterious.} When too many different entities claim the right to determine the standing of individuals in a population, the result can be ambiguity, uncertainty, and often conflict. And so a state, by asserting a ``monopoly on violence'' \citep{weber2015weber}, reduces ambiguity and allows individuals to prosper by aligning their behavior with the mandates of a singular, powerful entity. In this context, a natural extension of our work would consider the dynamics of political and affective polarization \citep{levendusky2009,iyengar2012,mason2016,mason2018,levin2021dynamics}, as they might be shaped by a tug of war between a tendency towards tribalism and a pull towards convergence on one social norm.

Our work contributes to a growing body of literature establishing \textit{Stern Judging} as a uniquely powerful social norm for behavioral judgment. Our analysis reveals that \textit{Stern Judging} enables a group to maximize in-group cooperation without engaging in unreciprocated out-group contributions, thereby making it attractive to individuals who seek to maximize payoff by switching norms. Consequently, we might expect moral norms like \textit{Stern Judging} to dominate in human societies. And yet there is substantial cross-cultural variation in how reputational judgments actually occur \citep{bouwmeester2017registered, henrich2001search}. Even within a society, there can be some degree of norm variation. The source of this variation in norms remains an active area of research across several fields. Within the context of the theory of indirect reciprocity, norm variation may persist if there are costs to switching social norms or switching group identity; or when out-group interactions are more rewarding (even if more risky) than in-group interactions; or if norms of moral judgment are linked to other social traits that experience different selection pressures in different groups.
Future research may help to resolve these open questions in the context of group-structured models, as well as empirically determine whether, and under what conditions, individuals and groups are willing to amend their social norms.

\bigskip
\subsubsection*{Acknowledgements}
JBP and TAK acknowledge support from the John Templeton Foundation (grant \#62281), the Simons Foundation (Math+X grant to the University of Pennsylvania), and the David \& Lucile Packard Foundation.

\clearpage

\begin{small}
\section*{Methods}

We consider a population of $N$ individuals who play a series of one-shot pairwise donation games \citep{rapoport1965prisoner} with each other.
Every round, each individual plays the donation game twice with everyone else, once as a donor and once as a recipient.
This game provides a minimal model of a social dilemma, in which a donor may either cooperate, paying a cost $c$ to convey a benefit $b > c$ to the recipient, or defect, paying no cost and conveying no benefit.
Each donor chooses an action based on their behavioral strategy: cooperators (denoted ALLC or $X$) always cooperate, defectors (ALLD, $Y$) always defect, and discriminators (DISC, $Z$) cooperate with those they consider to have good reputations and defect with those they consider to have bad reputations.
Following the round of all pairwise game interactions, the players update their views of each others' reputations, and then they update their strategies according to payoff-biased imitation, as described below.

\subsection{Reputations and gossip groups}

Each player belongs to one of $K$ distinct and disjoint ``gossip groups'', which comprise fractions $\nu_1, \nu_2, \ldots \nu_K$ of the total population.
An individual's group membership determines their view of the reputations of the other players: each group has a shared, consensus view of the reputation of every player in the population, but different groups may have different views of individuals' reputations.
This model characterizes a situation where individuals transmit information about reputations to other members of their group via rapid gossip \citep{sommerfeld2007gossip}, or, alternatively, each group has its own ``institution'' \citep{radzvilavicius2021nature} that broadcasts reputation assessments to the group.

Each round, everyone plays the pairwise donation game with everyone else, and then each group updates their (consensus) view of the reputation of each individual in the population, as follows.
For a given individual, the group samples a single random interaction from that round in which that individual acted as a donor.
Depending on the donor's action, the group's view of the recipient's reputation, and the group's social norm, the donor is assigned a new reputation by the group.
We consider a generalized social norm in which:
\begin{enumerate}
    \item cooperation with an individual with a good reputation is considered good,
    \item defection against an individual with a good reputation is considered bad.
    \item cooperation with an individual with a bad reputation is considered good with probability $\p$,
    \item defection against an individual with a bad reputation is considered good with probability $\q$.
\end{enumerate}
The social norm is thus parameterized by two probabilities, $\p$ and $\q$.
When $(\p,\q)=(0,1)$, for example, we recover the classic \emph{Stern Judging} norm \citep{kandori1992social,pacheco2006stern}, which stipulates that a donor interacting with a recipient of bad standing must defect to earn a good standing. Setting $(\p,\q) = (0,0),(1,0),$ or $(1,1)$ yields the other well-studied social norms \emph{Shunning}, \emph{Scoring}, and \emph{Simple Standing} \citep{sugden1986economics}. 

\subsection{Errors}
We include two types of errors: errors in social interaction and errors in reputation assessment. First, an individual who intends to cooperate may accidentally defect, which we call an \emph{execution error}; this occurs with probability $\act$.
Individuals who intend to defect never accidentally cooperate \citep{sasaki2017evolution}.
Second, an observer may erroneously assign an individual the wrong reputation, which we call an \emph{assessment error}; this occurs with probability $\ass$.
We also define the related parameter
\begin{equation}
    \epsilon = (1 - \act)(1 - \ass) + \act \ass,
\end{equation}
which quantifies the chance that an individual who intends to cooperate with someone of good reputation successfully does so and is correctly assigned a good reputation (first term) or accidentally defects but is erroneously assigned a good reputation nonetheless (second term).

Given the social norm and these error rates, we can characterize how a donor is assessed in terms of four probabilities:
\begin{itemize}
    \item $\pgc$, the probability that a donor who \emph{intends to cooperate} with a \emph{good} recipient  will be assigned a good reputation.
    \item $\pgd$, the probability that a donor who \emph{intends to defect} with a \emph{good} recipient  will be assigned a good reputation.
    \item $\pbc$, the probability that a donor who \emph{intends to cooperate} with a \emph{bad} recipient will be assigned a good reputation.
    \item $\pbd$, the probability that a donor who \emph{intends to defect} with a \emph{bad} recipient will be assigned a good reputation.
\end{itemize}

For an arbitrary social norm $(\p,\q)$ and error rates $\ass$ and $\act$, we can derive general expressions for these four probabilities that characterize reputation assessment (see Supplementary Information Section 1):
\begin{equation}
    \begin{split}
        \pgc & = \epsilon
        \\
        \pgd & = \ass
        \\
        \pbc & = \p(\epsilon - \ass) + \q(1 - \epsilon - \ass) + \ass
        \\
        \pbd & = \q(1 - 2\ass) + \ass.
    \end{split}
\end{equation}

\subsection{Mean-field reputation dynamics}

In the limit of a large population size, we consider an individual's expected reputation over many rounds of play, prior to any strategic changes in the population.
Let $g_s^{I,J}$ denote the probability that an individual with strategy $s$ in group $I$ has a good reputation in the eyes of an individual in group $J$.
(The first superscript index denotes ``who'', the donor; the second index denotes ``in whose eyes'', the observer.)
Furthermore, let $f_s^I$ be the frequency of individuals in group $I$ who have strategy $s$, so that $\sum_s f_s^I  = 1$.
We define
\begin{equation}
    g^{I,J} = \sum_{s \in \{X,Y,Z\}} f_s^I g_s^{I,J},
\end{equation}
which represents the expected fraction of individuals in group $I$ who are seen as good from the point of view of someone in group $J$. Note that the summation index $s$ in this expression, and all other such expressions below, denotes a sum over strategic types, namely $s \in \{X,Y,Z\}$.
We further define
\begin{equation}
    g^{\bullet,J} = \sum_{L=1}^K \nu_L g^{L,J},
\end{equation}
which represents the fraction of individuals in the whole population whom an individual in group $J$ sees as good.
Here, and throughout this presentation, capital letter summation indices (such as $L$, $J$, or $I$) denote a sum over all groups $\in \{1,2, \ldots, K\}$.

In the Supplementary Information, we show that the reputations associated with different strategic types satisfy
\begin{equation}
    \begin{split}
        g_X^{I,J} & = g^{\bullet,J} \pgc + (1 - g^{\bullet,J}) \pbc,
        \\
        g_Y^{I,J} & = g^{\bullet,J} \pgd + (1 - g^{\bullet,J}) \pbd,
        \\
        g_Z^{I,J} & = \delta_{I,J} \left[ g^{\bullet,J} \pgc + (1 - g^{\bullet,J}) \pbd \right]
        \\
        & \ \ \ + (1 - \delta_{I,J}) \left[ \Gij{I}{J} \pgc + (g^{\bullet,J} - \Gij{I}{J}) \pgd + (g^{\bullet,I} - \Gij{I}{J}) \pbc + (1 - g^{\bullet,J} - g^{\bullet,I} + \Gij{I}{J}) \pbd \right],
    \end{split}
\end{equation}
where the term $\Gij{I}{J}$ is defined as
\begin{equation}
    \Gij{I}{J} = \sum_{L=1}^K \nu_L \sum_{s \in \{X,Y,Z\}} f_s g_s^{L,I} g_s^{L,J},
\end{equation}
which reflects the probability that distinct groups $I \ne J$ agree that a randomly chosen individual in any group other than $I$ has a good reputation.

\subsection{Payoffs}
Payoffs accrue to each individual based on their performance in pairwise interactions.
An individual acquires a payoff $b$ for each interaction either with a cooperator ($X$) or with a discriminator ($Z$) who sees them as good.
A cooperator pays cost $c$ in each interaction, and a discriminator pays cost $c$ in each interaction with someone whom they see as good.
Thus, the average payoff for each of the three strategic types in an arbitrary group $I$ is
\begin{equation}
    \begin{split}
        \Pi_X^I & = (1 - \act) \left[ b \sum_{J=1}^K \nu_J (f_X^J + f_Z^J g_X^{I,J}) - c \right]
        \\
        \Pi_Y^I & = (1 - \act) \left[ b \sum_{J=1}^K \nu_J (f_X^J + f_Z^J g_Y^{I,J}) \right]
        \\
        \Pi_Z^I & = (1 - \act) \left[ b \sum_{J=1}^K \nu_J (f_X^J + f_Z^J g_Z^{I,J}) - c g^{\bullet, I} \right].
    \end{split}
    \label{eq:fit_multi_groupsB}
\end{equation}
Note that these payoffs are averaged over all pairwise interactions, i.e., they are already divided by the population size $N$.

\subsection{Insular social interactions}

Aside from restricting the flow of reputation information, group structure in a population may also influence partner choice for social interactions (game play). We extend the model to consider the case when individual prefer social interactions with in-group members, which we call \emph{insularity}.
We introduce the parameters $\out^{I,J} = \out^{J,I}$, which denote the probability that a potential interaction between members of groups $I$ and $J$ actually occurs.
In each round, each individual in the population considers a possible dyadic interaction with each other member of the population. If one member is from group $I$ and the other from group $J$, the interaction occurs with probability $\out^{I,J} \leq 1$.
In the Supplementary Information, we show that the resulting reputations then satisfy
\begin{equation}
    \begin{split}
        g_X^{I,J} & = \g^{I,J} \pgc^J + (1 - \g^{I,J}) \pbc^J,
        \\
        g_Y^{I,J} & = \g^{I,J} \pgd^J + (1 - \g^{I,J}) \pbd^J,
        \\
        g_Z^{I,J} & = \delta_{I,J} \left[ \g^{I,J} \pgc^J + (1 - \g^{I,J}) \pbd^J \right]
        \\
        & \ \ \ + (1 - \delta_{I,J}) \left[ \G^{I,J} (\pgc^J - \pgd^J - \pbc^J + \pbd^J) + \g^{I,J}(\pgd^J - \pbd^J) + \g^{I,I}(\pbc^J - \pbd^J) + \pbd^J \right],
    \end{split}
\end{equation}
in which
\begin{equation}
    \g^{I,J} = \frac{1}{\M^I} \sum_L \nu_L \out^{L,I} g^{L,J}
\end{equation}
is the probability that a member of group $J$ observes a member of group $I$ performing an action that yields a good reputation in $J$'s eyes (analogous to $g^{\bullet,J}$),
\begin{equation}
    \G^{I,J} = \frac{1}{\M^I} \sum_L \nu_L \out^{L,I} \sum_i f_s g_s^{L,I} g_s^{L,J}
\end{equation}
is the probability that a member of group $J$ observes a member of group $I$ interacting with a recipient in a non-$I$ group whom both $I$ and $J$ consider to be good (analogous to $G^{I,J}$), and
\begin{equation}
    \M^I = \sum_L \nu_L \out^{I,L}
\end{equation}
is a normalizing constant representing the total fraction of possible interactions that an $I$ group member actually engages in.
Setting all $\out^{I,J} = 1$ yields $\g^{I,J} = g^{\bullet,J}$, $\G^{I,J} = G^{I,J}$, and $\M^I = 1$.
In Supplementary Information Section $5.1$, we derive corresponding equations for fitnesses. We also derive mean fitnesses in \emph{groups} with different levels of insularity (SI section 5.7), and we consider the behavior of individuals with differing levels of insularity (SI section 5.8).

\subsection{Updating}
Each round, after all pairwise games have occurred and all reputations have been updated, a randomly chosen individual considers updating either their strategy or their group identity.
They compare their payoff, averaged over all games in which they have played, to that of another random individual in the population.
We consider two possibilities:
\begin{enumerate}
    \item imitating \emph{group membership}, in which an individual decides whether to change their group membership to that of their comparison partner.
    \item imitating \emph{behavioral strategy}, in which an individual decides whether to adopt the behavioral strategy of their comparison partner.
\end{enumerate}
If the focal individual has strategy $s$ and is a member of group $I$, and their comparison partner has strategy $s^\prime$ and is in group $J$, they copy their partner's comparison trait (either group identity or behavioral strategy) with a probability given by the Fermi function
\begin{equation}
    \phi(\Pi_s^I,\Pi_{s^\prime}^J) = \frac{1}{1 + \exp\left[ \beta (\Pi_s^I - \Pi_{s^\prime}^J) \right] }.
\end{equation}
Here, $\beta$ is a parameter known as the strength of selection \citep{traulsen2007pairwise,traulsen2010human}.
In the limit of small $\beta$ and large population size $N \to \infty$, the process of pairwise game-play, reputation assessment, and updating can be described by a deterministic replicator equation \citep{hofbauer1998} after an appropriate re-scaling of time (see derivation in Supplementary Information Section 8).

\subsection{Copying group membership}
When individuals copy each others' group membership, the resulting group sizes evolve according to the replicator equation
\begin{equation}
    \begin{split}
        \dot{\nu}_I & = \nu_I (\bar{\Pi}^I - \bar{\Pi}), \text{with}
        \\
        \bar{\Pi}^I & = \sum_s f_s \Pi_s^I,
        \\
        \bar{\Pi} & = \sum_J \nu_J \sum_s f_s \Pi_s^J.
    \end{split}
    \label{eq:group_switching_general}
\end{equation}
For most of our analysis of competing groups, we assume that all individuals are fixed for the discriminator strategy, meaning that they attend to their co-player's reputation when choosing whether to donate or not. Eq.~\eqref{eq:group_switching_general} then becomes simply
\begin{equation}
    \begin{split}
    \dot{\nu}_I & = \nu_I (\Pi_Z^I - \bar{\Pi}), \text{with}
    \\
    \bar{\Pi} & = \sum_J \nu_J \Pi_Z^J.
    \end{split}
    \label{eq:group_switching}
\end{equation}
In this case we can write $\Pi^I$, omitting both the bar (since there is no need to take an average) and the $Z$ subscript (since it is understood that the entire population has strategy $Z$).

We analyze the case of $K = 2$ competing groups. The two groups may follow different social norms for making consensus reputational judgments, so that group $I$ uses group-specific probabilities $\pgc^I, \pgd^I, \pbc^I, \pbd^I$ when assigning reputations.
We also consider the dynamics of group sizes when social interactions are insular, in which $\out^{I,J} = \delta_{I,J} + (1 - \delta_{I,J})\out$, i.e., interactions between in-group members happen with probability $1$, but interactions between out-group members happen with probability $0 \leq \out \leq 1$.
Finally, we develop and numerically solve equations for third-order social norms, including the remaining six of the so-called ``leading eight'' norms \citet{ohtsuki2006leading}.

\subsection{Copying behavioral strategies}

The replicator dynamics for behavioral strategies depend sensitively on how individuals choose their comparison partners. When group membership is fixed, we consider two possibilities for behavioral imitation. We focus on the first possibility in the main text, and we defer the second possibility to the Supplementary Information (Section 7):
\begin{enumerate}
    \item \emph{well-mixed strategic imitation}, in which an individual is equally likely to choose any other individual as a comparison partner.
    \item \emph{disjoint strategic imitation}, in which an individual must choose a member of their in-group as a comparison partner.
\end{enumerate}
In Supplementary Information Section 8.6, we also consider a more general model in which individuals choose members of their in-group with probability $1 - m$ and choose a random member of the population (irrespective of group membership) with probability $m$, and we show that this model reduces to the models above in the limits $m \to 1$ and $m \to 0$, respectively.

\subsubsection{Well-mixed strategic imitation}
If an individual in group $I$ is equally likely to choose anyone in the population as a comparison partner, then differences in strategy frequencies between groups do not persist; they converge to a value $f_s$ that is common to all groups, as we show in the Supplementary Information.
We have the resulting replicator equation for the frequencies of strategic types over time:
\begin{equation}
    \begin{split}
        \dot{f}_s & = f_s \sum_J \left(\nu_J [ \Pi_s^J - \bar{\Pi}^J ] \right) = f_s \sum_J \nu_J \Pi_s^J - \bar{\Pi}, \text{with}
        \\
        \bar{\Pi} & = \sum_J \nu_J \sum_i f_s^J \Pi_s^J.
    \end{split}
    \label{eq:replicator_mixed}
\end{equation}
Because the strategy frequencies cannot vary by group, the quantity that ultimately determines the change in the frequency of each strategy is the group-averaged fitness
\begin{equation}
    \bar{\Pi}_s = \frac{\sum_J \nu_J f_s^J \Pi_s^J}{\sum_J \nu_J f_s^J} = \sum_J \nu_J \Pi_s^J.
\end{equation}
As we show in the Supplementary Information, this formulation allows us to study the time-evolution and stability of strategies in terms of the average reputations,
\begin{equation}
    \bar{g}_s = \sum_I \sum_J \nu_I \nu_J g_s^{I,J},
\end{equation}
which represents the probability that a randomly chosen member of the population considers a random individual following strategy $i$ to have a good reputation. By averaging over groups and leveraging the fact that strategy frequencies do not differ by group, we can remove the fitness' dependence on an individual's group membership $I$.
Eq.~\eqref{eq:fit_multi_groupsB} thus simplifies to
\begin{equation}
    \begin{split}
        \Pi_X & = (1 - \act) \left[ b (f_X + f_Z \bar{g}_X) - c \right]
        \\
        \Pi_Y & = (1 - \act) \left[ b (f_X + f_Z \bar{g}_Y) \right]
        \\
        \Pi_Z & = (1 - \act) \left[ b (f_X + f_Z \bar{g}_Z) - c \bar{g} \right].
    \end{split}
    \label{eq:fit_well_mixed}
\end{equation}
We further show in the Supplementary Information that the average reputations of cooperators and defectors, $\bar{g}_X$ and $\bar{g}_Y$, can be expressed simply as
\begin{equation}
    \begin{split}
        \bar{g}_X & = \bar{g} \pgc + (1 - \bar{g}) \pbc
        \\
        \bar{g}_Y & = \bar{g} \pgd + (1 - \bar{g}) \pbd,
    \end{split}
\end{equation}
whereas the form of $\bar{g}_Z$ depends sensitively on the group structure of the population.

\subsubsection{Group-structured strategic imitation}
If an individual in group $I$ chooses only other individuals in group $I$ as potential comparison partners, then the frequency of strategy $i$ in group $I$ changes over time according to the following replicator equation:
\begin{equation}
    \begin{split}
        \dot{f}_s^I & = f_s (\Pi_s^I - \bar{\Pi}^I), \text{with}
        \\
        \bar{\Pi}^I & = \sum_i f_s \Pi_s^I.
    \label{eq:replicator_solo}
    \end{split}
\end{equation}
Note that, in this case, even though strategic imitation occurs only within each group, game play and payoff accumulation occur among all members of the population, so strategy frequencies are not independent across groups. In the main text we focus on the case of well-mixed strategic imitation, deferring to the Supplement an analysis of disjoint strategic imitation.

\subsection{Copying both behavioral strategy and group membership}

Finally, we consider an extension of our model in which an individual chooses anyone in the population as a comparison partner (i.e., well-mixed imitation) but may copy \emph{either} their partner's group membership (with probability $\tau$) \emph{or} their behavioral strategy (with probability $1 - \tau$).
We show that group sizes and strategy frequencies then co-evolve according to the equations
\begin{equation}
    \begin{split}
        \dot{\nu}_I & = \nu_I \tau (\bar{\Pi}^I - \bar{\Pi}),
        \\
        \dot{f}_s & = f_s (1 - \tau) (\bar{\Pi}_s - \bar{\Pi}),
    \end{split}
    \label{eq:copying_both}
\end{equation}
which we derive in Supplementary Information Section 8.6.

\end{small}

\clearpage
\bibliographystyle{apalike}
\bibliography{main}

\end{document}


\maketitle

\tableofcontents

\section{Generalized social norms}
\noindent In this section, we derive the expressions for $\pgc$, $\pgd$, $\pbc$ and $\pbd$ used in the main text.

We begin by generalizing the ``big four'' social norms, which occur as special cases of a two-parameter family of norms. We suppose that cooperation with a bad individual yields a good reputation with probability $\p$ (barring errors) and defecting with a bad individual yields a good reputation with probability $\q$ (again barring errors).
We recover \textit{Stern Judging}, \textit{Simple Standing}, \textit{Scoring}, and \textit{Shunning} with $(\p,\q) = (0, 1), (1, 1), (1, 0), (0, 0)$, respectively.

In the presence of errors of execution and assessment, an individual can obtain a good reputation in the following ways.
They may be observed:
\begin{enumerate}
    \item interacting with an individual with a good reputation and intending to cooperate.
    \begin{enumerate}
        \item With probability $1 - \act$, they successfully cooperate. With probability $1 - \ass$, they are successfully assigned a good reputation.
        \item With probability $\act$, they accidentally defect.
        With probability $\ass$, they are accidentally assigned a good reputation.
    \end{enumerate}
    Thus,
    \begin{equation*}
        \pgc = (1 - \act)(1 - \ass) + \act \ass = \epsilon.
    \end{equation*}
    \item interacting with an individual with a good reputation and intending to defect.
    This is always considered a bad action, so such an individual can only achieve a good reputation on accident.
    Thus,
    \begin{equation*}
        \pgd = \ass.
    \end{equation*}
    \item interacting with an individual with a bad reputation and intending to cooperate.
    \begin{enumerate}
        \item With probability $1 - \act$, they successfully cooperate.
        With probability $\p$, this is considered a ``good reputation'' action.
        With probability $1 - \ass$, they are successfully assigned a good reputation.
        \item With probability $1 - \act$, they successfully cooperate.
        With probability $1 - \p$, this is considered a ``bad reputation'' action.
        With probability $\ass$, they are accidentally assigned a good reputation.
        \item With probability $\act$, they accidentally defect.
        With probability $\q$, this is considered a ``good reputation'' action.
        With probability $1 - \ass$, they are successfully assigned a good reputation.
        \item With probability $\act$, they accidentally defect.
        With probability $1 - \q$, this is considered a ``bad reputation'' action.
        With probability $\ass$, they are accidentally assigned a good reputation.
    \end{enumerate}
    Thus, the total probability is 
    \begin{equation}
        \begin{split}
            \pbc & = (1 - \act)[\p(1 - \ass) + (1 - \p)\ass] + \act[\q(1 - \ass) + (1 - \q)\ass]
            \\
            & = (1 - \act)[\p - 2\p\ass + \ass] + \act[\q - 2\q\ass + \ass]
            \\
            & = \p - 2\p\ass + \ass - \p\act + 2\p\act\ass - \act\ass + \q\act - 2\q\act\ass + \act\ass
            \\
            & = \p(1 - 2\ass - \act + 2\ass\act) + \q(\act - 2\act\ass) + \ass
            \\
            & = \p(\epsilon - \ass) + \q(1 - \epsilon - \ass) + \ass.
        \end{split}
        \label{eq:coop}
    \end{equation}
    \item interacting with an individual with a bad reputation and intending to defect.
    \begin{enumerate}
        \item With probability $\q$, this is considered a ``good reputation'' action.
        They defect and successfully obtain a good reputation with probability $1 - \ass$.
        \item With probability $1 - \q$, this is considered a ``bad reputation'' action.
        They defect and accidentally obtain a good reputation with probability $\ass$.
    \end{enumerate}
    Thus, the total probability is
    \begin{equation}
        \pbd = \q(1 - \ass) + (1 - \q)\ass = \q(1 - 2\ass) + \ass.
        \label{eq:defect}
    \end{equation}
\end{enumerate}
We recover the traditional four social norms in the following limits:
\begin{enumerate}
    \item when $(\p, \q) = (0, 1)$ (stern judging), equation \eqref{eq:coop} becomes $1 - \epsilon$ and equation \eqref{eq:defect} becomes $1 - \ass$.
    \item when $(\p, \q) = (1, 1)$ (simple standing), equation \eqref{eq:coop} becomes $1 - \ass$ and equation \eqref{eq:defect} becomes $1 - \ass$.
    \item when $(\p, \q) = (1, 0)$ (scoring), equation \eqref{eq:coop} becomes $\epsilon$ and equation \eqref{eq:defect} becomes $\ass$.
    \item when $(\p, \q) = (0, 0)$ (shunning), equation \eqref{eq:coop} becomes $\ass$ and equation \eqref{eq:defect} becomes $\ass$.
\end{enumerate}
The values of $P_{BC}, P_{GC}, P_{GD}$, and $P_{BD}$ for these four norms are summarized in SI Table \ref{tab:reputation_probabilities}.

\begin{table}[!ht]
    \centering
    \begin{tabular}{|c|c|c|c|c|}
        \hline
        observer view of recipient & good & good & bad & bad
        \\
        donor intent & cooperate & defect & cooperate & defect
        \\
        good reputation probability & $\pgc$ & $\pgd$ & $\pbc$ & $\pbd$
        \\
        \hline
        general expression & $\epsilon$ & $\ass$ & $\p(\epsilon - \ass) + \q(1 - \epsilon - \ass) + \ass$ & $\q(1 - 2\ass) + \ass$
        \\
        \textit{Shunning}  ($p = 0, q = 0$) & $\epsilon$ & $\ass$ & $\ass$ & $\ass$
        \\
        \textit{Stern Judging} ($p = 0, q = 1$) & $\epsilon$ & $\ass$ & $1 - \epsilon$ & $1 - \ass$
        \\
        \textit{Scoring} ($p = 1, q = 0$) & $\epsilon$ & $\ass$ & $\epsilon$ & $\ass$
        \\
        \textit{Simple Standing} ($p = 1, q = 1$)  & $\epsilon$ & $\ass$ & $1 - \ass$ & $1 - \ass$
        \\
        \hline
    \end{tabular}
    \caption{
    \small{Probability that an observer will assign a donor a good reputation based on the donor's action and the observer's view of the recipient, under various social norms.
    Here, $\epsilon = (1 - \ass)(1 - \act) + \ass \act$ is the probability that an individual who intends to cooperate with a recipient who has a good reputation is ultimately themselves assigned a good reputation.
    They may either successfully cooperate and be correctly assigned a good reputation (first term) or accidentally defect and be wrongly assigned a good reputation (second term).}}
    \label{tab:reputation_probabilities}
\end{table}

\subsection{Reputation dynamics for cooperators, defectors, and discriminators}
Given the expressions for $P_{BC}, P_{GC}, P_{GD}$, and $P_{BD}$ derived above, we now consider what portion of each strategic type will be assigned a good reputation, and by whom.
We begin by defining
\begin{equation}
    \begin{split}
        g^{I,J} & = \sum_s f_s^I g_s^{I,J},
        \\
        g^{\bullet,J} & = \sum_L \nu_L g^{L,J}.
    \end{split}
\end{equation}

A cooperator in group $I$ can be assigned a good reputation in the eyes of group $J$ in two ways.
$J$ can observe the $I$ cooperator's interaction:
\begin{enumerate}
    \item with someone group $J$ sees as good (probability $g^{\bullet,J}$); the $I$ member intends to cooperate, which yields a good reputation with probability $\pgc$.
    \item with someone group $J$ sees as bad (probability $1 - g^{\bullet,J}$); the $I$ member intends to cooperate, which yields a good reputation with probability $\pbc$.
\end{enumerate}
We thus have
\begin{equation*}
    g_X^{I,J} = g^{\bullet,J} \pgc + (1 - g^{\bullet,J}) \pbc = g^{\bullet,J}(\pgc - \pbc) + \pbc.
\end{equation*}
Similar reasoning for defectors yields
\begin{equation*}
    g_Y^{I,J} = g^{\bullet,J} \pgd + (1 - g^{\bullet,J}) \pbd = g^{\bullet,J} (\pgd - \pbd) + \pbd.
\end{equation*}
Discriminators vary their behavior according to the reputation of the recipient, but discriminators in different groups are not guaranteed to have the same views of each recipient's reputation.
Thus, discriminators will be viewed differently by their in-group versus their out-group.
A discriminator in group $I$ can gain a good reputation in the eyes of group $I$ (their in-group)
in two ways.
$I$ can observe the $I$ discriminator's interaction:
\begin{enumerate}
    \item with someone group $I$ sees as good (probability $g^{\bullet,I}$); the $I$ discriminator intends to cooperate, which yields a good reputation with probability $\pgc$.
    \item with someone group $I$ sees as bad (probability $1 - g^{\bullet,I}$); the $I$ discriminator intends to defect, which yields a good reputation with probability $\pbd$.
\end{enumerate}
A discriminator in group $I$ can gain a good reputation in the eyes of group $J \neq I$ (their out-group) in four ways.
$J$ can observe the $I$ discriminator's interaction:
\begin{enumerate}
    \item with someone in an arbitrary group $L$ following strategy $s$ (probability $\nu_L f_s^L$) whom $I$ sees as good (probability $g_s^{L,I}$) and whom $J$ also sees as good (probability $g_s^{L,J}$); the $I$ discriminator intends to cooperate, which yields a good reputation with probability $\pgc$.
    \item with someone in an arbitrary group $L$ following strategy $s$ (probability $\nu_L f_s^L$) whom $I$ sees as bad (probability $1 - g_s^{L,I}$) but whom $J$ sees as good (probability $g_s^{L,J}$); the $I$ discriminator intends to defect, which yields a good reputation with probability $\pgd$.
    \item with someone in an arbitrary group $L$ following strategy $s$ (probability $\nu_L f_s^L$) whom $I$ sees as good (probability $g_s^{L,I}$) but whom $J$ sees as bad (probability $1 - g_s^{L,J}$); the $I$ discriminator intends to cooperate, which yields a good reputation with probability $\pbc$.
    \item with someone in an arbitrary group $L$ following strategy $s$ (probability $\nu_L f_s^L$) whom $I$ sees as bad (probability $1 - g_s^{L,I}$) and whom $J$ also sees as bad (probability $1 - g_s^{L,J}$); the $I$ discriminator intends to defect, which yields a good reputation with probability $\pbd$.
\end{enumerate}
Defining
\begin{equation}
    \Gij{I}{J} = \sum_L \nu_L \sum_s f_s g_s^{L,I} g_s^{L,J},
\end{equation}
we can sum over all groups and strategy combinations to obtain
\begin{equation*}
    \begin{split}
        \sum_L \nu_L \sum_s f_s^L g_s^{L,I} g_s^{L,J} & = \Gij{I}{J},
        \\
        \sum_L \nu_L \sum_s f_s^L (1 - g_s^{L,I}) g_s^{L,J} & = g^{\bullet,J} - \Gij{I}{J},
        \\
        \sum_L \nu_L \sum_s f_s^L g_s^{L,I}(1 - g_s^{L,J}) & = g^{\bullet,I} - \Gij{I}{J},
        \\
        \sum_L \nu_L \sum_s f_s^L (1 - g_s^{L,I})(1 - g_s^{L,J}) & = 1 - g^{\bullet,J} - g^{\bullet,I} + \Gij{I}{J}.
    \end{split}
\end{equation*}
Thus,
\begin{equation}
    \begin{split}
        g_Z^{I,J} & = \delta_{I,J} \big[ g^{\bullet,J} \pgc + (1 - g^{\bullet,J}) \pbd \big]
        \\
        & \ \ \ + (1 - \delta_{I,J}) \big[ \Gij{I}{J} \pgc + (g^{\bullet,J} - \Gij{I}{J}) \pgd + (g^{\bullet,I} - \Gij{I}{J}) \pbc + (1 - g^{\bullet,J} - g^{\bullet,I} + \Gij{I}{J}) \pbd \big]
        \\
        & = \delta_{I,J} \big[ g^{\bullet,J} \pgc + (1 - g^{\bullet,J}) \pbd \big]
        \\
        & \ \ \ + (1 - \delta_{I,J}) \big[ \Gij{I}{J} (\pgc - \pgd - \pbc + \pbd) + g^{\bullet,J}(\pgd - \pbd) + g^{\bullet,I}(\pbc - \pbd) + \pbd \big].
    \end{split}
    \label{eq:Z_reps}
\end{equation}

\subsection{Special case: \textit{Scoring}}

Under \textit{Scoring} ($p = 1, q = 0$), we have
\begin{equation*}
    \begin{split}
        \pgc & = \pbc = \epsilon,
        \\
        \pgd & = \pbd = \ass.
    \end{split}
\end{equation*}
In this case, the $I \neq J$ term of $g_Z^{I,J}$ becomes
\begin{equation*}
    \begin{split}
        & \Gij{I}{J}(\pgc - \pgd - \pbc + \pbd) + g^{\bullet,J}(\pgd - \pbd) + g^{\bullet,I}(\pbc - \pbd) + \pbd
        \\
        & = g^{\bullet,I}(\pbc - \pbd) + \pbd
        \\
        & = g^{\bullet,I} \epsilon + (1 - g^{\bullet,I}) \ass.
    \end{split}
\end{equation*}
Consequently, we have
\begin{equation*}
    \begin{split}
        g_X^{I,J} & = g^{\bullet,J} \epsilon + (1 - g^{\bullet,J}) \epsilon = \epsilon,
        \\
        g_Y^{I,J} & = g^{\bullet,J} \ass + (1 - g^{\bullet,J}) \ass = \ass,
        \\
        g_Z^{I,J} & = \delta_{I,J} \big[ g^{\bullet,J} \epsilon + (1 - g^{\bullet,J}) \ass \big] + (1 - \delta_{I,J}) \big[ g^{\bullet,I} \epsilon + (1 - g^{\bullet,I}) \ass \big]
        \\
        & = \delta_{I,J} \big[ g^{\bullet,I} \epsilon + (1 - g^{\bullet,I}) \ass \big] + (1 - \delta_{I,J}) \big[ g^{\bullet,I} \epsilon + (1 - g^{\bullet,I}) \ass \big]
        \\
        & = g^{\bullet,I} \epsilon + (1 - g^{\bullet,I}) \ass.
    \end{split}
\end{equation*}
The last line implies that $J$'s opinion of $I$ discriminators depends solely on whom $I$ sees as good, not whom $J$ sees as good.
This is reasonable; \textit{Scoring} is a first-order norm, in which \emph{any} cooperation is considered good and \emph{any} defection is considered bad, meaning that an $I$ discriminator will be considered good as a result of their interactions with those $I$ sees as good (with whom they therefore cooperate).
Likewise, an $I$ discriminator will be considered good as a result of their interactions with those $I$ sees as bad (with whom they therefore defect).
One may note that
\begin{equation*}
    \begin{split}
        g^{\bullet,I} & = \sum_L \nu_L g^{L,I} = \sum_L \nu_L \sum_s f_s^L g_s^{L,I}
        \\
        & = \sum_L \nu_L \big( f_X^L \epsilon + f_Y^L \ass + f_Z^L [g^{\bullet,I} \epsilon + (1 - g^{\bullet,I} \ass] \big)
        \\
        & = \epsilon \sum_L \nu_L f_X^L + \ass \sum_L \nu_L f_Y^L + g^{\bullet,I} \epsilon \sum_L \nu_L f_Z^L + (1 - g^{\bullet,I}) \ass \sum_L \nu_L f_Z^L
        \\
        \therefore g^{\bullet,I} & = \frac{\epsilon \sum_L \nu_L f_X^L + \ass \sum_L \nu_L (f_Y^L + f_Z^L)}{1 - \sum_L \nu_L f_Z^L (\epsilon - \ass)},
    \end{split}
\end{equation*}
which is independent of $I$.
In this way, under \textit{Scoring}, the reputation of discriminators does not depend on their group identity.
Moreover, if there is no difference in strategy frequency among groups, we have
\begin{equation*}
    \begin{split}
        g^{\bullet,I} & = \frac{\epsilon \sum_L \nu_L f_X^L + \ass \sum_L \nu_L (f_Y^L + f_Z^L)}{1 - \sum_L \nu_L f_Z^L (\epsilon - \ass)}
        \\
        & = \frac{\epsilon  f_X \sum_L \nu_L + (f_Y + f_Z) \ass \sum_L \nu_L }{1 - f_Z \sum_L \nu_L  (\epsilon - \ass)}
        \\
        & = \frac{\epsilon f_X + \ass (f_Y + f_Z)}{1 - f_Z (\epsilon - \ass)},
    \end{split}
\end{equation*}
which is independent of the number of groups and their relative sizes.
Thus, under \textit{Scoring}, if strategy frequencies are equal among groups, imposing a group structure on the population does not affect reputations at all, and hence it does not affect the strategy dynamics.

\subsection{The ``staying'' norm}

Under the ``staying'' norm \citep{sasaki2017evolution}, individuals do not change their opinions of a donor who interacts with a bad recipient at all, irrespective of the donor's action.
In our notation, this is tantamount to replacing $\pbc$ and $\pbd$ with $g_s^{I,J}$, where $s$ is the strategy whose reputation is being assessed, since, if the recipient has a bad reputation, the donor's reputation is unchanged.
Reputations are thus given by
\begin{equation}
    \begin{split}
        g_X^{I,J} & = g^{\bullet,J} \pgc + (1 - g^{\bullet,J}) g_X^{I,J}
        \\
        g_Y^{I,J} & = g^{\bullet,J} \pbc + (1 - g^{\bullet,J}) g_Y^{I,J}
        \\
        g_Z^{I,J} & = \delta_{I,J} \big[ g^{\bullet,J} \pgc + (1 - g^{\bullet,J}) g_Z^{I,J} \big]
        \\
        & \ \ \ + (1 - \delta_{I,J}) \big[ \Gij{I}{J} \pgc + (g^{\bullet,J} - \Gij{I}{J}) \pgd + (g^{\bullet,I} - \Gij{I}{J}) g_Z^{I,J} + (1 - g^{\bullet,J} - g^{\bullet,I} + \Gij{I}{J}) g_Z^{I,J} \big]
        \\
        & = \delta_{I,J} \big[ g^{\bullet,J} \pgc + (1 - g^{\bullet,J}) g_Z^{I,J} \big]
        \\
        & \ \ \ + (1 - \delta_{I,J}) \big[ \Gij{I}{J} \pgc + (g^{\bullet,J} - \Gij{I}{J}) \pgd + (1 - g^{\bullet,J}) g_Z^{I,J} \big]
        \\
        & = \delta_{I,J} \big[ g^{\bullet,J} \pgc + (1 - g^{\bullet,J}) g_Z^{I,J} \big]
        \\
        & \ \ \ + (1 - \delta_{I,J}) \big[ \Gij{I}{J} (\pgc - \pgd) + g^{\bullet,J}(\pgd - g_Z^{I,J}) + g^{\bullet,I}(\pbc - g_Z^{I,J}) + g_Z^{I,J} \big].
    \end{split}
\end{equation}
In equilibrium, this yields
\begin{equation}
    \begin{split}
        g_X^{I,J} = g_Z^{I,J}|_{I = J} & = \pgc,
        \\
        g_Y^{I,J} & = \pgd,
    \end{split}
\end{equation}
but the out-group version of $g_Z^{I,J}$ can still be very complicated.

\section{Invasibility of discriminators in a single group}

When $K = 1$, there are two stable equilibria: a population consisting entirely of defectors ($Y$) and a population consisting entirely of discriminators ($Z$).
Here, we consider the circumstances under which these equilibria are stable against invasion.

\subsection{Invasibility by defectors}

Let $f = f_Z$.
Defectors resist invasion by discriminators provided
\begin{equation*}
    \begin{split}
        \partial_{f} \dot{f}_Y|_{f = 0} & < 0
        \\
        \partial_{f} [f_Y (\Pi_Y - \bar{\Pi})]|_{f = 0} & < 0
        \\
        \partial_{f} [(1 - f) (\Pi_Y - (1 - f) \Pi_Y - f \Pi_Z)]|_{f  = 0} & < 0
        \\
        \partial_{f} [(1 - f) (\Pi_Y - (1 - f) \Pi_Y - f \Pi_Z)]|_{f = 0} & < 0
        \\
        \partial_{f} [(f - f^2) (\Pi_Z - \Pi_Y)]|_{f = 0} & < 0
        \\
        [(1 - 2 f)(\Pi_Z - \Pi_Y) + (f - f^2) \partial_{f}(\Pi_Z - \Pi_Y)]|_{f = 0} & < 0
        \\
        \Pi_Z|_{f = 0} & < \Pi_Y|_{f = 0}
        \\
        (b f g_Z - c g)|_{f = 0} & < b f g_Y |_{f = 0}
        \\
        -c g_Y & < 0.
    \end{split}
\end{equation*}
Since $c$ and $g_Y$ are both positive, this condition always obtains: discriminators can never invade a population of defectors.
Likewise, discriminators resist invasion by defectors provided
\begin{equation}
    \begin{split}
        \partial_{f} \dot{f}_Y|_{f = 1} & < 0
        \\
        \partial_{f} [f_Y (\Pi_Y - \bar{\Pi})]|_{f = 1} & < 0
        \\
        \partial_{f} [(1 - f) (\Pi_Y - (1 - f) \Pi_Y - f \Pi_Z)]|_{f  = 1} & < 0
        \\
        \partial_{f} [(1 - f) (\Pi_Y - (1 - f) \Pi_Y - f \Pi_Z)]|_{f = 1} & < 0
        \\
        \partial_{f} [(f - f^2) (\Pi_Z - \Pi_Y)]|_{f = 1} & < 0
        \\
        [(1 - 2 f)(\Pi_Z - \Pi_Y) + (f - f^2) \partial_{f}(\Pi_Z - \Pi_Y)]|_{f = 1} & < 0
        \\
        \Pi_Z|_{f = 1} & > \Pi_Y|_{f = 1}
        \\
        (b f g_Z - c g)|_{f = 1} & > b f g_Y |_{f = 1}
        \\
        (b g_Z - c g_Z)|_{f = 1} & > b g_Y |_{f = 1}
        \\
        b (g_Z - g_Y)|_{f = 1} & > c g_Z|_{f = 1}
        \\
        b g (\pgc - \pgd) & > c g
        \\
        \therefore \frac{b}{c} & > \frac{1}{\pgc - \pgd} = \frac{1}{\epsilon - \ass}.
    \end{split}
\end{equation}
This can also be written in terms of $g$: discriminators resist invasion by defectors provided
\begin{equation}
    \begin{split}
        \Pi_Z|_{f_Z = 1} & > \Pi_Y|_{f_Z = 1}
        \\
        b g_Z|_{f_Z = 1} - c g & > b g_Y|_{f_Z = 1}
        \\
        (b - c) g & > b [g \pgd + (1 - g) \pbd]
        \\
        g(b - c - b[\pgd - \pbd] & > b \pbd
        \\
        g(b[1 - \pgd + \pbd] - c) & > b \pbd
        \\
        \therefore g & > \frac{\pbd}{1 - \pgd + \pbd - c/b}.
    \end{split}
    \label{eq:coop_condition}
\end{equation}
We do not need to flip the inequality because $b > c$ and because $1 + \pbd - \pgd$ is guaranteed to be greater than or equal to $1$ for every social norm we consider.
Finally, there is a third equilibrium between the two which, by the mean value theorem, is unstable, at (letting $f = f_Z$ again)
\begin{equation}
    \begin{split}
        \dot{f} & = 0
        \\
        f (\Pi_Z - \bar{\Pi}) & = 0
        \\
        \Pi_Z - f \Pi_Z - (1 - f) \Pi_Y & = 0
        \\
        \Pi_Z & = \Pi_Y
        \\
        b f g_Z - c g & = b f g_Y
        \\
        b f (g \pgc + [1 - g] \pbd) - c g & = b f (g \pgd + [1 - g] \pbd)
        \\
        b f g (\pgc - \pgd) & = c g
        \\
        \therefore f & = \frac{c}{b} \frac{1}{\pgc - \pgd} = \frac{c}{b} \frac{1}{\epsilon - \ass}.
    \end{split}
\end{equation}
An equivalent way to express this is that discriminators rise in frequency provided
\begin{equation}
    f_Z (g_Z - g_Y) > c/b.
    \label{eq:growth_condition}
\end{equation}
If there is no value of $f_Z$ for which this is true, then discriminators do not rise in frequency; if it is not true for $f_Z = 1$ even when the inequality is relaxed, discriminators cannot resist invasion by defectors.

\subsection{Invasibility by cooperators}

Finally, we consider conditions under which \emph{cooperators} can invade a population of discriminators.
We proceed by reasoning similar to equation \eqref{eq:coop_condition}, noting that the stability of an equilibrium against invasion is determined by evaluating the fitnesses of the resident and the invader at that equilibrium.
Cooperators can invade discriminators when (letting $f = f_Z$)
\begin{equation}
    \begin{split}
        \Pi_X|_{f = 1} & > \Pi_Z|_{f = 1}
        \\
        (b f g_X - c)|_{f = 1} & > (b f g_Z - c g) |_{f = 1}
        \\
        b(g_X - g) & > c(1 - g)
        \\
        b(g \pgc + (1 - g) \pbc - g) & > c(1 - g)
        \\
        b(g[\pgc - \pbc - 1] + \pbc) & > c(1 - g)
        \\
        g (b [\pgc - \pbc - 1] + c) & > c - b \pbc
        \\
        \therefore &
        \begin{cases}
            g > \frac{c-b \pbc}{b(\pgc - \pbc - 1) + c} & \text{\textit{Shunning}, \textit{Stern Judging}},
            \\
            g < \frac{c-b \pbc}{b(\pgc - \pbc - 1) + c} & \text{\textit{Scoring}, \textit{Simple Standing}}.
        \end{cases}
    \end{split}
\end{equation}
For small error rates, this condition is never satisfied under \textit{Shunning} or \textit{Stern Judging} (the right hand side is generally greater than $1$), but it can be met under \textit{Scoring} and \textit{Simple Standing}.
With $\act = \ass = .02$ and $b = 2, c = 1$, the cutoff is about $0.92$ for both \textit{Simple Standing} and \textit{Scoring}; for $b = 5, c = 1$, the cutoff is about $0.95$.
This means that if discriminators do not view each other as having sufficiently good reputations, they become vulnerable to invasion \emph{by cooperators}!

\section{Multiple groups with well-mixed strategic imitation}

Under the assumption of well-mixed strategic imitation, the only interesting dynamical quantities are the ``total'' (group-averaged) strategy fitnesses $\Pi_s$, viz.:
\begin{equation*}
    \begin{split}
        \dot{f}_s = f_s \Big[ \Big(\sum_I \nu_I \Pi_s^I \Big) - \bar{\Pi} \Big] & = f_s [ \Pi_s - \bar{\Pi}].
    \end{split}
\end{equation*}
By summing over all groups, we obtain
\begin{equation}
    \begin{split}
        \Pi_Z & = \sum_I \nu_I \Pi_Z^I = \sum_I \Big\{ \nu_I (1 - \act) \Big[ b \sum_J \nu_J (f_X^J + f_Z^J g_Z^{I,J}) - c g^{\bullet, I} \Big] \Big\}
        \\
        & = (1 - \act) \sum_I \Big\{ \nu_I \Big[ b \sum_J \nu_J (f_X + f_Z g_Z^{I,J}) - c g^{\bullet, I} \Big] \Big\}
        \\
        & = (1 - \act) \Big[ b ( f_X + f_Z \sum_I \sum_J \nu_I \nu_J g_Z^{I,J}) - c \sum_I \nu_I g^{\bullet, I} \Big]
        \\
        & = (1 - \act) \Big[ b ( f_X + f_Z \sum_I \sum_J \nu_I \nu_J g_Z^{I,J}) - c \sum_I \sum_J \nu_I \nu_J g^{J,I} \Big]
        \\
        & = (1 - \act) \big[ b ( f_X + f_Z \bar{g}_Z) - c \bar{g} \big], \text{and likewise}
        \\
        \Pi_X & = \sum_I \nu_I \Pi_X^I = (1 - \act) \big[ b ( f_X + f_Z \bar{g}_X) - c \big],
        \\
        \Pi_Y & = \sum_I \nu_I \Pi_Y^I = (1 - \act) \big[ b ( f_X + f_Z \bar{g}_Y) \big].
    \end{split}
    \label{eq:average_reputations}
\end{equation}
Here, $\bar{g}_s$ is the probability that a randomly chosen individual of type $s$ is seen as good by a randomly chosen observer in the population; and likewise $\bar{g}$ is the probability that a randomly chosen individual is seen as good by a randomly chosen observer (by the linearity of averages we  have $\bar{g} = f_X \bar{g}_X + f_Y \bar{g}_Y + f_Z \bar{g}_Z$).
In general we will have
\begin{equation*}
    \begin{split}
        \bar{g}_X & = \sum_{I=1}^K \sum_{J=1}^K \nu_I \nu_J g_X^{I,J}
        \\
        & = \sum_{I=1}^K \sum_{J=1}^K \nu_I \nu_J  \big[ g^{\bullet,J} \pgc + (1 - g^{\bullet,J}) \pbc \big]
        \\
        & = \sum_{I=1}^K \sum_{J=1}^K \nu_I \nu_J g^{\bullet,J} \pgc + \sum_{I=1}^K \sum_{J=1}^K \nu_I \nu_J (1 - g^{\bullet,J}) \pbc
        \\
        & = \sum_{I=1}^K \nu_I \bar{g} \pgc + \sum_{I=1}^K \nu_I (1 - \bar{g}) \pbc
        \\
        & = \bar{g} \pgc + (1 - \bar{g}) \pbc, \text{and likewise}
        \\
        \bar{g}_Y & = \bar{g} \pgd + (1 - \bar{g}) \pbd.
    \end{split}
\end{equation*}
The form of $\bar{g}_Z$ will vary depending on the specific scenario, but we can obtain a couple of general relations.
First, note that equation \eqref{eq:coop_condition} becomes
\begin{equation}
    \begin{split}
        \Pi_Z|_{f_Z = 1} & > \Pi_Y|_{f_Z = 1}
        \\
        b \bar{g}_Z|_{f_Z = 1} - c \bar{g} & > b \bar{g}_Y|_{f_Z = 1}
        \\
        (b - c) \bar{g} & > b \big[ \bar{g} \pgd + (1 - \bar{g}) \pbd \big]
        \\
        \bar{g}(b - c - b[\pgd - \pbd]) & > b \pbd
        \\
        \bar{g}(b[1 - \pgd + \pbd] - c) & > b \pbd
        \\
        \therefore \bar{g} & > \frac{\pbd}{1 - \pgd + \pbd - c/b},
    \end{split}
    \label{eq:bar_g_condition}
\end{equation}
and equation \eqref{eq:growth_condition} becomes
\begin{equation*}
    f_Z (\bar{g}_Z - \bar{g}_Y) > c/b.
\end{equation*}
That is, under well-mixed strategic imitation, the conditions for discriminators to resist invasion by defectors and to increase in frequency over time can be written in terms of average reputations $\bar{g}_s$, though the value of those reputations will vary depending on the group structure.
An equivalent way to write equation \eqref{eq:bar_g_condition} is
\begin{equation}
    \begin{split}
        \frac{b}{c} & > \frac{\bar{g}}{\bar{g} - \bar{g}_Y}
        \\
        \therefore \frac{b}{c} & > \frac{\bar{g}}{\bar{g}(1+\pbd - \pgd) - \pbd}.
    \end{split}
\end{equation}

\subsection{Groups of identical size}\label{sec:identical_size}

When all $K$ groups have the same size $1/K$ and strategies spread via well-mixed copying, we can solve for $\bar{g}_Z$:
\begin{equation*}
    \begin{split}
        \bar{g}_Z & = \sum_{I=1}^K \sum_{J=1}^K \nu_I \nu_J g_Z^{I,J}
        \\
        & = \sum_{I=1}^K \sum_{J=1}^K \nu_I \nu_J \Big( \delta_{I,J} \big[ g^{\bullet,J} \pgc + (1 - g^{\bullet,J}) \pbd \big]
        \\
        & \ \ + (1 - \delta_{I,J}) \big[ \Gij{I}{J} \pgc + (g^{\bullet,J} - \Gij{I}{J}) \pgd + (g^{\bullet,I} - \Gij{I}{J}) \pbc + (1 - g^{\bullet,J} - g^{\bullet,I} + \Gij{I}{J}) \pbd \big] \Big)
         \\
         & = \sum_{J=1}^K \nu_J^2 \big[ g^{\bullet,J} \pgc + (1 - g^{\bullet,J}) \pbd \big]
         \\
         & \ \ + \mathop{\sum_{I=1}^K \sum_{J=1}^K}_{I \neq J} \nu_I \nu_J \big[ \Gij{I}{J} \pgc + (g^{\bullet,J} - \Gij{I}{J}) \pgd + (g^{\bullet,I} - \Gij{I}{J}) \pbc + (1 - g^{\bullet,J} - g^{\bullet,I} + \Gij{I}{J}) \pbd \big].
    \end{split}
\end{equation*}
The first term simplifies to
\begin{equation*}
    \begin{split}
        \sum_{J=1}^K \nu_J^2 \big[ g^{\bullet,J} \pgc + (1 - g^{\bullet,J}) \pbd \big] & = \frac{1}{K} \sum_{I=1}^K \nu_J \big[ g^{\bullet,J} \pgc + (1 - g^{\bullet,J}) \pbd \big]
        \\
        & = \frac{1}{K} \big[ \bar{g} \pgc + (1 - \bar{g}) \pbd \big].
    \end{split}
\end{equation*}
The second becomes
\begin{equation*}
    \begin{split}
        & \mathop{\sum_{I=1}^K \sum_{J=1}^K}_{I \neq J} \nu_I \nu_J \big[ \Gij{I}{J} \pgc + (\Gij{I}{J} - g^{\bullet,J}) \pgd + (\Gij{I}{J} - g^{\bullet,J}) \pbc + (1 - g^{\bullet,J} - g^{\bullet,I} + \Gij{I}{J}) \pbd \big]
        \\
        & = \mathop{\sum_{I=1}^K \sum_{J=1}^K}_{I \neq J} \nu_I \nu_J \big[ \Gij{I}{J} (\pgc - \pgd - \pbc + \pbd) + g^{\bullet,I} (\pgd - \pbd) + g^{\bullet,J} (\pbc - \pbd) + \pbd \big].
    \end{split}
\end{equation*}
Because all the groups are the same size and strategy frequencies are identical across groups, the values of $g_s^{I,J}$ can only vary depending on whether $I = J$ or not.
Define $g_s^{\en} = g_s^{I,I}$ and $g_s^{\ex} = g_s^{I,J}\big|_{I \neq J}$.
We exploit this symmetry to obtain
\begin{equation*}
    \begin{split}
        g^{\bullet,I} & = \sum_L^K \nu_L g^{L,I}
        \\
        & = \frac{1}{K} \sum_L^K g^{L,I}
        \\
        & = \frac{1}{K} g^{\en} + \frac{K - 1}{K} g^{\ex},
        \\
        g^{\bullet,J} & = \frac{1}{K} g^{\en} + \frac{K - 1}{K} g^{\ex}
    \end{split}
\end{equation*}
and
\begin{equation}
    \begin{split}
        \Gij{I}{J} & = \sum_L \nu_L \sum_s f_s^L g_s^{L,I} g_s^{L,J}
        \\
        & = \frac{1}{K} \sum_L \sum_s f_s^L g_s^{L,I} g_s^{L,J}
        \\
        & = \frac{1}{K} \sum_L \sum_s f_s g_s^{L,I} g_s^{L,J}
        \\
        & = \frac{1}{K} \sum_s f_s g_s^{\en} g_s^{\ex} + \frac{1}{K} \sum_s f_s g_s^{\ex} g_s^{\en} + \frac{K - 2}{K} \sum_s f_s g_s^{\ex} g_s^{\ex}
        \\
        & = \frac{2}{K} \sum_s f_s g_s^{\en} g_s^{\ex} + \frac{K - 2}{K} \sum_s f_s (g_s^{\ex})^2.
    \end{split}
    \label{eq:disagreement}
\end{equation}
Thus
\begin{equation}
    \begin{split}
        \bar{g}_Z & = \sum_{I=1}^K \sum_{J=1}^K \nu_I \nu_J g_Z^{I,J}
        \\
        & = \frac{1}{K} [ \bar{g} \pgc + (1 - \bar{g}) \pbd ]
        \\
        & \ \ \ + \frac{K-1}{K} \Big[ \Big( \frac{2}{K} \sum_s f_s g_s^{\en} g_s^{\ex} + \frac{K - 2}{K} \sum_s f_s (g_s^{\ex})^2 \Big) (\pgc - \pgd - \pbc + \pbd)
        \\
        & \ \ \ + \Big( \frac{1}{K} g^{\en} + \frac{K-1}{K} g^{\ex} \Big) (\pgd + \pbc - 2 \pbd) + \pbd \Big].
    \end{split}
    \label{eq:gZ_reputations}
\end{equation}
Observe that setting $f_Z = 1$ in equation \eqref{eq:gZ_reputations} yields exactly the system of equations one would need to solve in order to obtain $\bar{g}$ in a population of equally sized groups, viz.:
\begin{equation}
    \begin{split}
        \bar{g} & = \frac{g^{\en} + (K-1) g^{\ex}}{K},
        \\
        g^{\en} & = \bar{g} \pgc + (1 - \bar{g}) \pbd = \bar{g}(\pgc - \pbd) + \pbd,
        \\
        g^{\ex} & =  \Big( \frac{2}{K} g^{\en} g^{\ex} + \frac{K - 2}{K} (g^{\ex})^2 \Big) (\pgc - \pgd - \pbc + \pbd),
        \\
        & \ \ \ + \Big( \frac{1}{K} g^{\en} + \frac{K-1}{K} g^{\ex} \Big) (\pgd + \pbc - 2 \pbd) + \pbd.
    \end{split}
    \label{eq:gZ_no_insularity}
\end{equation}
For example, choosing \textit{Stern Judging} as the norm and setting $\act = 0$ yields $g^{\en} = 1 - \ass, g^{\ex} = 1/2$, consistent with the reputation expressions from \citet{nakamura2012favoritism}.

\subsection{Limit of many groups: private reputations}

As $K$ approaches infinity, the contribution of $g_s^{\en}$ to the total average reputation of $s$, $\bar{g}_s$, tends to zero, so that the entirety of $\bar{g}_s$ is due to the $g_s^{\ex} $ terms.
We thus have that
\begin{equation*}
    \lim_{K \to \infty} \bar{g}_Z = \sum_s f_s \bar{g}_s^2 (\pgc - \pgd - \pbc + \pbd) + 2\bar{g}(\pgd + \pbd - 2 \pbd) + \pbd.
\end{equation*}
Defining
\begin{equation*}
    \begin{split}
        g_2 & = \sum_s f_s \bar{g}_s^2,
        \\
        d_2 & = \sum_s f_s \bar{g}_s (1 - \bar{g}_s) = g - g_2,
        \\
        b_2 & = \sum_s f_s (1 - \bar{g}_s)^2 = 1 - 2g + g_2
    \end{split}
\end{equation*}
allows us to rewrite this as
\begin{equation*}
    \lim_{K \to \infty} \bar{g}_Z = g_2 \pgc + d_2 (\pgd + \pbc) + b_2 \pbd.
\end{equation*}
This is the bottom term of equation $5$ from \citet{radzvilavicius2019evolution} with empathy parameter $E = 0$, corresponding to fully private reputation assessment.
This result confirms that, when the number of groups goes to infinity, our model with separate groups is identical to everyone in the population following independent or private reputation assessment.
In this limit, the mean reputation in a population of discriminators is given by a solution to
\begin{equation*}
    \begin{split}
        0 & = g^2 (\pgc - \pgd - \pbc + \pbd) + g(\pgd + \pbc - 2 \pbd - 1) + \pbd
        \\
        \therefore g & =
        \begin{cases}
            \frac{1 - \sqrt{1 - 4 (\epsilon - \ass)\ass}}{2(\epsilon - \ass)}, & \text{shunning},
            \\
            \frac{1}{2}, & \text{stern judging},
            \\
            \frac{\ass}{1-\epsilon+\ass}, & \text{scoring},
            \\
            \frac{1 - \ass - \sqrt{(1 - \ass)(1 - \epsilon)}}{\epsilon - \ass}, & \text{simple standing}.
        \end{cases}
    \end{split}
\end{equation*}
We have picked out the solutions that are viable for $1 > \act > 0$ and $1 > \ass > 0$.
The result for \textit{Stern Judging} was previously obtained by \citet{uchida2013private}, which also showed that, in the presence of errors under private assessment, $g_s = 1/2$ for any strategy $s$, irrespective of the population's strategic composition.

\subsection{One large group and many small groups}

Without loss of generality, suppose that group $1$ has size $\nu$ and the remaining $K - 1$ groups each have size $(1 - \nu)/(K-1)$.
Starting with equation \eqref{eq:average_reputations}, we can unpack $\bar{g}_Z$.
We have
\begin{equation*}
    \begin{split}
        \bar{g}_Z & = \sum_I \sum_J \nu_I \nu_J g_Z^{I,J}
        \\
        & = \sum_{I=1}^K \sum_{J=1}^K \nu_I \nu_J \Big( \delta_{I,J} \big[ g^{\bullet,J} \pgc + (1 - g^{\bullet,J}) \pbd \big]
        \\
        & \ \ + (1 - \delta_{I,J}) \big[ \Gij{I}{J} \pgc + (g^{\bullet,J} - \Gij{I}{J}) \pgd + (g^{\bullet,I} - \Gij{I}{J}) \pbc + (1 - g^{\bullet,J} - g^{\bullet,I} + \Gij{I}{J}) \pbd \big] \Big)
        \\
        & = \nu^2 \big[ g^{\bullet,1} \pgc + (1 - g^{\bullet,1}) \pbd \big]
        \\
        & \ \ + \nu \frac{1 - \nu}{K - 1} \sum_{I=2}^K \big[ \Gij{I}{1} \pgc + (g^{\bullet,1} - \Gij{I}{1}) \pgd + (g^{\bullet,I} - \Gij{I}{1}) \pbc + (1 - g^{\bullet,1} - g^{\bullet,I} + \Gij{I}{1}) \pbd \big]
        \\
        & \ \ \ + \nu \frac{1 - \nu}{K - 1} \sum_{J=2}^K \big[ \Gij{1}{J} \pgc + (g^{\bullet,J} - \Gij{1}{J}) \pgd + (g^{\bullet,1} - \Gij{1}{J}) \pbc + (1 - g^{\bullet,J} - g^{\bullet,1} + \Gij{1}{J}) \pbd \big]
        \\
        & \ \ \ + \Big( \frac{1 - \nu}{K - 1} \Big) ^2 \Bigg( \sum_{J=2}^K \big[ g^{\bullet,J} \pgc + (1 - g^{\bullet,J}) \pbd \big]
        \\
        & \ \ \ \ + \mathop{\sum_{I=2}^K \sum_{J=2}^K}_{I \neq J} \big[ \Gij{I}{J} \pgc + (g^{\bullet,J} - \Gij{I}{J}) \pgd + (g^{\bullet,I} - \Gij{I}{J}) \pbc + (1 - g^{\bullet,J} - g^{\bullet,I} + \Gij{I}{J}) \pbd \big] \Bigg)
        \\
        & = \nu^2 \big[ g^{\bullet,1} \pgc + (1 - g^{\bullet,1}) \pbd \big]
        \\
        & \ \ \ + \nu (1 - \nu) \big[ \Gij{I}{1} \pgc + (g^{\bullet,1} - \Gij{I}{1}) \pgd + (g^{\bullet,I} - \Gij{I}{1}) \pbc + (1 - g^{\bullet,1} - g^{\bullet,I} + \Gij{I}{1}) \pbd \big] \Big |_{I \neq 1}
        \\
        & \ \ \ + \nu (1 - \nu) \big[ \Gij{1}{J} \pgc + (g^{\bullet,J} - \Gij{1}{J}) \pgd + (g^{\bullet,1} - \Gij{1}{J}) \pbc + (1 - g^{\bullet,J} - g^{\bullet,1} + \Gij{1}{J}) \pbd \big] \Big |_{J \neq 1}
        \\
        & \ \ \ + \frac{(1 - \nu)^2}{K - 1} \Big( \big[ g^{\bullet,J} \pgc + (1 - g^{\bullet,J}) \pbd \big] \Big) \Big|_{J \neq 1}
        \\
        & \ \ \ + \Big( \frac{(1 - \nu)(K-2)}{K-1}\Big) ^2 \big[ \Gij{I}{J} \pgc + (g^{\bullet,J} - \Gij{I}{J}) \pgd
        \\
        & \ \ \ \ \ + (g^{\bullet,I} - \Gij{I}{J}) \pbc + (1 - g^{\bullet,J} - g^{\bullet,I} + \Gij{I}{J}) \pbd \big] \Big|_{I \neq J \neq 1}.
    \end{split}
\end{equation*}

As $K \to \infty$, the $I = J$ elements drop out.
What remains is a special case of equation 12 of \citet{radzvilavicius2021nature}, in which part of the population consists of adherents to a single institution of reputation assessment and the remainder consists of private assessors.

\section{Switching group membership}

We now consider a variant of our model in which individuals can switch gossip group identity depending on their difference in fitness (with probability given by a Fermi function, as in the rest of our analysis).
We assume $K = 2$ groups, both of which are fixed for strategy $Z$.

\subsection{Same norm and payoffs}

When both groups follow the same social norm and payoffs are group-independent, we have
\begin{equation}
    \begin{split}
        \dot{\nu}_1 & = \nu_1 (\Pi_Z^1 - \bar{\Pi}),
        \\
        \dot{\nu}_2 & = \nu_2 (\Pi_Z^2 - \bar{\Pi}), \text{with}
        \\
        \bar{\Pi} & = \nu_1 \Pi_Z^1 + \nu_2 \Pi_Z^2.
    \end{split}
\end{equation}
We expect $\nu_1$ to grow if
\begin{equation}
    \begin{split}
        \dot{\nu}_1 & > 0
        \\
        \therefore \nu_1 (\Pi_Z^1 - \bar{\Pi}) & > 0
        \\
        \therefore \nu_1 (\Pi_Z^1 - \nu_1 \Pi_Z^1 - (1 - \nu_1) \Pi_Z^2) & > 0
        \\
        \therefore (\nu_1 - \nu_1^2) (\Pi_Z^1 - \Pi_Z^2) & > 0
        \\
        \therefore \Pi_Z^1 & > \Pi_Z^2
        \\
        \therefore b( \nu_1 g_Z^{1,1} + \nu_2 g_Z^{1,2}) - c g^{\bullet,1} & > b( \nu_1 g_Z^{2,1} + \nu_2 g_Z^{2,2}) - c g^{\bullet,2}
        \\
        \therefore b( \nu_1 g_Z^{1,1} + \nu_2 g_Z^{1,2}) - c (\nu_1 g_Z^{1,1} + \nu_2 g_Z^{2,1}) & > b( \nu_1 g_Z^{2,1} + \nu_2 g_Z^{2,2}) - c (\nu_1 g_Z^{1,2} + \nu_2 g_Z^{2,2})
        \\
        \therefore \nu_1 (b [g_Z^{1,1} - g_Z^{2,1}] - c [g_Z^{1,1} - g_Z^{1,2}]) & > \nu_2 (b [g_Z^{2,2} - g_Z^{1,2}] - c [g_Z^{2,2} - g_Z^{2,1}])
        \\
        \therefore \frac{\nu_1}{\nu_2} & > \frac{b (g_Z^{2,2} - g_Z^{1,2}) - c (g_Z^{2,2} - g_Z^{2,1})}{b (g_Z^{1,1} - g_Z^{2,1}) - c (g_Z^{1,1} - g_Z^{1,2})}.
    \end{split}
    \label{eq:nu_1/nu_2}
\end{equation}
When both groups follow the same social norm and are of the same size, there cannot be any difference between $g_Z^{1,1}$ and $g_Z^{2,2}$, nor between $g_Z^{1,2}$ and $g_Z^{2,1}$.
The last line of equation \eqref{eq:nu_1/nu_2} thus simplifies to $1$, which at least suggests $\nu_1 = \nu_2 = 1/2$ is significant.
We can show that $\nu_1 > 1/2$ by explicitly solving for the ratio $\nu_1/\nu_2$:
\begin{equation*}
    \begin{split}
        \frac{\nu_1}{\nu_2} & > \frac{b (g_Z^{2,2} - g_Z^{1,2}) - c (g_Z^{2,2} - g_Z^{2,1})}{b (g_Z^{1,1} - g_Z^{2,1}) - c (g_Z^{1,1} - g_Z^{1,2})}
        \\
        & > \frac{b(1 + \nu_1(\pbd - \pgc) + \nu_2(\pbc - \pgd)) - c(1 - \nu_1(\pgc + \pgd - \pbc - \pbd))}{b(1 + \nu_2(\pbd - \pgc) + \nu_1(\pbc - \pgd)) - c(1 - \nu_2(\pgc + \pgd - \pbc - \pbd))}
        \\
        \therefore \frac{\nu_1}{1 - \nu_1} & > \frac{b(1 + \nu_1(\pbd - \pgc) + (1 - \nu_1)(\pbc - \pgd)) - c(1 - \nu_1(\pgc + \pgd - \pbc - \pbd))}{b(1 + (1 - \nu_1)(\pbd - \pgc) + \nu_1(\pbc - \pgd)) - c(1 - (1 - \nu_1)(\pgc + \pgd - \pbc - \pbd))}.
    \end{split}
\end{equation*}
We can collect powers of $\nu_1$ to obtain
\begin{equation*}
    \begin{split}
        & \nu_1^2 (b[\pgc - \pgd + \pbc - \pbd] - c[\pgc + \pgd - \pbc - \pbc])
        \\
        & \ \ \ + \nu_1 (b[1 - \pgc + \pbd] - c[1 - \pgc - \pgd + \pbc + \pbd])
        \\
        & > \nu_1^2 (b [\pbc - \pbd + \pgc - \pgd] - c[\pgc + \pgd - \pbc - \pbd])
        \\
        & \ \ \ + \nu_1(b[2 \pgd - \pgc + -2 \pbc + \pbd - 1] - c[\pbc + \pbd - \pgc - \pgd - 1] )
        \\
        & \ \ \ + b(1 + \pbc - \pgd) - c.
    \end{split}
\end{equation*}
The quadratic terms cancel, leaving
\begin{equation}
    \begin{split}
        \therefore \nu_1(b[2 + 2 \pbc - 2 \pgd] - 2 c) & > b(1 + \pbc - \pgd) - c
        \\
        \therefore \nu_1 & > 1/2.
    \end{split}
\end{equation}

\subsection{Different norms and payoffs}

We now allow social norms and payoffs to differ between groups.
Suppose group $J$ uses a social norm with reputation probabilities $\pgc^J, \pgd^J, \pbc^J, \pbd^J$ when assessing others' reputations, and suppose an individual in group $I$ who cooperates with an individual in group $J$ conveys a benefit $b^{I,J}$ but pays a cost $c^{I,J}$.
(A natural application of variable benefits and costs is to allow in-group and out-group interactions to differ, so that $b^{I,J} = \delta_{I,J} b^{\en} + (1 - \delta_{I,J}) b^{\ex}$ and $c^{I,J} = \delta_{I,J} c^{\en} + (1 - \delta_{I,J}) c^{\ex}$.)
For completeness, we present payoffs for all three strategic types:
\begin{equation}
    \begin{split}
        \Pi_X^I & = (1 - \act) \Big\{ \sum_J \nu_J \big[ b^{J,I} (f_X^J + f_Z^J g_X^{I,J}) \big] - c^{I,J} \Big\}
        \\
        \Pi_Y^I & = (1 - \act) \Big\{\sum_J \nu_J \big[ b^{J,I} (f_X^J + f_Z^J g_Y^{I,J})\big] \Big\}
        \\
        \Pi_Z^I & = (1 - \act) \Big\{ \sum_J \nu_J \big[ b^{J,I} (f_X^J + f_Z^J g_Z^{I,J}) \big] - c^{I,J} g^{\bullet,I} \Big\}.
    \end{split}
    \label{eq:vary_norm_fitnesses}
\end{equation}
Group-dependent social norms mean that reputation equations change:
\begin{equation}
    \begin{split}
        g_X^{I,J} & = g^{\bullet,J} \pgc^J + (1 - g^{\bullet,J}) \pbc^J
        \\
        g_Y^{I,J} & = g^{\bullet,J} \pgd^J + (1 - g^{\bullet,J}) \pbd^J
        \\
        g_Z^{I,J} & = \delta_{I,J} \big[ g^{\bullet,J} \pgc^J + (1 - g^{\bullet,J}) \pbd^J \big]
        \\
        & \ \ \ + (1 - \delta_{I,J}) \big[ \Gij{I}{J} \pgc^J + (g^{\bullet,J} - \Gij{I}{J}) \pgd^J + (g^{\bullet,I} - \Gij{I}{J}) \pbc^J + (1 - g^{\bullet,J} - g^{\bullet,I} + \Gij{I}{J}) \pbd^J \big], \text{with}
        \\
        g^{I,J} & = \sum_s f_s^I g_s^{I,J},
        \\
        g^{\bullet,J} & = \sum_L \nu_L g^{L,J},
        \\
        \Gij{I}{J} & = \sum_L \nu_L \sum_s f_s g_s^{L,I} g_s^{L,J}.
    \end{split}
    \label{eq:vary_norm_reputations}
\end{equation}
In the main text, we do not vary the $b^{I,J}$ and $c^{I,J}$, but we simultaneously solve equations \eqref{eq:vary_norm_fitnesses} and \eqref{eq:vary_norm_reputations} for cases where $I$ and $J$ follow different social norms and both populations are fixed for $Z$.
The dynamics are at most bistable, with $\nu_1$ shrinking unless it is above a critical frequency $\nu_1^*$.
For \emph{Stern Judging}, this critical frequency is almost always less than $1/2$, meaning groups that follow \emph{Stern Judging} are likely to grow over a larger region of phase space than any of the second-order norms we consider.

\begin{figure}
    \centering
    \includegraphics[width=0.5\textwidth]{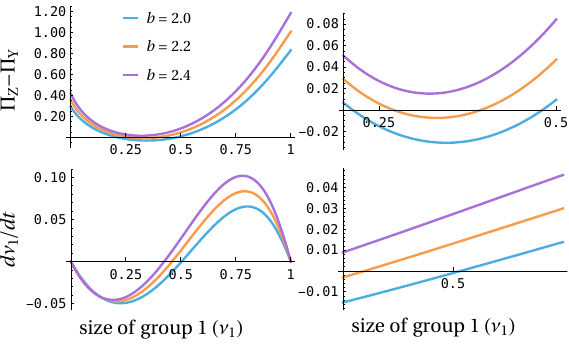}
    \caption{\small{Competition between \textit{Stern Judging} (group $1$) and \textit{Shunning} (group $2$) in $K=2$ groups.
    For low values of $b$ we have $\nu_1^* > 1/2$, so that \textit{Shunning} (not \textit{Stern Judging}) can take over the population even when starting from a minority.
    In this regime, however, as the \textit{Shunning} group grows, the population passes through a regime where it becomes vulnerable to invasion by pure defectors (top two plots). Increasing $b$ to the point where this instability no longer occurs is sufficient to push the $\nu_1^*$ below $1/2$, so that \textit{Stern Judging} will take over the population when starting from a minority. And so, in summary, in all regimes where the population resists invasion by defectors, \textit{Stern Judging} out-competes \textit{Shunning}, even when starting in the minority.}}
    \label{fig:two_norms_grid}
\end{figure}

The sole exception is when \emph{Stern Judging} competes against \emph{Shunning}, for which $\nu_1^* > 1/2$.
In SI figure \ref{fig:two_norms_grid}, we show that this case is distinctive because, as the \emph{Shunning} group expands, the population becomes invasible by \emph{defectors}.
This means that, in a model with both group and strategy evolution (i.e., $0 < \tau < 1$), it cannot be guaranteed that the population would continue to consist of discriminators under this model; a lucky defector mutant that invades at the right time might take over the entire population.

\subsection{Private reputations}
\label{sec:two_groups_private}

We briefly consider what happens when two groups each adhere to \emph{private} reputation assessment but follow different norms.
We assume both groups are fixed for discriminators.
In that case,
\begin{equation*}
    \begin{split}
        g^{I,J} & = \sum_L \nu_L \big( g^{I,L} g^{J,L} \pgc^J + g^{I,L} [1 - g^{J,L}] \pbc^J + [1 - g^{I,L}] g^{J,L} \pgd^J + [1 - g^{I,L}][1 - g^{J,L}] \pbd^J \big)
        \\
        & = \Gij{I}{J} \pgc^J + (g^{\bullet,I} - \Gij{I}{J}) \pbc^J + (g^{\bullet,J} - \Gij{I}{J}) \pgd^J + (1 - g^{\bullet,I} - g^{\bullet,J} + \Gij{I}{J}) \pbd^J.
    \end{split}
\end{equation*}
An important difference between this expression and $g_Z^{I,J}$ in equation \eqref{eq:vary_norm_reputations} is that individuals \emph{in the same group} are not guaranteed to share reputational views of anyone else in the population, meaning that the $I = J$ term does not have a different form.
Thus, for two groups, we have
\begin{equation*}
    \begin{split}
        g^{1,1} & = \Gij{1}{1} \pgc^1 + (g^{\bullet,1} - \Gij{1}{1}) (\pgd^1 + \pbc^1) + (1 - 2 g^{\bullet,1} + \Gij{1}{1}) \pbd^1
        \\
        g^{1,2} & = \Gij{1}{2} \pgc^2 + (g^{\bullet,1} - \Gij{1}{2}) \pbc^2 + (g^{\bullet,2} - \Gij{1}{2}) \pgd^2 + (1 - g^{\bullet,1} - g^{\bullet,2} + \Gij{1}{2}) \pbd^2
        \\
        g^{2,1} & = \Gij{2}{1} \pgc^1 + (g^{\bullet,2} - \Gij{2}{1}) \pbc^1 + (g^{\bullet,1} - \Gij{2}{1}) \pgd^1 + (1 - g^{\bullet,2} - g^{\bullet,1} + \Gij{2}{1}) \pbd^1
        \\
        g^{2,2} & = \Gij{2}{2} \pgc^2 + (g^{\bullet,2} - \Gij{2}{2}) (\pgd^2 + \pbc^2) + (1 - 2 g^{\bullet,2} + \Gij{2}{2}) \pbd^1.
    \end{split}
\end{equation*}
When $\nu_1 \to 1$ (i.e., there is only one group), this reduces to
\begin{equation}
    g = g_2 \pgc + (g - g_2) (\pgd + \pbc) + (1 - 2g + g_2) \pbd,
\end{equation}
which is the standard expression for one group with private reputations.
When information about reputations is shared within a group, individuals can benefit both by choosing a more socially beneficial norm (one that minimizes their risk of unreciprocated cooperation) and by aligning themselves with the reputational assessments of a larger group.
When both groups rely solely on private assessments, the second advantage is weakened, because merely following the same norm as a given group is not sufficient to ensure a good reputation in the eyes of that group: this is especially true of \textit{Stern Judging} and \textit{Shunning}, which are relatively intolerant of disagreement.

\begin{figure}
    \begin{center}
        \includegraphics[width=6.2in]{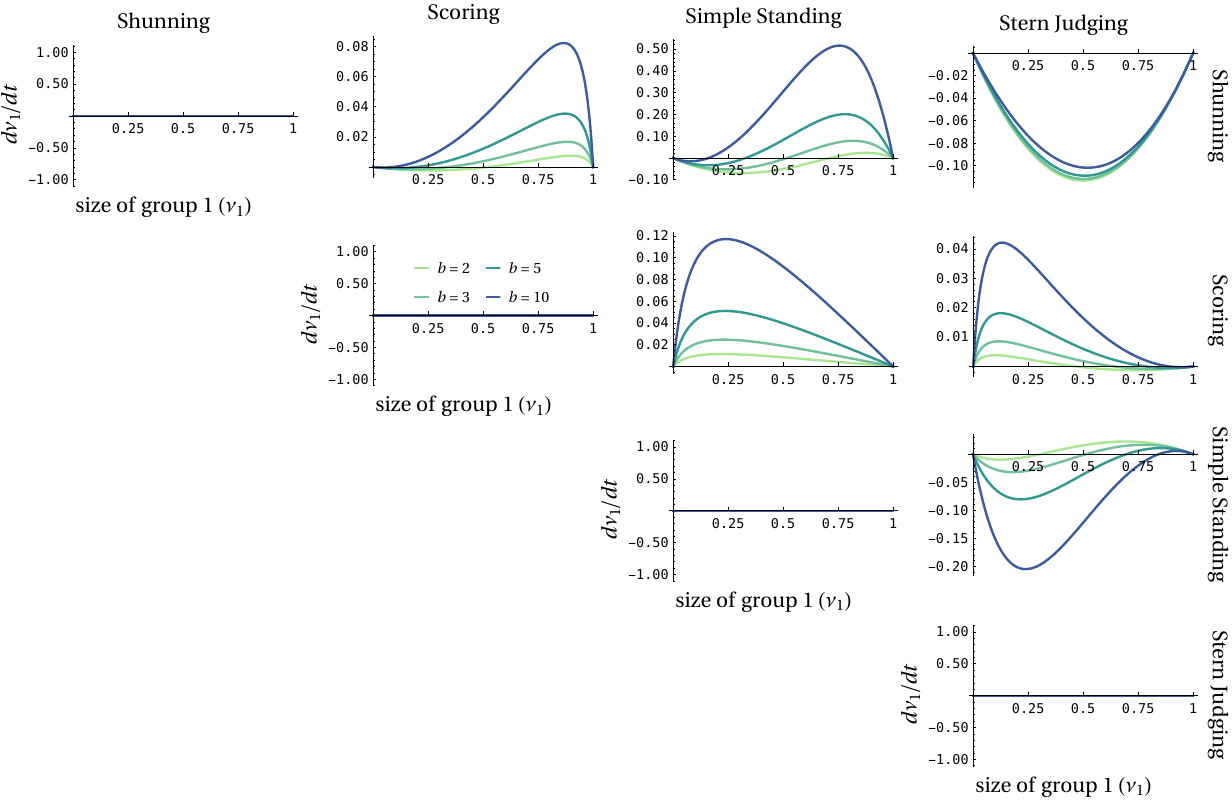}
    \end{center}
    \caption{\small{Group size dynamics for $K = 2$ groups and varying values of the benefit $b$, with \emph{private} (individually held) reputations rather than public reputations shared among group members.
    The norm used in group $1$ is along the top: the norm used in group $2$, along the right.
    In all plots, $c  = 1$, $\ass = \act = 0.02$.
    Values of $b$ are as inset in the \textit{Scoring}-\textit{Scoring} figure.}}
    \label{fig:group_switch_b_private}
\end{figure}

We consider competition between social norms under private assessment in SI figure \ref{fig:group_switch_b_private}, by allowing individuals to switch group identity and, thus, which norm they use in assessing others.
A murkier picture emerges than under group-wise public assessment.
\textit{Stern Judging} and \textit{Shunning} are both capable of ``beating'' other norms, in the sense of growing in size even when their group is less than half the population, under certain circumstances.
However, the fact that they are themselves vulnerable to invasion by defectors (whereas \textit{Simple Standing} is not) means it is difficult to draw a general conclusion about the ``strength'' of these norms.
Simple standing emerges as a ``strong'' norm that can generally outcompete other norms, especially for high values of $b$, and is itself capable of fomenting a high level of cooperation under private assessment.

\section{Insular social interactions}

We now consider the possibility that, instead of interacting equally with everyone in the population, individuals have different interaction rates with their in-group versus their out-group.
With probability $\out^{I,J}$, a possible interaction between individuals in groups $I$ and $J$ happens (for simplicity we assume $\out^{I,J} = \out^{J,I}$).
We average fitnesses over all interactions that actually happen; for an individual in group $I$, this will be given by
\begin{equation}
    \M^I = \sum_L \nu_L \out^{I,L},
\end{equation}
and the total number of interactions an individual in group $I$ engages in with someone in group $J$ will be given by $\nu_J \out^{I,J}$ (times $N$, which we divide out).

\subsection{Reputations and fitnesses}

Before writing down fitnesses, it is instructive to consider how reputations change.
We need to consider the fact that interactions that do not happen cannot be observed and therefore cannot factor into updating someone's reputation.
We thus assume that an observer is equally likely to observe any of the donor's actions \emph{that actually happened}, which means that when they consider a random interaction of $I$'s, it is with an individual in arbitrary group $L$ with probability $\nu_L \out^{I,L}/\M^I$.

\subsubsection{Cooperator reputation}

A cooperator in group $I$ can be assigned a good reputation in the eyes of group $J$ in two ways.
$J$ can observe the $I$ cooperator's interaction:
\begin{enumerate}
    \item with someone in arbitrary group $L$ (probability $\nu_L \out^{L,I}/\M^I$) whom group $J$ sees as good (probability $g^{L,J}$); the $I$ member cooperates, which yields a good reputation with probability $\pgc^J$.
    \item with someone in arbitrary group $L$ (probability $\nu_L \out^{L,I}/\M^I$) whom group $J$ sees as bad (probability $1 - g^{L,J}$); the $I$ member cooperates, which yields a good reputation with probability $\pbc^J$.
\end{enumerate}
If we define
\begin{equation*}
    \g^{I,J} = \frac{1}{\M^I} \sum_L \nu_L \out^{L,I} g^{L,J},
\end{equation*}
the average reputation (in $J$'s eyes) of the component of the population $I$ interacted with, then
\begin{equation*}
    g_X^{I,J} = \g^{I,J} \pgc^J + (1 - \g^{I,J}) \pbc^J = \g^{I,J}(\pgc - \pbc) + \pbc.
\end{equation*}
This is different from the classical case of uniform population-wide interaction, because how many interactions, and with whom, a member of $I$ engages in now depends on $I$.
That is, $\g^{I,J}$ could be thought of as $g^{\bullet,J}$, corrected for the fact that $I$ no longer interacts uniformly with the entire population: $J$ can only judge $I$ based on whom $I$ actually interacted with.
Setting all $\out^{I,J} = 1$ yields $\g^{I,J} = g^{\bullet,J}$, $\M_I = 1$, and $\G^{I,J} = \Gij{I}{J}$.

\subsubsection{Defector reputation}

A defector in group $I$ can be assigned a good reputation in the eyes of group $J$ in two ways.
$J$ can observe the $I$ defector's interaction:
\begin{enumerate}
    \item with someone in arbitrary group $L$ (probability $\nu_L \out^{L,I}/\M^I$) whom group $J$ sees as good (probability $g^{L,J}$); the $I$ member defects, which yields a good reputation with probability $\pgd^J$.
    \item with someone in arbitrary group $L$ (probability $\nu_L \out^{L,I}/\M^I$) whom group $J$ sees as bad (probability $1 - g^{L,J}$); the $I$ member defects, which yields a good reputation with probability $\pbd^J$.
\end{enumerate}
Thus,
\begin{equation}
    g_Y^{I,J} = \g^{I,J} \pgd^J + (1 - \g^{I,J}) \pbd^J = \g^{I,J}(\pgd - \pbd) + \pbd.
    \label{eq:insularity_defector}
\end{equation}

\subsubsection{Discriminator reputation}

A discriminator in group $I$ can gain a good reputation in the eyes of group $I$ (their in-group)
in two ways.
$I$ can observe the $I$ discriminator's interaction:
\begin{enumerate}
    \item with someone in arbitrary group $L$ (probability $\nu_L \out^{L,I}/\M^I$) whom group $I$ sees as good (probability $g^{L,I}$); the $I$ discriminator cooperates, which yields a good reputation with probability $\pgc^J$.
    \item with someone in arbitrary group $L$ (probability $\nu_L \out^{L,I}/\M^I$) whom group $I$ sees as bad (probability $1 - g^{L,I}$); the $I$ discriminator defects, which yields a good reputation with probability $\pbd^J$.
\end{enumerate}
A discriminator in group $I$ can gain a good reputation in the eyes of group $J \neq I$ (their out-group) in four ways.
$J$ can observe the $I$ discriminator's interaction:
\begin{enumerate}
    \item with someone in an arbitrary group $L$ following strategy $s$ (probability $\nu_L \out^{L,I} f_s^L/\M^I$) whom $I$ sees as good (probability $g_s^{L,I}$) and whom $J$ also sees as good (probability $g_s^{L,J}$); the $I$ discriminator cooperates, which yields a good reputation with probability $\pgc$.
    \item with someone in an arbitrary group $L$ following strategy $s$ (probability $\nu_L \out^{L,I} f_s^L/\M^I$) whom $I$ sees as bad (probability $1 - g_s^{L,I}$) but whom $J$ sees as good (probability $g_s^{L,J}$); the $I$ discriminator defects, which yields a good reputation with probability $\pgd$.
    \item with someone in an arbitrary group $L$ following strategy $s$ (probability $\nu_L \out^{L,I} f_s^L/\M^I$) whom $I$ sees as good (probability $g_s^{L,I}$) but whom $J$ sees as bad (probability $1 - g_s^{L,J}$); the $I$ discriminator cooperates, which yields a good reputation with probability $\pbc$.
    \item with someone in an arbitrary group $L$ following strategy $s$ (probability $\nu_L \out^{L,I} f_s^L/\M^I$) whom $I$ sees as bad (probability $1 - g_s^{L,I}$) and whom $J$ also sees as bad (probability $1 - g_s^{L,J}$); the $I$ discriminator defects, which yields a good reputation with probability $\pbd$.
\end{enumerate}
We can sum over all groups and strategy combinations to obtain
\begin{equation*}
    \begin{split}
        \frac{1}{\M^I} \sum_L \nu_L \out^{L,I} \sum_s f_s g_s^{L,I} g_s^{L,J} & = \G^{I,J},
        \\
        \frac{1}{\M^I} \sum_L \nu_L \out^{L,I} \sum_s f_s (1 - g_s^{L,I}) g_s^{L,J} & = \g^{I,J} - \G^{I,J},
        \\
        \frac{1}{\M^I} \sum_L \nu_L \out^{L,I} \sum_s f_s g_s^{L,I}(1 - g_s^{L,J}) & = \g^{I,I} - \G^{I,J},
        \\
        \frac{1}{\M^I} \sum_L \nu_L \out^{L,I} \sum_s f_s (1 - g_s^{L,I})(1 - g_s^{L,J}) & = 1 - \g^{I,J} - \g^{I,I} + \G^{I,J}.
    \end{split}
\end{equation*}
Thus,
\begin{equation}
    \begin{split}
        g_Z^{I,J} & = \delta_{I,J} \big[ \g^{I,J} \pgc^J + (1 - \g^{I,J}) \pbd^J \big]
        \\
        & \ \ \ + (1 - \delta_{I,J}) \big[ \G^{I,J} \pgc^J + (\g^{I,J} - \G^{I,J}) \pgd^J + (\g^{I,I} - \G^{I,J}) \pbc^J + (1 - \g^{I,J} - \g^{I,I} + \G^{I,J}) \pbd^J \big]
        \\
        & = \delta_{I,J} \big[ \g^{I,J} \pgc^J + (1 - \g^{I,J}) \pbd^J \big]
        \\
        & \ \ \ + (1 - \delta_{I,J}) \big[ \G^{I,J} (\pgc^J - \pgd^J - \pbc^J + \pbd^J) + \g^{I,J}(\pgd^J - \pbd^J) + \g^{I,I}(\pbc^J - \pbd^J) + \pbd^J \big].
    \end{split}
    \label{eq:Z_reps_insularity}
\end{equation}

\subsubsection{Fitnesses}

We can now write down fitnesses.
An individual in group $I$ acquires a payoff $b^{J,I}$ for each group $J$ interaction either with a cooperator or with a discriminator who sees them as good.
In group $I$, a cooperator pays cost $c^{I,J}$ in each interaction they engage in, and a discriminator pays cost $c^{I,J}$ in each interaction with someone whom they see as good.
An arbitrary individual in group $I$ engages in $\M^I$ interactions.
If the individual is a discriminator, then of these, $\nu_J \out^{J,I}$ interactions will be with someone in group $J$, and the discriminator will regard them as good with probability $g^{J,I}$.

Finally, we average the payoffs differently.
In the no-insularity case, we divide payoffs by all $N$ interactions an individual engages in.
With insularity, individuals engage in $\M^I$ interactions (times $N$), so we normalize by $\M^I$ to obtain their interaction-averaged payoff.
Thus, the average payoff for each of the three strategic types in an arbitrary group $I$ is
\begin{equation}
    \begin{split}
        \Pi_X^I & = \frac{1}{\M^I} (1 - \act) \Big\{ \sum_J \nu_J \out^{I,J} \big[ b^{J,I} (f_X^J + f_Z^J g_X^{I,J}) - c^{I,J} \big] \Big\}
        \\
        \Pi_Y^I & = \frac{1}{\M^I} (1 - \act) \Big\{ \sum_J \nu_J \out^{I,J} \big[ b^{J,I} (f_X^J + f_Z^J g_Y^{I,J})\big] \Big\}
        \\
        \Pi_Z^I & = \frac{1}{\M^I} (1 - \act) \Big\{ \sum_J \nu_J \out^{I,J} \big[ b^{J,I} (f_X^J + f_Z^J g_Z^{I,J}) - c^{I,J} g^{J,I} \big] \Big\}.
    \end{split}
\end{equation}

\subsubsection{Hybrid strategies}

The above treatment makes it trivial to write down reputations for strategies that distinguish explicitly between in-group and out-group, for example, cooperate with one's in-group and discriminate with one's out-group.
These ``hybrid'' strategies were prominently featured in, e.g., \citet{masuda2012ingroup}.
As an example, we present the strategy of discriminating with one's in-group and defecting with one's out-group, which we denote $ZY$.
We have
\begin{equation}
    \begin{split}
        g_{ZY}^{I,J} & = \delta_{I,J} \big[ \g^{I,J} \pgc^J + (1 - \g^{I,J}) \pbd^J \big]
        \\
        & \ \ \ + (1 - \delta_{I,J}) \big[ \g^{I,J} \pgd^J + (1 - \g^{I,J}) \pbd^J \big]
        \\
        & = \delta_{I,J} \big[ \g^{I,J} \pgc^J + (1 - \g^{I,J}) \pbd^J \big]
        \\
        & \ \ \ + (1 - \delta_{I,J}) \big[ \g^{I,J} \pgd^J + (1 - \g^{I,J}) \pbd^J \big].
    \end{split}
\end{equation}
The fitness term is easily written down depending on the other strategies in the population; a $ZY$ individual accrues a benefit from cooperators, from discriminators who see them as good, and from other $ZY$ individuals \emph{in the same group} who see them as good, and they pay the cost for any individuals \emph{in the same group} whom their group sees as good.

\subsubsection{Favoring of in-group interactions}

Suppose we have
\begin{equation}
    \out^{I,J} = \delta_{I,J} + (1 - \delta_{I,J}) \out,
\end{equation}
i.e., individuals always interact with their in-group, but out-group interactions only happen with probability $\out$.
Suppose, also, that benefits and costs do not vary by group.
We can study this numerically: see main text figure $3$.

When $\out \to 0$, we have
\begin{equation}
    \begin{split}
        \M^I & = \sum_L \nu_L \out^{I,L} = \nu_I,
        \\
        \g^{I,J} & = \frac{1}{\M^I} \sum_L \nu_L \out^{L,I} g^{L,J} = \frac{1}{\M^I} \nu_I g^{I,J} = g^{I,J},
        \\
        \Gamma^{I,J} & = \frac{1}{\M^I} \sum_L \nu_L \out^{L,I} \sum_i f_s g_s^{L,I} g_s^{L,J} = \frac{1}{\M^I} \nu_I \sum_i f_s g_s^{L,I} g_s^{L,J} = \Gij{I}{J},
        \\
        \Pi_X^I & = \frac{1}{\M^I} (1 - \act) \Big\{ \sum_J \nu_J \out^{I,J} \big[ b (f_X^J + f_Z^J g_X^{I,J}) - c \big] \Big\} = (1 - \act) [b (f_X^I + f_Z^I g_X^{I,I}) - c],
        \\
        \Pi_Y^I & = \frac{1}{\M^I} (1 - \act) \Big\{ \sum_J \nu_J \out^{I,J} \big[ b (f_X^J + f_Z^J g_Y^{I,J})\big] \Big\} = (1 - \act) [b (f_X^I + f_Z^I g_Y^{I,I})],
        \\
        \Pi_Z^I & = \frac{1}{\M^I} (1 - \act) \Big\{ \sum_J \nu_J \out^{I,J} \big[ b (f_X^J + f_Z^J g_Z^{I,J}) - c g^{J,I} \big] \Big\} = (1 - \act) [b (f_X^I + f_Z^I g_Z^{I,I}) - c g^{I,I}].
    \end{split}
\end{equation}
Thus, when groups are completely insular, they only accrue payoffs from (and pay costs for) interactions with their own group: $K$ groups effectively behave as $K$ completely disjoint, independent populations.

\subsection{Weighted average reputations under well-mixed copying}

Under well-mixed copying, it is possible to study the dynamics of each strategic type using a sort of weighted average.
We have
\begin{equation*}
    \begin{split}
        \Pi_Z & = \sum_I \nu_I \Pi_Z^I
        \\
        & = (1 - \act) \sum_I \nu_I \frac{1}{\M^I} \sum_J \nu_J \out^{I,J} \big[ b (f_X + f_Z g_Z^{I,J}) - c g^{J,I} \big]
        \\
        & = (1 - \act) \Big\{ b \sum_I \nu_I \frac{1}{\M^I} \sum_J \nu_J \out^{I,J} (f_X + f_Z g_Z^{I,J}) - \sum_I \nu_I \frac{1}{\M^I} \sum_J \nu_J \out^{I,J} g^{J,I} \Big\}
        \\
        & = (1 - \act) \Big[ b f_X \sum_I \nu_I \frac{1}{\M^I} \sum_J \nu_J \out^{I,J} + b f_Z \sum_I \nu_I \frac{1}{\M^I} \sum_J \nu_J \out^{I,J} g_Z^{I,J} - \sum_I \nu_I \frac{1}{\M^I} \sum_J \nu_J \out^{I,J} g^{J,I} \Big]
        \\
        & = (1 - \act) \Big[ b f_X \sum_I \nu_I \frac{1}{\sum_J \nu_J \out^{I,J}} \sum_J \nu_J \out^{I,J} + b f_Z \sum_I \nu_I \frac{1}{\sum_J \nu_J \out^{I,J}} \sum_J \nu_J \out^{I,J} g_Z^{I,J}
        \\
        & \ \ \ - \sum_I \nu_I \frac{1}{\sum_J \nu_J \out^{I,J}} \sum_J \nu_J \out^{I,J} g^{J,I} \Big]
        \\
        & = (1 - \act) \Big[ b (f_X + f_Z \tilde{g}_Z) - c \hat{g} \Big].
    \end{split}
\end{equation*}
Here,
\begin{equation}
    \tilde{g}_s = \sum_I \nu_I \frac{1}{\M^I} \sum_J \nu_J \out^{I,J} g_s^{I,J}
    \label{eq:tilde_g}
\end{equation}
is a weighted average of the population's view of an individual following strategy $s$; it corresponds to the possible interactions an individual with reputation $g_s^{I,J}$ may engage in, weighted by the probability  $\out^{I,J}$ that those interactions actually occur.
Note that we can also write
\begin{equation}
    \tilde{g} = \sum_I \nu_I \g^{I,I}.
\end{equation}
Likewise,
\begin{equation}
    \hat{g} = \sum_I \nu_I \frac{1}{\M^I} \sum_J \nu_J \out^{I,J} g^{J,I}
\end{equation}
is a weighted average of an arbitrary individual's view of the rest of the population.
Note that $\hat{g} = \tilde{g}$ if all the $\nu_I$ are equal or all the $\out^{I,J}$ are equal.
By identical reasoning we will have
\begin{equation}
    \begin{split}
        \Pi_X & = (1 - \act) \Big[ b ( f_X + f_Z \tilde{g}_X ) - c \Big],
        \\
        \Pi_Y & = (1 - \act) \Big[ b ( f_X + f_Z \tilde{g}_Y ) \Big].
    \end{split}
\end{equation}

\subsection{Equally sized groups}

Suppose now that every group has the same size $1/K$ and the insularities are given by $\out^{I,J} = \delta_{I,J} + (1 - \delta_{I,J}) \out$, i.e., individuals always interact with fellow in-group members but only interact with out-group individuals with probability $\out$.
By symmetry, reputational views will only differ depending on whether the observer is in the donor's in-group or out-group.
Define $g^\en = g^{I,I}$ and $g^\ex = g^{I,J}|_{I \neq J}$, i.e., $g^\en$ is an individual's view of their in-group and $g^\ex$ their out-group.
The weighted average reputation $\tilde{g}$ (Eq.~\eqref{eq:tilde_g}) can be expanded out thus:
\begin{equation}
    \begin{split}
        \tilde{g} = \sum_I \nu_I \frac{1}{\M^I} \sum_J \nu_J \out^{I,J} g^{I,J} & = \Big( \frac{1}{K} \Big)^2 \frac{1}{1/K + \out (K-1)/K} \sum_I \sum_J \out^{I,J} g^{I,J}
        \\
        & = \Big( \frac{1}{K} \Big)^2 \frac{K}{1 + \out (K-1)} (K g^\en + \out K(K-1) g^\ex)
        \\
        & = \frac{ g^\en + \out (K - 1)g^\ex }{1 + \out (K-1)}.
    \end{split}
    \label{eq:g_tilde_equal_size}
\end{equation}
In a population of discriminators, $g^{\en}$ and $g^{\ex}$ can be expanded out:
\begin{equation}
    \begin{split}
        g^{\en} & = \g^{I,I} (\pgc - \pbd) + \pbd
        \\
        & = \frac{1}{\M^I} \sum_L \nu_L \out^{L,I} g^{L,I} (\pgc - \pbd) + \pbd
        \\
        & = \frac{1}{1+\out(K-1)} \sum_L \out^{L,I} g^{L,I} (\pgc - \pbd) + \pbd
        \\
        & = \frac{1}{1+\out(K-1)} \big[ g^{\en} + (K-1) \out g^{\ex} \big] (\pgc - \pbd) + \pbd
    \end{split}
    \label{eq:g_in_equal_size}
\end{equation}
and
\begin{equation}
    \begin{split}
        g^{\ex} & = \G^{I,J} (\pgc - \pgd - \pbc + \pbd) + \g^{I,J}(\pgd - \pbd) + \g^{I,I}(\pbc - \pbd) + \pbd
        \\
        & = \frac{1}{\M^I} \sum_L \nu_L \out^{L,I} g^{L,I} g^{L,J} (\pgc - \pgd - \pbc + \pbd)
        \\
        & \ \ \ + \frac{1}{\M^I} \sum_L \nu_L \out^{L,I} g^{L,J} (\pgd - \pbd)
        \\
        & \ \ \ + \frac{1}{\M^I} \sum_L \nu_L \out^{L,I} g^{L,I} (\pbc - \pbd)
        \\
        & \ \ \ + \pbd
        \\
        & = \frac{1}{1 + \out(K-1)} \big[ (1 + \out) g^{\en} g^{\ex} + (K-2) \out (g^{\ex})^2) \big] (\pgc - \pgd - \pbc + \pbd)
        \\
        & \ \ \ + \frac{1}{1 + \out(K-1)} \big[ \out g^{\en} + \big( 1 + (K-2) \out \big) g^{\ex} \big] (\pgd - \pbd)
        \\
        & \ \ \ + \frac{1}{1+\out(K-1)} \big[ g^{\en} + (K-1) \out g^{\ex} \big] (\pbc - \pbd)
        \\
        & \ \ \ + \pbd.
    \end{split}
    \label{eq:g_out_equal_size}
\end{equation}
These equations can be solved, but their general solution is not very informative.
Note that sending $\act \to 0$
and setting $\theta = 1/[1+\out(K-1)]$ yields the corresponding expressions from \citet{nakamura2012favoritism}; see SI table \ref{tab:ingroup_outgroup_reps}.
A useful simplification is to send $\out \to 0$:
\begin{equation}
    \begin{split}
        g^{\en}|_{\out \to 0} & = \frac{\pbd}{1 - \pgc + \pbd}
        \\
        & =
        \begin{dcases}
            \frac{\ass}{1 - \epsilon + \ass} = \dfrac{\ass}{2 \ass + \act - 2 \act \ass} & \text{\emph{Shunning}, \emph{Scoring}},
            \\
            \frac{1 - \ass}{2 - \epsilon + \ass} = \dfrac{1 - \ass}{1 + \act - 2 \act \ass} & \text{\emph{Stern Judging}, \emph{Simple Standing}},
        \end{dcases}
        \\
        g^{\ex}|_{\out \to 0} & = \frac{\pbd(1 + \pbc - \pgc)}{\pbd(2 + \pbc - 2 \pgc) + (1 - \pgc)(1 - \pgd)}
        \\
        & =
        \begin{dcases}
            \frac{\ass(1 + \ass - \epsilon)}{\ass (2 + \ass - 2 \epsilon) + (1 - \epsilon)(1 - \ass)} = \frac{\ass[2\ass(1 - \act) + \act]}{\ass(1 + 2\ass)(1 - \act) + \act} & \text{\emph{Shunning}},
            \\
            \frac{\ass}{1 - \epsilon + \ass} = \frac{\ass}{2\ass(1 - \act) + \act} & \text{\emph{Scoring}},
            \\
            \frac{(1 - \ass)(2 - 2\epsilon)}{(1 - \ass)(4 - 4 \epsilon)} = \frac{1}{2} & \text{\emph{Stern Judging}},
            \\
            \frac{(1 - \ass)(2 - \ass - \epsilon)}{(1 - \ass)(3 - \ass - 2 \epsilon) + (1 - \epsilon)(1 - \ass)} = \frac{1 + \act - 2\ass\act}{1 + 2\ass + 3 \act - 6 \ass \act} & \text{\emph{Simple Standing}}.
        \end{dcases}
    \end{split}
    \label{eq:w=0}
\end{equation}
The expressions for $g^{\en}|_{\out \to 0}$ are the same as the main text expressions for $g$ with $K = 1$ and $f_Z = 1$.

\begin{table}[!ht]
    \centering
    \begin{tabular}{|Sc|Sc|Sc|}
        \hline
        norm & $g^{\en}$ & $g^{\ex}$
        \\
        \hline
        \makecell{\textit{Stern Judging}
        \\
        ($K \geq 3$)
        } & $\dfrac{2(1 - \ass)(1 + [K-1] \out) - \act (2 + [K-1] \out)}{2(1 + [K-1] \out)}$ & $\dfrac{K - 2 -\act}{2 (K-2)}$
        \\
        \hline
        \makecell{\textit{Stern Judging}
        \\
        ($K = 2, \act > 0$)
        }& $1 - \dfrac{2\act + (2 + \out) \ass}{2\sqrt{1 + \out}}$ &
        \makecell{ $\dfrac{1}{4\out \sqrt{1 + \out}}$ \\ $ \times \Big\{ 4(1 + \out - \sqrt{1 + \out}) + 2\act(2 + \out - 2\sqrt{1 + \out})$ \\ $ + \ass[8 - 8\sqrt{1 + \out} + \out(8 + \out - 4\sqrt{1 + \out})]\Big\}$}
        \\
        \hline
        \makecell{\textit{Stern Judging}
        \\
        ($K = 2, \act = 0$)
        } & $1 - \ass$ & $1/2$
        \\
        \hline
        \textit{Simple Standing} & $1 - \ass - \act$ & $1 - (\ass + \act) [2 + (K-1) \out] $
        \\
        \hline
        \textit{Shunning} & $\dfrac{\ass(1 + 2[K-1] \out)}{(K-1) \out}$ & $\ass$
        \\
        \hline
    \end{tabular}
    \caption{
    \small{In-group and out-group reputations for all-discriminator populations consisting of $K$ equally sized groups, to \emph{first order} in $\ass$ and $\act$.
    These agree with the expressions provided in \citet{nakamura2012favoritism} given $\theta = 1/[1 + \out(K-1)]$ and $\act \to 0$.
    The expression for \emph{Shunning} is novel to our analysis: in their notation, the \emph{Shunning} in-group reputation is $(2 - \theta) \ass / (1 - \theta)$, and both the in-group and out-group reputations include $\mathcal{O}(\ass^2)$ terms that we ignore.
    The \emph{Shunning} in-group approximation breaks down as $\out \to 0$; it appears to be valid only for $\out_r \gg \act$.
    Otherwise, the exact $\out = 0$ expression (equation \eqref{eq:w=0}) is a useful approximations for $g^{\en}$.
    Likewise, the \emph{Stern Judging} ($K = 2, \act > 0$) out-group term fails when $\out \gg \act$.
    The $\act = 0$ out-group expression is a better approximation.
    Finally, \emph{Scoring} is omitted from this table because the exact expression $g = \ass/(1 - \epsilon + \ass)$ is always valid irrespective of group structure and insularity.}
    }
    \label{tab:ingroup_outgroup_reps}
\end{table}

\subsection{Invasibility of equally sized groups by defectors}

Given equations \eqref{eq:g_in_equal_size} and \eqref{eq:g_out_equal_size}, we can determine when discriminators resist invasion by defectors.
We require

\begin{equation}
    \begin{split}
        \Pi_Z & > \Pi_Y
        \\
        b \tilde{g}_Z - c \hat{g} & > b \tilde{g}_Y
        \\
        (b - c) \tilde{g} & > b \tilde{g}_Y
        \\
        \therefore \frac{b}{c} & > \frac{\tilde{g}}{\tilde{g} - \tilde{g}_Y}, \text{with}
        \\
        \tilde{g}_Y & = \sum_I \nu_I \frac{1}{\M^I} \sum_J \nu_J \out^{I,J} \big[ \g^{I,J}(\pgd - \pbd) + \pbd \big].
    \end{split}
    \label{eq:bc_condition_gY}
\end{equation}
Solving for $\tilde{g}_Y$ requires that we compute the sum at the end of equation \eqref{eq:bc_condition_gY}.
Let $\out^{\ex}$ be the out-group interaction parameter and $\out^{\en}$ the probability of in-group interactions (these are $\out$ and $1$ respectively; these terms are used solely for bookkeeping).
The sum is
\begin{equation}
    \begin{split}
        \sum_I \nu_I \frac{1}{\M^I} \sum_J \nu_J \out^{I,J} \g^{I,J} & = \sum_I \nu_I \frac{1}{\M^I} \sum_J \nu_J \out^{I,J} \sum_I \nu_I \frac{1}{\M^I} \sum_L \nu_L \out^{L,I} g^{L,J}
        \\
        & = \sum_I \sum_I \sum_J \sum_L (\nu_I)^2 \Big(\frac{1}{\M^I} \Big)^2 \nu_J \nu_L \out^{I,J} \out^{L,I} g^{L,J}
        \\
        & = \Big( \frac{1}{K} \Big)^4 \Big( \frac{1}{1/K + \out (K-1)/K} \Big)^2 \sum_I \sum_I \sum_J \sum_L \out^{I,J} \out^{L,I} g^{L,J}
        \\
        & = \Big( \frac{1}{K} \Big)^4 \Big( \frac{K}{1 + \out (K-1)} \Big)^2
        \\
        & \ \ \ \times \sum_I \Big[ K(K-1)(K-2) \out^{\ex} \out^{\ex} g^{\ex} + K(K-1) \out^{\en} \out^{\ex} g^{\ex}
        \\
        & \ \ \ + K(K-1) \out^{\ex} \out^{\en} g^{\ex} + K(K-1) \out^{\ex} \out^{\ex} g^{\en} + K \out^\en \out^\en g^\en
        \\
        & = \Big( \frac{1}{K} \Big)^4 \Big( \frac{K}{1 + \out (K-1)} \Big)^2
        \\
        & \ \ \ \times \sum_I \Big[ K(K-1)(K-2) \out^2 g^\ex + 2K(K-1) \out g^\ex
        \\
        & \ \ \ + K(K-1) \out^2 g^\en + K g^\en \Big]
        \\
        & = \Big( \frac{1}{K} \Big)^3 \Big( \frac{K}{1 + \out (K-1)} \Big)^2
        \\
        & \ \ \ \times \Big[ K(K-1)(K-2) \out^2 g^\ex + 2K(K-1) \out g^\ex + K(K-1) \out^2 g^\en + K g^\en \Big]
        \\
        & = \Big( \frac{1}{1 + \out (K-1)} \Big)^2
        \\
        & \ \ \ \times \Big[ (K-1)(K-2) \out^2 g^\ex + 2(K-1) \out g^\ex + (K-1) \out^2 g^\en +  g^\en \Big]
        \\
        & = \frac{g^{\ex} \out (K-1) [(K-2)\out + 2] + g^{\en}[(K-1)\out^2 + 1]}{[1 + \out (K-1)]^2}.
    \end{split}
    \label{eq:gamma_IJ_sum}
\end{equation}
Thus,
\begin{equation}
    \tilde{g}_Y = \frac{g^{\ex} \out (K-1) [(K-2)\out + 2] + g^{\en}[(K-1)\out^2 + 1]}{[1 + \out (K-1)]^2}(\pgd - \pbd) + \pbd.
    \label{eq:tilde_gY}
\end{equation}
Substituting equations \eqref{eq:tilde_gY}, \eqref{eq:g_tilde_equal_size}, \eqref{eq:g_in_equal_size}, and \eqref{eq:g_out_equal_size} into equation \eqref{eq:bc_condition_gY} yields the condition that $b/c$ must be greater than a fraction whose numerator is given by
\begin{equation*}
    [1+(K-1)\out][g^{\en} + g^{\ex}(K-1)\out]
\end{equation*}
and denominator by
\begin{equation*}
    \begin{split}
    & g^{\en} \big[ \big( (K-1) \out^2+1\big) (\pbd - \pgd)  + (K-1)\out + 1 \big]
   \\
   & \ \ \ +  g^{\ex} (K-1) \out \big[ \big( (K-2) \out+2\big) (\pbd - \pgd)+(K-1)\out + 1 \big]
   \\
   & \ \ \ - \pbd \big[(K-1)\out + 1 \big]^2.
   \end{split}
\end{equation*}
These are consistent with the expressions from \citet{nakamura2012favoritism}.
The $b/c$ condition can also be expressed in terms of the weighted average reputation $\tilde{g}$.
We solve for $g^\en$ and $g^\ex$ self-consistently via
\begin{equation}
    \begin{split}
        g^\en|_{f_Z = 1} & = \g^{I,I}(\pgc - \pbd) + \pbd
        \\
        & = \Big( \frac{1}{\M^I} \sum_L \nu_L \out^{L,I} g^{L,I} \Big)(\pgc - \pbd) + \pbd
        \\
        & = \frac{K}{1 + \out(K-1)} \Big(\frac{1}{K} g^\en + \out \frac{K-1}{K} g^\ex \Big)(\pgc - \pbd) + \pbd
        \\
        & = \frac{g^\en + \out (K-1) g^\ex}{1 + \out(K-1)}(\pgc - \pbd) + \pbd
        \\
        \therefore g^\en \Big(1 - \frac{1}{1 + \out (K-1)} \Big) & = \frac{\out (K-1) g^\ex}{1 + \out(K-1)}(\pgc - \pbd) + \pbd
        \\
        \therefore g^\en \frac{\out(K-1)}{1 + \out (K-1)} & = \frac{\out (K-1) g^\ex}{1 + \out(K-1)}(\pgc - \pbd) + \pbd
        \\
        \therefore g^\en  & = g^\ex(\pgc - \pbd) + \frac{1 + \out(K-1)}{\out(K-1)}\pbd
    \end{split}
    \label{eq:gamma_II_Z}
\end{equation}
or, equivalently,
\begin{equation*}
    g^\en = \tilde{g} (\pgc - \pbd) + \pbd.
\end{equation*}
Likewise
\begin{equation*}
    \begin{split}
        \frac{g^\en + \out(K-1) g^\ex}{1 + \out(K-1)} & = \tilde{g}
        \\
        \therefore \frac{\out(K-1) g^\ex}{1 + \out(K-1)} & = \tilde{g} - \frac{g^\en}{1 + \out(K-1)}
        \\
        \therefore g^\ex & = \frac{\tilde{g}[1 + \out(K-1)] - g^\en}{\out(K-1)}
        \\
        & = \frac{\tilde{g}[1 + \out(K-1) - \pgc + \pbd] - \pbd}{\out(K-1)}
    \end{split}
\end{equation*}
Combining equations \eqref{eq:gamma_II_Z}, \eqref{eq:gamma_IJ_sum}, and \eqref{eq:insularity_defector} yields
\begin{equation*}
    \begin{split}
        (b - c) \tilde{g} & > b \Big[ \frac{g^{\ex} \out (K-1) [(K-2)\out + 2] + g^{\en}[(K-1)\out^2 + 1]}{[1 + \out (K-1)]^2}(\pgd - \pbd) + \pbd \Big]
        \\
        \therefore \frac{b-c}{b} \tilde{g} & > \frac{\{\tilde{g}[1 + \out(K-1) - \pgc + \pbd] - \pbd \}[(K-2)\out + 2] + [\tilde{g}(\pgc - \pbd) + \pbd][(K-1)\out^2 + 1]}{[1 + \out (K-1)]^2}
        \\
        & \ \ \ \times \big(\pgd - \pbd \big) + \pbd
        \\
        \therefore \frac{b-c}{b} \tilde{g} & > \tilde{g}\frac{[2 - \pgc + \pbd + \out(K - 2 + \pgc - \pbd)](\pgd - \pbd)}{1 + \out(K-1)}+ \frac{(1 - \out) \pbd (\pgd - \pbd)}{1 + \out(K-1)} + \pbd.
    \end{split}
\end{equation*}
This can be rearranged to yield
\begin{equation*}
    \tilde{g}\Big(1 - \frac{c}{b} + \frac{[2 - \pgc + \pbd + \out(K - 2 + \pgc - \pbd)](\pbd - \pgd)}{1 + \out(K-1)} \Big) > \frac{(1 - \out) \pbd (\pgd - \pbd)}{1 + \out(K-1)} + \pbd.
\end{equation*}
Thus, discriminators resist invasion by defectors provided
\begin{equation*}
    \begin{split}
        \tilde{g} & > \frac{(1 - \out)\pbd(\pbd - \pgd) + [1 + \out(K-1)] \pbd}{[2 - \pgc + \pbd + \out(K - 2 + \pgc - \pbd)](\pbd - \pgd) + (1 - c/b)[1 + \out(K-1)]}, \text{or}
        \\
        \tilde{g} & > \frac{ \pbd[K \out + (1 - \out)(1 - \pgd + \pbd)] }{[2 - \pgc + \pbd + \out(K - 2 + \pgc - \pbd)](\pbd - \pgd) + (1 - c/b)[1 + \out(K-1)]}.
    \end{split}
\end{equation*}
Setting $\out = 1$ yields equation \eqref{eq:coop_condition}.

\subsection{Invasibility of equally sized groups by cooperators}

Bolstered by our preceding analysis, we also consider when discriminators resist invasion by cooperators; such invasion can occur, e.g., under \textit{Simple Standing}.
We require
\begin{equation*}
    \begin{split}
        \Pi_Z & > \Pi_X
        \\
        (b-c) \tilde{g} & > b \tilde{g}_X - c
        \\
        \frac{b}{c} & > \frac{\tilde{g}-1}{\tilde{g} - \tilde{g}_X}, \text{with}
        \\
        \tilde{g}_X & = \sum_I \nu_I \frac{1}{\M^I} \sum_J \nu_J \out^{I,J} \big[ \g^{I,J}(\pgc - \pbc) + \pbc \big]
        \\
        & = \frac{g^{\ex} \out (K-1) [(K-2)\out + 2] + g^{\en}[(K-1)\out^2 + 1]}{[1 + \out (K-1)]^2}(\pgc - \pbc) + \pbc.
    \end{split}
\end{equation*}
The critical $b/c$ value thus simplifies to a fraction whose numerator is given by
\begin{equation*}
    [1+(K-1)\out][g^{\en} + (g^{\ex} - 1)(K-1)\out - 1]
\end{equation*}
and denominator by
\begin{equation*}
    \begin{split}
    & g^{\en} \big[ \big( (K-1) \out^2+1\big) (\pbc - \pgc)  + (K-1)\out + 1 \big]
   \\
   & \ \ \ +  g^{\ex} (K-1) \out \big[ \big( (K-2) \out+2\big) (\pbc - \pgc)+(K-1)\out + 1 \big]
   \\
   & \ \ \ - \pbc \big[(K-1)\out + 1 \big]^2,
   \end{split}
\end{equation*}
which again is consistent with \citet{nakamura2012favoritism}.
We can likewise express this condition in terms of $\tilde{g}$:
\begin{equation*}
    \begin{split}
        (b-c) \tilde{g} & > b \Big[ \frac{g^{\ex} \out (K-1) [(K-2)\out + 2] + g^{\en}[(K-1)\out^2 + 1]}{[1 + \out (K-1)]^2}(\pgc - \pbc) + \pbc \Big] - c
        \\
        \therefore \frac{b-c}{b} \tilde{g} & > \frac{\{\tilde{g}[1 + \out(K-1) - \pgc + \pbd] - \pbd \}[(K-2)\out + 2] + [\tilde{g}(\pgc - \pbd) + \pbd][(K-1)\out^2 + 1]}{[1 + \out (K-1)]^2}
        \\
        & \ \ \ \times \big(\pgc - \pbc \big) + \pbc - \frac{c}{b}
        \\
        \therefore \frac{b-c}{b} \tilde{g} & > \tilde{g}\frac{[2 - \pgc + \pbd + \out(K - 2 + \pgc - \pbd)](\pgc - \pbc)}{1 + \out(K-1)}+ \frac{(1 - \out) \pbd (\pgc - \pbc)}{1 + \out(K-1)} + \pbc - \frac{c}{b}.
    \end{split}
\end{equation*}
This can be rearranged to yield
\begin{equation*}
    \tilde{g}\Big(1 - \frac{c}{b} + \frac{[2 - \pgc + \pbd + \out(K - 2 + \pgc - \pbd)](\pbc - \pgc)}{1 + \out(K-1)} \Big) > \frac{(1 - \out) \pbd (\pgc - \pbc)}{1 + \out(K-1)} + \pbc - \frac{c}{b}.
\end{equation*}
Thus, discriminators resist invasion by cooperators provided
\begin{equation*}
    \begin{split}
        \tilde{g} & > \frac{(1 - \out)\pbd(\pgc - \pbc) + [1 + \out(K-1)](\pbc - c/b)}{[2 -\pgc + \pbd + \out(K - 2 + \pgc - \pbd)](\pbc-\pgc) + (1 - c/b)[1 + \out(K-1)]}
    \end{split}
\end{equation*}
for \textit{Stern Judging} and \textit{Shunning} (an inequality that, in general, never obtains--the right hand side is greater than $1$).
For \textit{Scoring} and \textit{Simple Standing}, the sign of the inequality is reversed, as the denominator is negative.

\subsection{Norm competition with insularity and variable costs and benefits}

If benefits and costs vary, so that $\bin$ and $\cin$ are the benefit and cost associated with an intra-group interaction and $\bout$ and $\cout$ are the cost associated with an inter-group interaction, then $\nu_1$ grows if

\begin{equation}
    \begin{split}
        \dot{\nu}_1 & > 0
        \\
        \therefore \Pi_Z^1 & > \Pi_Z^2
        \\
        \therefore \frac{1}{\M^1} \big[ \nu_1 ( \bin g^{1,1} - \cin g^{1,1}) + \nu_2 \out ( \bout g^{1,2} - \cout g^{2,1}) \big] & > \frac{1}{\M^2} \big[ \nu_1 \out ( \bout g^{2,1} - \cout g^{1,2}) + \nu_2 ( \bin g^{2,2} - \cin g^{2,2}) \big]
        \\
        \therefore \nu_1 \big[ \frac{1}{\M^1} (\bin g^{1,1} - \cin g^{1,1}) - \frac{1}{\M^2} \out (\bout g^{2,1} - \cout g^{1,2}) \big] & > \nu_2 \big[ \frac{1}{\M^2} (\bin g^{2,2} - \cin g^{2,2}) - \frac{1}{\M^1} \out (\bout g^{1,2} - \cout g^{2,1}) \big]
        \\
        \therefore \nu_1 \big[ \M^2 (\bin g^{1,1} - \cin g^{1,1}) - \M^1 \out (\bout g^{2,1} - \cout g^{1,2}) \big] & > \nu_2 \big[ \M^1 (\bin g^{2,2} - \cin g^{2,2}) - \M^2 \out (\bout g^{1,2} - \cout g^{2,1}) \big]
        \\
        \therefore \frac{\nu_1}{\nu_2} & > \frac{\M^1 (\bin g^{2,2} - \cin g^{2,2}) - \M^2 \out (\bout g^{1,2} - \cout g^{2,1})}{\M^2 (\bin g^{1,1} - \cin g^{1,1}) - \M^1 \out (\bout g^{2,1} - \cout g^{1,2})}
    \end{split}
\end{equation}
When we hold $\cin$ and $\cout$ constant but set $\bout > \bin$, it becomes \emph{harder} for \textit{Stern Judging} (group $1$) to beat \textit{Shunning} (group $2$).
When $\nu = 1/2$, we have (for $\ass = \act = .02$) $g^{1,1} = .96, g^{2,1} = .42, g^{1,2} = .09, g^{2,2} = .13$.
What this means is that group $1$ cooperates with group $2$ more than the reverse, so increases in $\bout$ are not reciprocated.
Thus, raising $\bout$ rather than $\bin$ increases the rate of unreciprocated fitness gain by group $2$, allowing it to outcompete group $1$.

\subsection{Dependence of group fitness on insularity}

In this section, we consider how the fitness of a group of discriminators depends on the insularity parameter $\out$, under the assumptions of well-mixed copying, $K$ equally sized groups, and $\out^{I,J} = \delta_{I,J} + (1 - \delta_{I,J}) \out$.
This can shine light on how insularity might be expected to evolve in a group-level selection scenario, in which an entire group can be replaced by a group with a different level of insularity.
(The problem of how insularity might evolve at the individual level is deferred to section \ref{sec:insular_invasibility}.)
We have
\begin{equation*}
    \begin{split}
        \M^I & = \sum_L \nu_L \out^{I,L}
        \\
        & = \frac{1 + \out (K-1)}{K} = \M \text{~(no dependence on $I$})
        \\
        \therefore \Pi_Z & = \sum_I \nu_I \Pi_Z^I
        \\
        & = \sum_I \nu_I \frac{1}{\M^I} \Big\{ \sum_J \nu_J \out^{I,J} \big[ b g^{I,J} - c g^{J,I} \big] \Big\} \text{~(dropping the $(1-\act)$ prefactor)}
        \\
        & = \frac{1}{\M} \frac{1}{K} \big[ b g^{\en} - c g^{\en} + (K-1) \out (b g^{\ex} - c g^{\ex}) \big]
        \\
        & = \frac{1}{1 + \out (K-1)} (b-c) \big[ g^{\en} + \out (K-1) g^{\ex} \big]
        \\
        & = (b-c) g^{\en} \frac{1 + \out (K-1) g^{\ex}/g^{\en}}{1 + \out (K-1)}.
    \end{split}
\end{equation*}
We have written this in a suggestive form.
Because the $\out (K-1)$ term in the numerator has a factor $g^{\ex}/g^{\en}$ attached to it, and because this factor is (for every social norm besides \textit{Scoring}) less than $1$, the numerator will generally shrink relative to the denominator as $\out$ increases and grow as $\out$ decreases.
The fitness of a group thus generally increases with decreasing $\out$.
This sensitively depends on our decision to normalize by dividing by $\M$, which is necessary to ensure that interactions that do not happen make no contribution to fitness (instead of, e.g., contributing zero fitness, which would be indistinguishable from mutual defection).
If we did not divide by $\M$, we would instead have
\begin{equation*}
    \begin{split}
        \Pi_Z & = \frac{b-c}{K} \big[ g^{\en} + \out (K-1) g^{\ex} \big]
        \\
        & = \Big( \frac{(b-c) g^{\en}}{K} \Big) \big[ 1 + \out (K-1) g^{\ex}/g^{\en} \big],
    \end{split}
\end{equation*}
which is monotonically increasing in $\out$.
Finally, if we allow out-group and in-group interactions to have different payoffs, we have
\begin{equation*}
    \begin{split}
        \Pi_Z & = \frac{1}{\M} \frac{1}{K} \big[ b^{\en} g^{\en} - c^{\en} g^{\en} + \out (K-1) (b^{\ex} g^{\ex} - c^{\ex} g^{\ex}) \big]
        \\
        & = \frac{1}{1 + \out (K-1)}\big[ (b^{\en} - c^{\en}) g^{\en} + \out (K-1) (b^{\ex} - c^{\ex}) g^{\ex} \big]
        \\
        & = \frac{(b^{\en} - c^{\en}) g^{\en} + \out (K-1) (b^{\ex} - c^{\ex}) g^{\ex}}{1 + \out (K-1)}
        \\
        & = (b^{\en} - c^{\en}) g^{\en} \frac{ 1 + \out (K-1) (b^{\ex} - c^{\ex})g^{\ex}/[(b^{\en} - c^{\en})g^{\en}]}{1 + \out (K-1)}
    \end{split}
\end{equation*}
The relevant ratio is now $(b^{\ex} - c^{\ex})g^{\ex}/[(b^{\en} - c^{\en})g^{\en}]$.
If this ratio is greater than $1$ (for example, because out-group interactions are more rewarding than in-group interactions), it is possible for insularity to be selected against at the group level, as higher $\out$ results in increased fitness.

\subsection{Invasion by insular mutants}
\label{sec:insular_invasibility}

Here we consider a population fixed for discriminators, but the resident population is facing potential invasion by a mutant discriminator with a different level of insularity.
Residents have out-group interaction parameter $\out_r$, and mutants have out-group interaction parameter $\out_m$; both mutants always interact with their in-groups.
We set $\out_m < \out_r$, i.e., the mutant is more insular (less likely to engage in out-group interactions) than the resident, and we posit that potential interactions between out-group mutants and residents occur only with probability $\psi(\out_r,\out_m)$.
Two natural forms for $\psi(\out_r,\out_m)$ are
\begin{equation}
    \begin{split}
        \psi(\out_r,\out_m) & = \frac{\out_r + \out_m}{2} \text{~(the arithmetic mean)}
        \\
        \psi(\out_r,\out_m) & = \sqrt{\out_r \out_m} \text{~(the geometric mean)}.
    \end{split}
\end{equation}
The geometric mean formulation has the advantage that $\sqrt{\out_s}$ can be thought of as the probability that an individual of type $s$ proposes an out-group interaction and $\sqrt{\out_{s^\prime}}$ the probability that their out-group partner accepts; however, it is more difficult to work with than the arithmetic mean.
Thus, we use the arithmetic mean for the remainder of this analysis; using the geometric mean instead effects minor quantitative but not qualitative changes.

\begin{table}[!ht]
    \centering
    \begin{tabular}{|Sc|Sc|}
        \hline
        norm & $g_m^{\en}$
        \\
        \hline
        \makecell{\textit{Stern Judging}
        \\
        $(K \geq 3)$}& $\dfrac{ 2(1 - \act - \ass) + (K-1)(2 - 2\ass - \act) \psi(\out_r,\out_m) }{2(1 + [K-1] \psi(\out_r,\out_m))}$
        \\
        \hline
        \makecell{\textit{Stern Judging}
        \\
        $(K = 2, \act > 0)$}
        & 
        $\dfrac{2(1 - \ass - \act)\out_r + [2(\act(1 + \sqrt{1 + \out_r} + \out_r ) + \ass(2 - (w + \out_r)\sqrt{1 + \out_r})]\psi(\out_r,\out_m)}{2 \out_r (1 + \psi(\out_r,\out_m))}$
        \\
        \hline
        \makecell{\textit{Stern Judging}
        \\
        $(K = 2, \act = 0)$}
        & $1 - \ass$
        \\
        \hline
        \textit{Simple Standing} & $1 - \ass - \act$
        \\
        \hline
        \textit{Shunning} & $\dfrac{\ass(1 + 3[K-1] \out_r + 2 \out_r \psi(\out_r,\out_m)[K-1]^2)}{(K-1) \out_r(1 + [K-1] \psi(\out_r,\out_m))}$
        \\
        \hline
    \end{tabular}
    \caption{
    \small{Approximate in-group reputations for a \emph{mutant} with different out-group interaction parameter $\out_m$ from the resident; in this scenario, the mutant is invading a population consisting of $K$ equally sized groups under well-mixed strategic imitation.
    Expressions are to first order in $\ass$ and $\act$.
    See the caveats in the caption of table \ref{tab:ingroup_outgroup_reps}.
    }}
    \label{tab:mutant_ingroup_reps}
\end{table}

\begin{table}[!ht]
    \centering
    \begin{tabular}{|Sc|Sc|}
        \hline
        norm & $g_m^{\ex}$
        \\
        \hline
        \makecell{
        \textit{Stern Judging}
        \\
        $(K \geq 3)$
        }& $\dfrac{K - 2 -\act}{2 (K-2)}$
        \\
        \hline
        \makecell{\textit{Stern Judging}
        \\
        $(K = 2, \act > 0)$}
        & 
        \makecell{ $\dfrac{1}{4\out \sqrt{1 + \out_r}(1 + \psi(\out_r,\out_m))}$ \\ $ \times \Big\{ 4(1 + \out - \sqrt{1 + \out}) + \ass(8(1 - \sqrt{1 + \out_r}) + \out_r(4 - \out_r))$
        \\
        $ +\act(4(1 - \sqrt{1 + \out_r} + \out_r (4\sqrt{1 + \out_r} - 2)$
        \\
        $ + \big[ 4(1 + \out - \sqrt{1 + \out}) +  2\act(4(1 - \sqrt{1 + \out_r}) + \out_r)$
        \\
        $ + \ass \out_r (10 + \out_r - 4\sqrt{1 + \out_r}) + 12\ass(1 - \sqrt{\out_r}) \big] \psi(\out_r,\out_m) \Big\} $}
        \\
        \hline
        \makecell{\textit{Stern Judging}
        \\
        $(K = 2, \act = 0)$}
        & $1/2$
        \\
        \\
        \hline
        \textit{Simple Standing} & $\dfrac{1 - 2(\ass + \act) + (K-1)[1 - 3\act - 3\ass - (K-1)(\act + \ass) \out_r]\psi(\out_r,\out_m)}{1 + (K-1) \psi(\out_r,\out_m)} $
        \\
        \hline
        \textit{Shunning} & $\ass$
        \\
        \hline
    \end{tabular}
    \caption{
    \small{Approximate out-group reputations for a \emph{mutant} with different out-group interaction parameter $\out_m$ from the resident; in this scenario, the mutant is invading a population consisting of $K$ equally sized groups under well-mixed strategic imitation.
    Expressions are to first order in $\ass$ and $\act$.
    See the caveats in the caption of table \ref{tab:ingroup_outgroup_reps}.
    }}
    \label{tab:mutant_outgroup_reps}
\end{table}

We assume well-mixed copying and $K$ groups of equal size $1/K$.
Let $f_r$ be the frequency of the resident and $f_m$ the frequency of the mutant; because we are concerned with the invasibility of the mutant, we will set $f_r = 1$.
The total number of interactions the two types engage in is
\begin{equation*}
    \begin{split}
        \M_r^I & = \sum_J \nu_J \big[ f_r \out_r^{I,J} + f_m \psi (\out_r^{I,J},\out_m^{I,J} ) \big]
        \\
        & = \frac{1 + \out_r (K-1)}{K},
        \\
        \M_m^I & = \sum_J \nu_J \big[ f_r \Big[ \psi (\out_r^{I,J},\out_m^{I,J} ) + f_m \out_m^{I,J} \big]
        \\
        & = \frac{1 + \psi(\out_r, \out_m) (K-1)}{K}.
    \end{split}
\end{equation*}
We will drop the $I$ superscript moving forward, as there is no dependence on $I$.
The fitnesses of the resident and mutant are given respectively by
\begin{equation}
    \begin{split}
        \Pi_r|_{f_r = 1} &  = (1 - \act) \sum_I \frac{1}{\M_r} \Big\{ \sum_J \nu_J \sum_s f_s^J \psi ( \out_r^{I,J} \out_s^{I,J} ) \big[ b^{J,I} g_r^{I,J} - c g_s^{J,I} \big] \Big\} \Big|_{f_r = 1}
        \\
        & = (1 - \act) \frac{1}{1 + \out_r (K-1)} \sum_I \Big\{ \sum_J \out_r^{I,J} \big[ b^{J,I} g_r^{I,J} - c g_r^{J,I} \big] \Big\}
        \\
        & = (1 - \act) \frac{b^{\en} g_r^{\en} - c^{\en} g_r^{\en} + (K-1) \out_r (b^{\ex} g_r^{\ex} - c^{\ex} g_r^{\ex})}{1 + \out_r (K-1)}
        \\
        \Pi_m|_{f_r = 1} & = (1 - \act)\sum_I \frac{1}{\M_m} \Big\{ \sum_J \nu_J \sum_s f_s^J \psi( \out_m^{I,J}, \out_s^{I,J} ) \big[ b^{J,I} g_m^{I,J} - c^{I,J} g_s^{J,I} \big] \Big\} \Big|_{f_r = 1}
        \\
        & = (1 - \act)\frac{1}{1 + \psi( \out_m, \out_r )(K-1)} \sum_I \Big\{ \sum_J \psi( \out_m^{I,J}, \out_s^{I,J} ) \big[ b^{J,I} g_m^{I,J} - c^{I,J} g_r^{J,I} \big] \Big\}
        \\
        & = (1 - \act) \frac{b^{\en} g_m^{\en} - c^{\en} g_r^{\en} + (K-1) \psi( \out_m, \out_r ) (b^{\ex} g_m^{\ex} - c^{\ex} g_r^{\ex}) }{1 + \psi (\out_m, \out_r) (K-1) }.
    \end{split}
    \label{eq:mutant_resident_fitness}
\end{equation}
When $f_r = 1$, reputations are given by
\begin{equation*}
    \begin{split}
        g_r^{\en} & = \frac{1}{\M_r} \frac{1}{K} (g_r^{\en} + \out_r (K - 1) g_r^{\ex}) (\pgc - \pbd) + \pbd,
        \\
        g_r^{\ex} & = \frac{1}{\M_r} \frac{1}{K} \Big\{ \big[ (1 + \out_r) g_r^{\en} g_r^{\ex} + (K - 2) \out_r (g_r^{\ex})^2\big] (\pgc - \pgd - \pbc + \pbd)
        \\
        & \ \ \ + \big[ \out_r g_r^{\en} + (1 + (K - 2) \out_r) g_r^{\ex} \big] (\pgd - \pbd)
        \\
        & \ \ \ + \big[ g_r^{\en} + (K - 1) \out_r g_r^{\ex}\big] (\pbc - \pbd) \Big\} + \pbd,
        \\
        g_m^{\en} & = \frac{1}{\M_m} \frac{1}{K} (g_r^{\en} + \psi (\out_m, \out_r)(K-1)  g_r^{\ex}) (\pgc - \pbd) + \pbd,
        \\
        g_m^{\ex} & = \frac{1}{\M_m} \frac{1}{K} \Big\{ \big[ (1 + \psi (\out_m, \out_r)(K-1) ) g_r^{\en} g_r^{\ex}
        \\
        & \ \ \ \ \ + (K - 2) \psi (\out_m, \out_r) (g_r^{\ex})^2\big] (\pgc - \pgd - \pbc + \pbd)
        \\
        & \ \ \ + \big[ \psi (\out_m, \out_r) g_r^{\en} + (1 + (K - 2) \psi (\out_m, \out_r) ) g_r^{\ex} \big] (\pgd - \pbd)
        \\
        & \ \ \ + \big[ g_r^{\en} + (K-1) \psi (\out_m, \out_r)  g_r^{\ex}\big] (\pbc - \pbd) \Big\} + \pbd.
    \end{split}
\end{equation*}
Mutant invasibility requires that $\Pi_m|_{f_r = 1} - \Pi_r|_{f_r = 1} > 0$.
Dropping the $(1 - \act)$ prefactor allows us to rewrite this invasibility condition as
\begin{equation}
    \begin{split}
        \Pi_m|_{f_r = 1} - \Pi_r|_{f_r = 1} & > 0
        \\
        b^{\ex} (g_m^{\ex} - g_r^{\ex}) + \frac{(b^{\ex} - c^{\ex}) g_r^{\ex} - (b^{\en} - c^{\en}) g_r^{\en}}{1 + \out_r (K-1)} + \frac{b^{\en} g_m^{\en} - b^{\ex} g_m^{\ex} - c^{\en} g_r^{\en} + c^{\ex} g_r^{\ex} }{1 + \psi(\out_r,\out_m) (K-1)} & > 0.
    \end{split}
    \label{eq:mutant_invasibility_condition}
\end{equation}
The first term in equation \eqref{eq:mutant_invasibility_condition} is the difference in mutant and resident fitness due to being on the receiving end of out-group cooperation events.
The second term is the difference in the fitness the \emph{resident} accrues as a result of \emph{out-group versus in-group} interactions.
The third is the difference in the fitness the \emph{mutant} accrues as a result of \emph{in-group versus out-group} interactions.
And so in summary, for the mutant to invade the resident, some combination of the following must be true:
\begin{enumerate}
    \item The mutant must be targeted by more (and potentially more rewarding) out-group cooperation events than the resident.
    \item The resident must, on net, suffer as a result of out-group cooperation events (which may be more rewarding but are less likely to be reciprocated: consider that, in general, $g_r^{\ex} < g_r^{\en}$).
    \item The mutant must, on net, benefit as a result of forgoing out-group cooperation events.
\end{enumerate}

These three conditions establish that, in general, more rewarding out-group interactions favor less insularity (i.e., higher $\out$) and thus make it harder for more insular mutants to invade.
How much harder will depend on the social norm, in particular the values of $g_r^{\en}$, $g_r^{\ex}$, $g_m^{\en}$, and $g_m^{\ex}$.
The first two can be read off of SI table \ref{tab:ingroup_outgroup_reps} (with $\out = \out_r$); the provided expressions are valid for $g_r^{\en}$ and $g_r^{\ex}$.
The latter two can be found in SI tables \ref{tab:mutant_ingroup_reps} and \ref{tab:mutant_outgroup_reps}.

Simplifying equation \eqref{eq:mutant_invasibility_condition} in a useful way is difficult, but we can show that, in general, it will be possible for insular mutants to invade \emph{unless} out-group cooperation is much more rewarding than in-group cooperation.
This is intuitive, as insular mutants forgo out-group interactions; for insularity to be selected against, the interactions they forgo must be especially rewarding.
Moreover, because it is generally easier to be seen as good by one's in-group than by one's out-group, we expect that if $b^{\en} = b^{\ex}$ and $c^{\en} = c^{\ex}$, it will not be possible for populations to resist invasion by more insular mutants.

We demonstrate that, when interactions' costs and benefits do not depend on group identity, insular mutants can essentially always invade.
To do so, we expand \eqref{eq:mutant_invasibility_condition} in $\ass$ and $\act$; we drop terms that are $\mathcal{O}(\ass^2)$, $\mathcal{O}(\act^2)$, and $\mathcal{O}(\ass \act)$.
(We do not address \emph{Scoring} here, as insularity does not affect reputations, and therefore it does not affect fitnesses.)
For Shunning, we obtain the following condition fora  mutant with out-group interaction parameter $\out_m$ to invade a resident with parameter $\out_r$:
\begin{equation*}
    \begin{split}
        0 < & \frac{\ass (\out_r - \out_m)}{\out_r [1 + (K-1) \out_r][2 + (K-1)(\out_r + \out_m)]^2}
        \\
        & \ \ \ \times \Big( (\out_r - \out_m) \ass \big[ (b^{\en} - c^{\en})[1 + 2(K-1)\out_r][2 + (K-1)(\out_r + \out_m)]
        \\
        & \ \ \ + 2 b^{\en}[1 + (K-1) \out_r] - (b^{\ex} - c^{\ex})(K-1)\out_r [2 + (K-1)(\out_r + \out_m)] \Big).
    \end{split}
\end{equation*}
When in-group and out-group interactions are indistinguishable, we have
\begin{equation*}
    \frac{\ass (\out_r - \out_m) \big[ (b - c)[2 + (K-1)(\out_r + \out_m)] + 2 b \big]}{\out_r [2 + (K-1)(\out_r + \out_m)]^2},
\end{equation*}
which is always positive for $\out_r > \out_m$.
For Simple Standing,
\begin{equation*}
    \begin{split}
        0 & < \frac{(K-1)(\out_r - \out_m)}{[1 + (K-1) \out_r][2 + (K-1)(\out_r + \out_m)]^2}
        \\
        & \ \ \ \times \Big( (b^{\en} - c^{\en})(1 - \ass - \act)[2 + (K-1)(\out_r + \out_m)]
        \\
        & \ \ \ - (b^{\ex} - c^{\ex})[2 + (K-1)(\out_r + \out_m)][1 - 2 \act - 2 \ass - (\act + \ass)(K-1)\out_r]
        \\
        & \ \ \ + b^{\ex}(K-1)[1 + (K-1)\out_r](\ass + \act)(\out_m + \out_r) \Big).
    \end{split}
\end{equation*}
For identical in- and out-group interactions, this becomes
\begin{equation}
    \frac{(\out_r - \out_m) (K-1) (\ass + \act) \big[(b-c) [2 + (K-1)(\out_r + \out_m)] + b (K-1)(\out_m + \out_r) \big]}{[2 + (K-1)(\out_r + \out_m)]^2}
\end{equation}
This, too, is always positive provided $\out_r > \out_m$.

For Stern Judging, a general expression can be obtained for $K > 2$:
\begin{equation}
    \begin{split}
        0 & < \frac{(K-1)(\out_r - \out_m)}{2[1 + (K-1)\out_r]^2 [2 + (K-1)(\out_r + \out_m)]^2(K-2)} \times
        \\
        & \ \ \ \Big( (b^{\en} - c^{\en})(K-2)[2+(K-1)(\out_r + \out_m)][(K-1)\out_r(2 - 2\ass - \act) + 2(1 - \ass - \act)]
        \\
        & \ \ \ - 2 b^{\en}(K-2) \act [1 + (K-1)\out_r]
        \\
        & \ \ \ - (b^{\ex} - c^{\ex})(K - \act - 2)[2 + (K-1)(\out_r + \out_m)] \Big).
    \end{split}
    \label{eq:SJ_insularity_K3}
\end{equation}
We can obtain a better understanding of how in-group and out-group interactions affect the invasion of insular mutants by taking the limit $\act \to 0$:
\begin{equation}
    \frac{[2 (1 - \ass)(b^{\en} - c^{\en}) - (b^{\ex} - c^{\ex})] (K-1) (\out_r - \out_m )}{2[1 + (K-1) \out_r][2 + (K-1)(\out_r + \out_m)]} > 0,
    \label{eq:insular_invasibility}
\end{equation}
which is always satisfied for $\out_r > \out_m$ \emph{unless}
\begin{equation}
    b^{\ex} - c^{\ex} > 2(1 - \ass) (b^{\en} - c^{\en}).
    \label{eq:SJ_insularity_condition}
\end{equation}
That is, residents resist invasion by higher-insularity mutants only when out-group interactions are much more rewarding than in-group interactions.
It is intriguing to note that this agrees perfectly with the condition in SI section $5.11$ for group fitness, i.e., $(b^{\ex} - c^{\ex})g^{\ex}/[(b^{\en} - c^{\en})g^{\en}] > 1$, since (under \textit{Stern Judging} and with $\act = 0$) we have $g^{\en} = 1 - \ass$ and $g^{\ex} = 1/2$.
(Evaluating the left hand side of equation \eqref{eq:mutant_invasibility_condition} numerically reveals that this approximation slightly underestimates the value of the ratio $(b^{\ex} - c^{\ex})/(b^{\en} - c^{\en})$ required for a population to resist invasion.)

Setting in-group and out-group costs and benefits equal to each other in equation \eqref{eq:SJ_insularity_K3} yields
\begin{equation*}
    \begin{split}
        0 & < \frac{(K-1)(\out_r - \out_m)}{2[1 + (K-1)\out_r]^2 [2 + (K-1)(\out_r + \out_m)]^2 (K-2)} \times
        \\
        & \ \ \ \Big( (b - c ) [2 + (K-1) (\out_r + \out_m)] \big[ (K-1) \out_r \{ (K-2)(1 - 2\ass) - (K-3) \act \} 
        \\
        & \ \ \ \ \ + K(1 - 2 \act - 2 \ass) + 4\ass + 5\act - 2 \big] - 2 b \act[1 + (K-1) \out_r] \Big).
    \end{split}
\end{equation*}
Sending $\act \to 0$ reduces this to a specific form of equation \eqref{eq:insular_invasibility}:
\begin{equation*}
    \frac{(b-c)(K-1)(1 - 2\ass)(\out_r - \out_m)}{2[1 + (K-1) \out_r][2 + (K-1)(\out_r + \out_m)]} > 0.
\end{equation*}
This condition is always satisfied provided $\out_r > \out_m$ and $\ass < 1/2$, meaning that, when interactions' costs and benefits do not depend on group membership, a higher-insularity mutant (i.e., one that interacts less with its out-group) can always invade the resident population under \textit{Stern Judging}, just as we found with \textit{Shunning} and \textit{Simple Standing}.

Finally, for \textit{Stern Judging}, the condition for an insular mutant to invade when $K = 2$ is different from the general case:
\begin{equation}
    \begin{split}
    0 & < \frac{\out_r - \out_m}{4 \out_r (1 + \out_r)^{3/2} (2 + \out_r + \out_m)^2}
    \\
    & \ \ \ \times \Big( 2 \out_r (b^{\en} - c^{\en})(2 + \out_r + \out_m)(2 \sqrt{1 + \out_r} - \ass(2 + \out_r) - 2 \act)
    \\
    & \ \ \ + 4 b^{\en} (1 + \out_r) [\ass \out_r - 2(\ass + \act)(\sqrt{1 + \out_r}-1)]
    \\
    & \ \ \ - (b^{\ex} - c^{\ex})(2 + \out_r + \out_m)\big[\ass (\out_r^2 - 4 \out_r \sqrt{1 + \out_r} + 8 \{1 + \out_r - \sqrt{1 + \out_r} \} )
    \\
    & \ \ \ \ \ + 2 \act (2 + \out_r - 2 \sqrt{1 + \out_r}) + 4 (1 + \out_r - \sqrt{1 + \out_r}) \big]
    \\
    & \ \ \ - 2 b^{\ex} (1 + \out_r)(\out_r + \out_m) (\ass \out_r - 2 (\act + \ass)(\sqrt{1 + \out_r} - 1) \Big).
    \end{split}
    \label{eq:SJ_insularity_K2}
\end{equation}
Sending both error rates to zero is informative:
\begin{equation*}
    \frac{(\out_r - \out_m) \big[ (b^{\en} - c^{\en}) \out_r - (b^{\ex} - c^{\ex})(\sqrt{1 + \out_r} - 1) \big]}{\out_r (1 + \out_r) (2 + \out_m + \out_r)} > 0.
\end{equation*}
This condition is satisfied for $\out_r > \out_m$ provided
\begin{equation*}
    \begin{split}
    (b^{\en} - c^{\en}) \out_r & > (b^{\ex} - c^{\ex}) (\sqrt{1 + \out_r} - 1)
    \\
    \therefore \frac{b^{\en} - c^{\en}}{b^{\ex} - c^{\ex}} & > \frac{\sqrt{1 + \out_r} - 1}{\out_r}.
    \end{split}
\end{equation*}
The ratio on the right is between $1/2$ (for $\out_r \to 0$) and $\sqrt{2} - 1 \approx .41$ (for $\out_r \to 1$), meaning that insular mutants can invade unless out-group cooperation is a little more than twice as beneficial as in-group cooperation, a condition similar to the $K > 2$ case.

For equal in-group and out-group costs and benefits, equation \eqref{eq:SJ_insularity_condition} becomes
\begin{equation*}
    \begin{split}
        0 & < \frac{\out_r - \out_m}{4 \out_r (1 + \out_r)^{3/2} (2 + \out_r + \out_m)^2} \times
        \\
        & \ \ \ \Big( (b - c) (2 + \out_r + \out_m)\big[ 4 (1 + \out_r) (\sqrt{1 + \out_r} - 1) - \ass( 3 \out_r^2 - 4 \out_r \sqrt{1 + \out_r} + 8 \out_r - 8 \sqrt{1 + \out_r} + 8)
        \\
        & \ \ \ \ \ - \act( 6 \out_r - 4 \sqrt{1 + \out_r} + 4) \big]
        \\
        & \ \ \ + 2 b (1 + \out_r)(2 - \out_r - \out_m)(\ass \out_r - 2(\ass + \act)(\sqrt{1 + \out_r} - 1) \Big)
    \end{split}
\end{equation*}
Sending the error rates to zero allows us to show that this is, in general, positive for $\out_r > \out_m$:
\begin{equation*}
    \frac{(\out_r - \out_m)(b - c)(\sqrt{1 + \out_r} - 1)}{\out_r \sqrt{1 + \out_r}(2 + \out_r + \out_m)} > 0.
\end{equation*}

\begin{figure}
    \centering
    \includegraphics[width=0.9\textwidth]{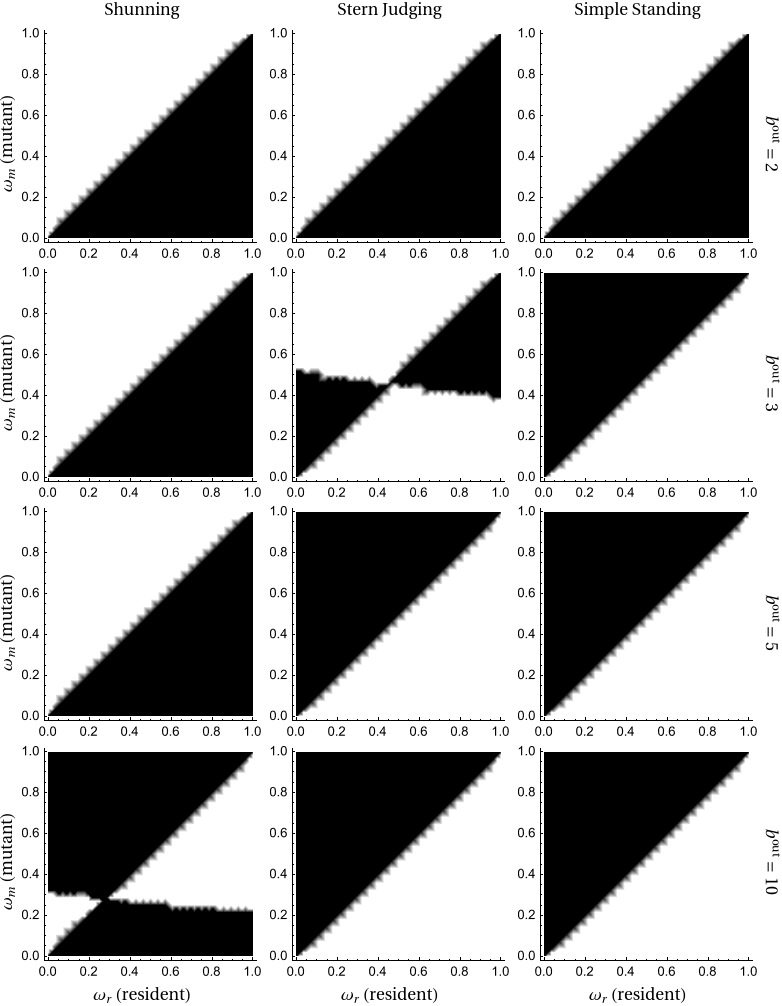}
    \caption{\small{The invasibility of a resident population with out-group interaction parameter $\out_r$ by a mutant with parameter $\out_m$ under different norms and out-group benefits.
    The value of $b^{\ex}$ is along the right side of each row: other parameter values are $K = 2$, $\ass = \act = 0.02$, $b^{\en} = 2$, and $c^{\en} = c^{\ex} = 1$.
    White corresponds to the mutant being able to invade: black corresponds to the resident resisting invasion.
    For \emph{Stern Judging} at $b^{\ex} = 3$, an intermediate value of $\out$ can be achieved by successive invasion of mutants.
    For \emph{Shunning} at $b^{\ex} = 10$, there is bistability.
    For all other norms and parameter combinations shown, the population evolves either toward full mixing ($\out = 1$) or full insularity ($\out = 0$).
    }}
    \label{fig:insular_inv_K2}
\end{figure}

In general, out-group reputations compared to in-group reputations are very low, middling, and fairly high under \emph{Shunning}, \emph{Stern Judging}, and \emph{Simple Standing} respectively.
We thus expect that, if we hold $b^{\en}$, $c^{\en}$, and $c^{\ex}$ constant, we will need to raise $b^{\ex}$ more under \emph{Shunning} than under \emph{Stern Judging}, and more under \emph{Stern Judging} than under \emph{Simple Standing}, for a population to resist invasion by insular mutants.
We verify this numerically by checking the sign of equation \eqref{eq:mutant_invasibility_condition}.
Results are shown in SI figure \ref{fig:insular_inv_K2}. In white areas, the difference between mutant and resident fitness is positive, meaning that the mutant can invade; in black areas, the difference is negative.
We find that when $b^{\ex} = b^{\en}$, it is impossible to resist invasion by successively more insular mutants, irrespective of the social norm.
For \emph{Simple Standing}, this situation reverses quickly with increased $b^{\ex}$; it reverses more slowly for \emph{Stern Judging} and \emph{Shunning}.
Some norm and parameter combinations can support stable intermediate values of the out-group interaction parameter $\out$, but in most cases the population will either evolve to be fully mixed or fully insular ($\out = 1$ or $0$, respectively).

\section{Third-order norms and the remaining ``leading eight'' norms}

All of our analysis hitherto has focused on second-order norms.
We now turn to the interesting question of third-order norms, in which the reputation of a donor may be updated according to not only the donor's action and the recipient's reputation, but also the donor's current reputation.
In this space there are not $4$ but $16$ possible behavioral strategies, which can be represented as four bits, corresponding to the donor's action when both their reputation and the recipient's is good, when their own reputation is good but the donor's is bad, and so on.
We first consider the simplest case of publicly shared reputations, then we generalize to group-wise reputations.

\subsection{Properties of the leading eight}

The ``leading eight'' social norms consist of reputation dynamics (assessment rules, i.e., rules for assigning ``good'' and ``bad'' reputations) and behavioral strategies (action rules, i.e., rules for deciding whether to defect or cooperate based on one's own reputation and that of the recipient) that satisfy the following conditions \citep{ohtsuki2006leading}:
\begin{enumerate}
    \item Either \emph{good} or \emph{bad} cooperating with \emph{good} is \emph{good}.
    \item Either \emph{good} or \emph{bad} defecting with \emph{good} is \emph{bad}.
    \item \emph{Good} defecting with \emph{bad} is \emph{good}.
    \item Either \emph{good} or \emph{bad} individuals will cooperate with \emph{good}.
    \item \emph{Good} individuals will defect with \emph{bad}.
    \item Iff \emph{bad} cooperating with \emph{bad} is \emph{good} and \emph{bad} defecting with \emph{bad} is \emph{bad}, then \emph{bad} will cooperate with \emph{bad}: otherwise, \emph{bad} will defect with \emph{bad}.
\end{enumerate}

There are eight such combinations of reputation dynamics and action rule (behavioral strategy), summarized in table \ref{tab:leading_eight}.
The action rule is represented by $c_{UV}$, which are the probabilities that an individual with reputation $U$ cooperates against an individual with reputation $V$.
These values are always either $1 - \act$ (because individuals who intend to cooperate can accidentally defect with probability $\act$) or $0$ (because individuals who intend to defect can never accidentally cooperate).
The reputation dynamics are specified by the values $n_{UAV}$, which are the probability of earning a good reputation by having reputation $U$ and performing action $A$ against an individual with reputation $V$.
These values are always either $1 - \ass$ or $\ass$, allowing for assessment error.
(This notation is slightly different from our treatment of second-order social norms, where $P_{AV}$ is the probability of earning a good reputation by \emph{intending} to perform action $A$ against a recipient with reputation $V$.
The difference is that $P_{CV}$ includes the possibility of both successful cooperation \emph{and} accidental defection against an individual with reputation $V$, whereas $n_{UCV}$ does not: successful cooperation and accidental defection are treated separately.)

\subsection{Public reputations}

When the whole population follows the same reputation dynamics and is fixed for the same action rule, the mean proportion of individuals with good reputations is given by
\begin{equation}
    \begin{split}
        g & = g^2 (\cgg \ngcg + [1 - \cgg] \ngdg) + g (1 - g) (\cgb \ngcb + [1 - \cgb] \ngdb + \cbg \nbcg + [1 - \cbg] \nbdg)
        \\
        & \ \ \ + (1 - g)^2 (\cbb \nbcb + [1 - \cbb] \nbdb).
    \end{split}
    \label{eq:third_order_one_group}
\end{equation}
Under \textit{Stern Judging}, we will have
\begin{equation}
    \begin{split}
        \cgg & = \cbg = 1 - \act,
        \\
        \cgb & = \cbb = 0,
        \\
        \ngcg & = \nbcg = 1 - \ass,
        \\
        \ngcb & = \nbcb = \ass,
        \\
        \ngdg & = \nbdg = \ass,
        \\
        \ngdb & = \nbdb = 1 - \ass,
    \end{split}
\end{equation}
so equation \eqref{eq:third_order_one_group} simplifies to
\begin{equation}
    \begin{split}
        g & = g^2 ([1 - \act] [1 - \ass] + \act \ass) + g (1 - g) (1 - \ass + [1 - \act] [1 - \ass] + \act \ass) + (1 - g)^2 (1 - \ass)
        \\
        & = g ([1 - \act][1 - \ass] + \act \ass) + (1 - g) (1 - \ass),
    \end{split}
\end{equation}
as expected; this equation is also the same under \textit{Simple Standing} (where $\ngcb = \nbcb = 1 - \ass$ rather than $\ass$, but this term does not appear in the reputation dynamics, because $\cgb = \cbb = 0$).
Note that \textit{Stern Judging} and \textit{Simple Standing} are norms $s_6$ and $s_3$ (respectively) in table \ref{tab:leading_eight}.

Under \textit{Scoring}, we have
\begin{equation}
    \begin{split}
        \cgg & = \cbg = 1 - \act,
        \\
        \cgb & = \cbb = 0,
        \\
        \ngcg & = \nbcg = 1 - \ass,
        \\
        \ngcb & = \nbcb = 1 - \ass,
        \\
        \ngdg & = \nbdg = \ass,
        \\
        \ngdb & = \nbdb = \ass,
    \end{split}
\end{equation}
so equation \eqref{eq:third_order_one_group} is
\begin{equation}
    \begin{split}
        g & = g^2 ([1 - \act] [1 - \ass] + \act \ass) + g (1 - g) ( \ass + [1 - \act] [1 - \ass] + \act \ass ) + (1 - g)^2 \ass 
        \\
        & = g ([1 - \act][1 - \ass] + \act \ass) + (1 - g) \ass,
    \end{split}
\end{equation}
again as expected: it would be the same under \textit{Shunning}, for which the only difference is $\ngcb = \nbcb = \ass$.
(Scoring and \textit{Shunning} are not part of the ``leading eight'': we present those equations only for completeness.)

\begin{table}
    \begin{tabular}{c|cccccccc|cccc}
        norm & $\ngcg$ & $\ngdg$ & $\ngcb$ & $\ngdb$ & $\nbcg$ & $\nbdg$ & $\nbcb$ & $\nbdb$ & $\cgg$ & $\cgb$ & $\cbg$ & $\cbb$
        \\
        \hline
        $s_1$ & $1 - \ass$ & $\ass$ & $1 - \ass$ & $1 - \ass$ & $1 - \ass$ & $\ass$ & $1 - \ass$ & $\ass$ & $1 - \act$ & $0$ & $1 - \act$ & $1 - \act$
        \\
        $s_2$ & $1 - \ass$ & $\ass$ & $\ass$ & $1 - \ass$ & $1 - \ass$ & $\ass$ & $1 - \ass$ & $\ass$ & $1 - \act$ & $0$ & $1 - \act$ & $1 - \act$
        \\
        \hline
        $s_3$ & $1 - \ass$ & $\ass$ & $1 - \ass$ & $1 - \ass$ & $1 - \ass$ & $\ass$ & $1 - \ass$ & $1 - \ass$ & $1 - \act$ & $0$ & $1 - \act$ & $0$
        \\
        $s_4$ & $1 - \ass$ & $\ass$ & $1 - \ass$ & $1 - \ass$ & $1 - \ass$ & $\ass$ & $\ass$ & $1 - \ass$ & $1 - \act$ & $0$ & $1 - \act$ & $0$
        \\
        $s_5$ & $1 - \ass$ & $\ass$ & $\ass$ & $1 - \ass$ & $1 - \ass$ & $\ass$ & $1 - \ass$ & $1 - \ass$ & $1 - \act$ & $0$ & $1 - \act$ & $0$
        \\
        $s_6$ & $1 - \ass$ & $\ass$ & $\ass$ & $1 - \ass$ & $1 - \ass$ & $\ass$ & $\ass$ & $1 - \ass$ & $1 - \act$ & $0$ & $1 - \act$ & $0$
        \\
        \hline
        $s_7$ & $1 - \ass$ & $\ass$ & $1 - \ass$ & $1 - \ass$ & $1 - \ass$ & $\ass$ & $\ass$ & $\ass$ & $1 - \act$ & $0$ & $1 - \act$ & $0$
        \\
        $s_8$ & $1 - \ass$ & $\ass$ & $\ass$ & $1 - \ass$ & $1 - \ass$ & $\ass$ & $\ass$ & $\ass$ & $1 - \act$ & $0$ & $1 - \act$ & $0$
    \end{tabular}
    \caption{\small{The ``leading eight'' social norms and action rules, modeled after table $2$ of \citet{podder2021}.
    Here, $n_{UAV}$ is the probability that a donor with reputation $U$, who performs action $A$ against a recipient with reputation $V$, earns a good reputation, and $c_{UV}$ is the probability that a donor with reputation $U$ will cooperate with a recipient with reputation $V$.
    As elsewhere in our model, we allow for assessment error with probability $\ass$ and asymmetric execution error with probability $\act$ (individuals can accidentally defect but not accidentally cooperate).
    Norm $s_3$ is \textit{Simple Standing}, and norm $s_6$ is \textit{Stern Judging}; both are symmetric with respect to the reputation of the donor and, thus, are second-order norms.}}
    \label{tab:leading_eight}
\end{table}

\subsection{Group-wise reputations}

We now consider the possibility that the population is divided into groups, which each follow their own third-order social norm, i.e., each group has its own rule for assigning reputations and is fixed for its own particular behavioral strategy.
We assume that an individual in group $I$ acts according to their own view of themselves and the recipient, as well as their own action rule, but that $J$ judges them according to $J$'s reputation dynamics and $J$'s view of the recipient.
Thus, when $I = J$, we will have
\begin{equation}
    \begin{split}
        g^{I,J} & = g^{I,I} g^{\bullet,J} (\cgg^I \ngcg^J + [1 - \cgg^I] \ngdg^J)+ g^{I,I} (1 - g^{\bullet,J}) (\cgb^I \ngcb^J + [1 - \cgb^I] \ngdb^J)
        \\
        & \ \ \ + (1 - g^{I,I}) g^{\bullet,J} (\cbg^I \nbcg^J + [1 - \cbg^I] \nbdg^J) + (1 - g^{I,I}) (1 - g^{\bullet,J}) (\cbb^I \nbcb^J + [1 - \cbb^I] \nbdb^J).
    \end{split}
\end{equation}
When $I \neq J$, we account for the fact that $I$ and $J$ may have different views of both the donor $I$ and recipient $L$.
We enumerate these possibilities.
When group $J$ observes an interaction by an individual from group $I$, they can form a good reputation of $I$ in the following ways.
With probability $\nu_L$, an interaction between $I$ and $L$ is observed, and:
\begin{enumerate}
    \item with probability $g^{I,I} g^{L,I}$, the donor sees themselves and the recipient as good.
    \begin{enumerate}
        \item with probability $g^{I,J} g^{L,J}$, the observer thinks the donor and recipient are both good.
        If the donor cooperates (probability $\cgg^I$), the observer considers that good with probability $\ngcg^J$.
        If the donor defects (probability $1 - \cgg^I$), the observer considers that good with probability $\ngdg^J$.
        \item with probability $g^{I,J}(1 - g^{L,J})$, the observer thinks the donor is good but the recipient is bad.
        If the donor cooperates (probability $\cgg^I$), the observer considers that good with probability $\ngcb^J$.
        If the donor defects (probability $1 - \cgg^I$), the observer considers that good with probability $\ngdb^J$.
        \item with probability $(1 - g^{I,J})g^{L,J}$, the observer thinks the donor is bad but the recipient is good.
        If the donor cooperates (probability $\cgg^I$), the observer considers that good with probability $\nbcg^J$.
        If the donor defects (probability $1 - \cgg^I$), the observer considers that good with probability $\nbdg^J$.
        \item with probability $(1 - g^{I,J}) (1 - g^{L,J})$, the observer thinks the donor and recipient are both bad.
        If the donor cooperates (probability $\cgg^I$), the observer considers that good with probability $\nbcb^J$.
        If the donor defects (probability $1 - \cgg^I$), the observer considers that good with probability $\nbdb^J$.
    \end{enumerate}
    \item with probability $g^{I,I}(1 - g^{L,I})$, the donor sees themselves as good and the recipient as bad.
    \begin{enumerate}
        \item with probability $g^{I,J} g^{L,J}$, the observer thinks the donor and recipient are both good.
        If the donor cooperates (probability $\cgb^I$), the observer considers that good with probability $\ngcg^J$.
        If the donor defects (probability $1 - \cgb^I$), the observer considers that good with probability $\ngdg^J$.
        \item with probability $g^{I,J} (1 - g^{L,J})$, the observer thinks the donor is good but the recipient is bad.
        If the donor cooperates (probability $\cgb^I$), the observer considers that good with probability $\ngcb^J$.
        If the donor defects (probability $1 - \cgb^I$), the observer considers that good with probability $\ngdb^J$.
        \item with probability $(1 - g^{I,J})g^{L,J}$, the observer thinks the donor is bad but the recipient is good.
        If the donor cooperates (probability $\cgb^I$), the observer considers that good with probability $\nbcg^J$.
        If the donor defects (probability $1 - \cgb^I$), the observer considers that good with probability $\nbdg^J$.
        \item with probability $(1 - g^{I,J})(1 - g^{L,J})$, the observer thinks the donor and recipient are both bad.
        If the donor cooperates (probability $\cgb^I$), the observer considers that good with probability $\nbcb^J$.
        If the donor defects (probability $1 - \cgb^I$), the observer considers that good with probability $\nbdb^J$.
    \end{enumerate}
    \item with probability $(1 - g^{I,I}) g^{L,I}$, the donor sees themselves as bad and the recipient as good.
    \begin{enumerate}
        \item with probability $g^{I,J} g^{L,J}$, the observer thinks the donor and recipient are both good.
        If the donor cooperates (probability $\cbg^I$), the observer considers that good with probability $\ngcg^J$.
        If the donor defects (probability $1 - \cbg^I$), the observer considers that good with probability $\ngdg^J$.
        \item with probability $g^{I,J} (1 - g^{L,J})$, the observer thinks the donor is good but the recipient is bad.
        If the donor cooperates (probability $\cbg^I$), the observer considers that good with probability $\ngcb^J$.
        If the donor defects (probability $1 - \cbg^I$), the observer considers that good with probability $\ngdb^J$.
        \item with probability $(1 - g^{I,J})g^{L,J}$, the observer thinks the donor is bad but the recipient is good.
        If the donor cooperates (probability $\cbg^I$), the observer considers that good with probability $\nbcg^J$.
        If the donor defects (probability $1 - \cbg^I$), the observer considers that good with probability $\nbdg^J$.
        \item with probability $(1 - g^{I,J})(1 - g^{L,J})$, the observer thinks the donor and recipient are both bad.
        If the donor cooperates (probability $\cbg^I$), the observer considers that good with probability $\nbcb^J$.
        If the donor defects (probability $1 - \cbg^I$), the observer considers that good with probability $\nbdb^J$.
    \end{enumerate}
    \item with probability $(1 - g^{I,I})(1 - g^{L,I})$, the donor sees themselves and the recipient as bad.
    \begin{enumerate}
        \item with probability $g^{I,J} g^{L,J}$, the observer thinks the donor and recipient are both good.
        If the donor cooperates (probability $\cbb^I$), the observer considers that good with probability $\ngcg^J$.
        If the donor defects (probability $1 - \cbb^I$), the observer considers that good with probability $\ngdg^J$.
        \item with probability $g^{I,J} (1 - g^{L,J})$, the observer thinks the donor is good but the recipient is bad.
        If the donor cooperates (probability $\cbb^I$), the observer considers that good with probability $\ngcb^J$.
        If the donor defects (probability $1 - \cbb^I$), the observer considers that good with probability $\ngdb^J$.
        \item with probability $(1 - g^{I,J})g^{L,J}$, the observer thinks the donor is bad but the recipient is good.
        If the donor cooperates (probability $\cbb^I$), the observer considers that good with probability $\nbcg^J$.
        If the donor defects (probability $1 - \cbb^I$), the observer considers that good with probability $\nbdg^J$.
        \item with probability $(1 - g^{I,J})(1 - g^{L,J})$, the observer thinks the donor and recipient are both bad.
        If the donor cooperates (probability $\cbb^I$), the observer considers that good with probability $\nbcb^J$.
        If the donor defects (probability $1 - \cbb^I$), the observer considers that good with probability $\nbdb^J$.
    \end{enumerate}
\end{enumerate}
Summing over all possible groups $L$ yields
\begin{equation}
    \begin{split}
        g^{I,J} & = \delta_{I,J} \Bigg\{ g^{I,I} g^{\bullet,J} (\cgg^I \ngcg^J + [1 - \cgg^I] \ngdg^J)+ g^{I,I} (1 - g^{\bullet,J}) (\cgb^I \ngcb^J + [1 - \cgb^I] \ngdb^J)
        \\
        & \ \ \ + (1 - g^{I,I}) g^{\bullet,J} (\cbg^I \nbcg^J + [1 - \cbg^I] \nbdg^J) + (1 - g^{I,I}) (1 - g^{\bullet,J}) (\cbb^I \nbcb^J + [1 - \cbb^I] \nbdb^J) \Bigg\}
        \\
        & + (1 - \delta_{I,J}) \Bigg\{ \sum_L \nu_L \Bigg( g^{I,I} g^{L,I} \Bigg[ g^{I,J} g^{L,J} (\cgg^I \ngcg^J + [1 - \cgg^I] \ngdg^J) + g^{I,J} (1 - g^{L,J}) (\cgg^I \ngcb^J + [1 - \cgg^I] \ngdb^J)
        \\
        & \ \ \ \ \ + (1 - g^{I,J}) g^{L,J} (\cgg^I \nbcg^J + [1 - \cgg^I] \nbdg^J) + (1 - g^{I,J}) (1 - g^{L,J}) (\cgg^I \nbcb^J + [1 - \cgg^I] \nbdb^J) \Bigg]
        \\
        & \ \ \ + g^{I,I} (1 - g^{L,I}) \Bigg[ g^{I,J} g^{L,J} (\cgb^I \ngcg^J + [1 - \cgb^I] \ngdg^J) + g^{I,J} (1 - g^{L,J}) (\cgb^I \ngcb^J + [1 - \cgb^I] \ngdb^J)
        \\
        & \ \ \ \ \ + (1 - g^{I,J}) g^{L,J} (\cgb^I \nbcg^J + [1 - \cgb^I] \nbdg^J) + (1 - g^{I,J}) (1 - g^{L,J}) (\cgb^I \nbcb^J + [1 - \cgb^I] \nbdb^J) \Bigg]
        \\
        & \ \ \ + (1 - g^{I,I}) g^{L,I} \Bigg[ g^{I,J} g^{L,J} (\cbg^I \ngcg^J + [1 - \cbg^I] \ngdg^J) + g^{I,J} (1 - g^{L,J}) (\cbg^I \ngcb^J + [1 - \cbg^I] \ngdb^J)
        \\
        & \ \ \ \ \ + (1 - g^{I,J}) g^{L,J} (\cbg^I \nbcg^J + [1 - \cbg^I] \nbdg^J) + (1 - g^{I,J}) (1 - g^{L,J}) (\cbg^I \nbcb^J + [1 - \cbg^I] \nbdb^J) \Bigg]
        \\
        & \ \ \ + (1 - g^{I,I}) (1 - g^{L,I}) \Bigg[ g^{I,J} g^{L,J} (\cbb^I \ngcg^J + [1 - \cbb^I] \ngdg^J) + g^{I,J} (1 - g^{L,J}) (\cbb^I \ngcb^J + [1 - \cbb^I] \ngdb^J)
        \\
        & \ \ \ \ \ + (1 - g^{I,J}) g^{L,J} (\cbb^I \nbcg^J + [1 - \cbb^I] \nbdg^J) + (1 - g^{I,J}) (1 - g^{L,J}) (\cbb^I \nbcb^J + [1 - \cbb^I] \nbdb^J) \Bigg] \Bigg) \Bigg\}.
    \end{split}
    \label{eq:third_order_multi_groups}
\end{equation}
We now allow individuals to switch between gossip groups.
We assume that, when an individual switches groups, it adopts both the reputation dynamics \emph{and} action rule (behavioral strategy) of that group.
The fitness of group $I$ can be expressed as
\begin{equation}
    \begin{split}
        \Pi^I & = b \sum_J \nu_J \big[ g^{J,J} g^{I,J} \cgg^J + g^{J,J} (1 - g^{I,J}) \cgb^J + (1 - g^{J,J}) g^{I,J} \cbg^J + (1 - g^{J,J}) (1 - g^{I,J}) \cbb^J \big]
        \\
        \ \ \ & - c \sum_J \nu_J \big[ g^{I,I} g^{J,I} \cgg^I + g^{I,I} (1 - g^{J,I}) \cgb^I + (1 - g^{I,I}) g^{J,I} \cbg^I + (1 - g^{I,I}) (1 - g^{J,I}) \cbb^I \big].
    \end{split}
\end{equation}
When there are two equally sized groups following different norms, fitnesses are given by (dropping the $1/2$ prefactor)
\begin{equation}
    \begin{split}
        \Pi^1 & = b \big[ (g^{1,1})^2 \cgg^1 + g^{1,1}(1 - g^{1,1}) (\cgb^1 + \cbg^1) + (1 - g^{1,1})^2 \cbb^1
        \\
        & \ \ \ + g^{2,2} g^{1,2} \cgg^2 + g^{2,2}(1 - g^{1,2}) \cgb^2 + (1 - g^{2,2}) g^{1,2} \cbg^2 + (1 - g^{2,2})(1 - g^{1,2}) \cbb^2 \big]
        \\
        & \ - c \big[ (g^{1,1})^2 \cgg^1 + g^{1,1} (1 - g^{1,1}) (\cgb^1 + \cbg^1) + (1 - g^{1,1})^2 \cbb^1
        \\
        & \ \ \ + g^{1,1} g^{2,1} \cgg^1 + g^{1,1} (1 - g^{2,1}) \cgb^1 + (1 - g^{1,1}) g^{2,1} \cbg^1 + (1 - g^{1,1})(1 - g^{2,1}) \cbb^1 \big]
        \\
        \Pi^2 & = b \big[ g^{1,1} g^{2,1} \cgg^1 + g^{1,1}(1 - g^{2,1}) \cgb^1 + (1 - g^{1,1}) g^{2,1} \cbg^1 + (1 - g^{1,1})(1 - g^{2,1} \cbb^1
        \\
        & \ \ \ + (g^{2,2})^2 \cgg^2 + g^{2,2}(1 - g^{2,2}) (\cgb^2 + \cbg^2) + (1 - g^{2,2})^2 \cbb^2 \big]
        \\
        & \ - c \big[ g^{2,2} g^{1,2} \cgg^2 + g^{2,2} (1 - g^{1,2}) \cgb^1 + (1 - g^{2,2}) g^{1,2} \cbg^1 + (1 - g^{2,2}) (1 - g^{1,2}) \cbb^1
        \\
        & \ \ \ + (g^{2,2})^2 \cgg^2 + g^{2,2} (1 - g^{2,2}) (\cgb^2 + \cbg^2 + (1 - g^{2,2})^2 \cbb^2 \big].
    \end{split}
\end{equation}
The fitness gain of group $1$ due to self-interactions is given by
\begin{equation}
    \begin{split}
        \pi^{1,1} & = b \big[ (g^{1,1})^2 \cgg^1 + g^{1,1}(1 - g^{1,1}) (\cgb^1 + \cbg^1) + (1 - g^{1,1})^2 \cbb^1 \big]
        \\
        & \ \ \ - c \big[ (g^{1,1})^2 \cgg^1 + g^{1,1} (1 - g^{1,1}) (\cgb^1 + \cbg^1) + (1 - g^{1,1})^2 \cbb^1 \big]
        \\
        & = (b - c) \big[ (g^{1,1})^2 \cgg^1 + g^{1,1}(1 - g^{1,1}) (\cgb^1 + \cbg^1) + (1 - g^{1,1})^2 \cbb^1 \big].
    \end{split}
\end{equation}
The fitness gain of group $1$ due to interactions with group $2$ is
\begin{equation}
    \begin{split}
        \pi^{1,2} & = b \big[ (g^{2,2} g^{1,2} \cgg^2 + g^{2,2}(1 - g^{1,2}) \cgb^2 + (1 - g^{2,2}) g^{1,2} \cbg^2 + (1 - g^{2,2})(1 - g^{1,2}) \cbb^2 \big]
        \\
        & \ \ \ - c \big[ g^{1,1} g^{2,1} \cgg^1 + g^{1,1} (1 - g^{2,1}) \cgb^1 + (1 - g^{1,1}) g^{2,1} \cbg^1 + (1 - g^{1,1})(1 - g^{2,1}) \cbb^1 \big].
    \end{split}
\end{equation}
Likewise,
\begin{equation}
    \begin{split}
        \pi^{2,2} & = (b - c) \big[ (g^{2,2})^2 \cgg^2 + g^{2,2}(1 - g^{2,2}) (\cgb^2 + \cbg^2) + (1 - g^{2,2})^2 \cbb^2 \big],
        \\
        \pi^{2,1} & = b \big[ g^{1,1} g^{2,1} \cgg^1 + g^{1,1}(1 - g^{2,1}) \cgb^1 + (1 - g^{1,1}) g^{2,1} \cbg^1 + (1 - g^{1,1})(1 - g^{2,1} \cbb^1 \big]
        \\
        & \ \ \ - c \big[ g^{2,2} g^{1,2} \cgg^2 + g^{2,2} (1 - g^{1,2}) \cgb^1 + (1 - g^{2,2}) g^{1,2} \cbg^1 + (1 - g^{2,2}) (1 - g^{1,2}) \cbb^1 \big]
    \end{split}
\end{equation}
We thus have
\begin{equation}
    \begin{split}
        \Pi_1 & > \Pi_2
        \\
        \Pi_1 - \Pi_2 & > 0
        \\
        \pi^{1,1} + \pi^{1,2} - \pi^{2,1} - \pi^{2,2} & > 0
        \\
        \therefore (b - c) \Big( \big[ (g^{1,1})^2 \cgg^1 + g^{1,1}(1 - g^{1,1}) (\cgb^1 + \cbg^1) + (1 - g^{1,1})^2 \cbb^1 \big] &
        \\
        - \big[ (g^{2,2})^2 \cgg^2 + g^{2,2}(1 - g^{2,2}) (\cgb^2 + \cbg^2) + (1 - g^{2,2})^2 \cbb^2 \big] \Big) &
        \\
        + \  (b + c) \Big( \big[ (g^{2,2} g^{1,2} \cgg^2 + g^{2,2}(1 - g^{1,2}) \cgb^2 + (1 - g^{2,2}) g^{1,2} \cbg^2 + (1 - g^{2,2})(1 - g^{1,2}) \cbb^2 \big] &
        \\
        - \big[ g^{1,1} g^{2,1} \cgg^1 + g^{1,1}(1 - g^{2,1}) \cgb^1 + (1 - g^{1,1}) g^{2,1} \cbg^1 + (1 - g^{1,1})(1 - g^{2,1} \cbb^1 \big] \Big) & > 0.
    \end{split}
\end{equation}
This is a form very similar to equation (6) from the main text: the $(b - c)$ term is the difference in the fitnesses of groups $1$ and $2$ due to within-group interactions, and the $(b + c)$ term is the difference in their fitnesses due to between-group interactions, i.e., fitness differences due to unreciprocated between-group cooperation.

\begin{figure}
    \centering
    \includegraphics[width=2.8in]{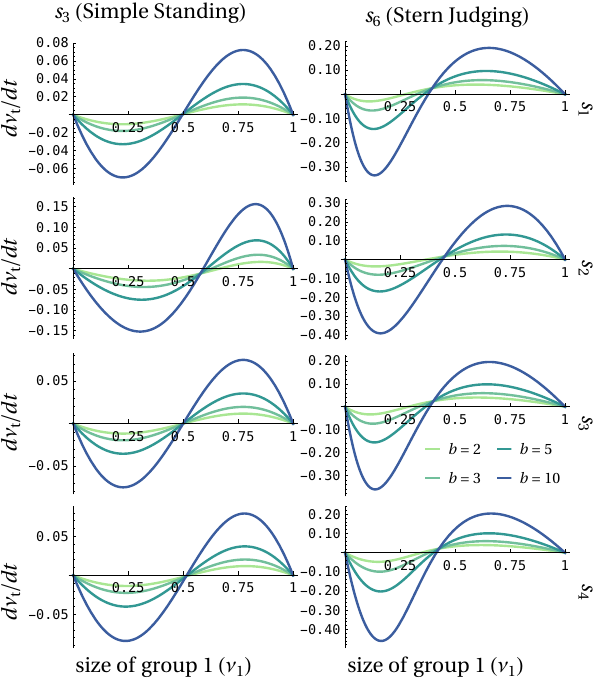}
    \includegraphics[width=2.8in]{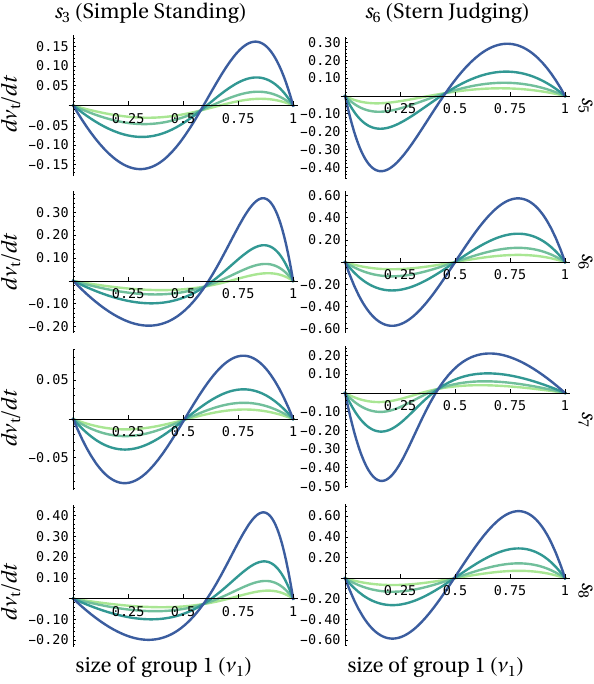}
    \caption{\small{Group size dynamics for $K = 2$ groups and varying values of the benefit $b$, when one group follows a second-order norm (either \textit{Stern Judging} or \textit{Simple Standing}) and the other follows a different ``leading eight'' norm, which may be third-order.
    In each pair of columns, the norm used in group $1$ is along the top: the norm used in group $2$, along the right.
    Values of $b$ are as inset in the $s_6 - s_3$ figure.
    In all plots, $c  = 1$, $\ass = \act = 0.02$.}}
    \label{fig:group_switch_third_order}
\end{figure}

Next, we consider competition between groups following different norms in a manner similar to the main text.
We focus on the two ``leading eight'' norms that happen to be second-order, namely \textit{Stern Judging} and \textit{Simple Standing}.
The results can be seen in figure \ref{fig:group_switch_third_order}.
While \textit{Simple Standing} is readily out-competed by most of the other leading eight, \textit{Stern Judging} is not: for \textit{Stern Judging}, we have $\nu_1^* < 1/2$ in almost every scenario, with the exception of competition against $s_8$, for which it is greater than $1/2$ for small values of $b$ and never moves far below $1/2$.
$s_8$ is very similar to \emph{Stern Judging}, with the exception that it regards bad individuals who defect against other bad individuals as bad, not good.

\subsection{The many-group limit: private reputations}

If the number of groups gets large ($K \to \infty$), all groups follow the same norm, and all groups are of the same size $1/K$, then we can reason similarly to subsection \ref{sec:identical_size}: self-interactions almost never occur, so the $I = J$ term of equation \eqref{eq:third_order_multi_groups} drops out, and the remaining $g^{I,J}$ converge to a common value $g$.
This yields
\begin{equation}
    \begin{split}
        g & = g^2 \big( g^2 [\cgg \ngcg + (1  -\cgg) \ngdg] + g (1-g) [\cgg (\ngcb + \nbcg)
        \\
        & \ \ \ + (1 - \cgg) (\ngdb + \nbdg) ] + (1 - g)^2 [\cgg \nbcb + (1 - \cgg) \nbdb] \big)
        \\
        & \ \ + g (1-g) \big( g^2 [(\cgb + \cbg) \ngcg + (2 - \cgb - \cbg) \ngdg] + g (1-g) [(\cgb + \cbg) (\ngcb + \nbcg)
        \\
        & \ \ \  + (2 - \cgb - \cbg) (\ngdb + \nbdg) ] + (1 - g)^2 [(\cgb + \cbg) \nbcb + (2 - \cgb - \cbg) \nbdb] \big)
        \\
        & \ \ + (1 - g)^2 \big( g^2 [\cbb \ngcg + (1  - \cbb) \ngdg] + g (1-g) [\cbb (\ngcb + \nbcg) 
        \\
        & \ \ \ + (1 - \cbb) (\ngdb + \nbdg) ] + (1 - g)^2 [\cbb \nbcb + (1 - \cbb) \nbdb] \big).
    \end{split}
\end{equation}
If we specify that the action rule does not depend on the reputation of the actor, we have $\cgg = \cbg = c_G$ and $\cgb = \cbb = c_B$.
We obtain a familiar expression:
\begin{equation}
    \begin{split}
        g & = g^2 \big( g^2 [c_G \ngcg + (1  - c_G) \ngdg] + g (1-g) [c_G (\ngcb + \nbcg)
        \\
        & \ \ \ + (1 - c_G) (\ngdb + \nbdg) ] + (1 - g)^2 [c_G \nbcb + (1 - c_G) \nbdb] \big)
        \\
        & \ \ + g (1-g) \big( g^2 [(c_B + c_G) \ngcg + (2 - c_B - c_G) \ngdg] + g (1-g) [(c_B + c_G) (\ngcb + \nbcg)
        \\
        & \ \ \  + (2 - c_B - c_G) (\ngdb + \nbdg) ] + (1 - g)^2 [(c_B + c_G) \nbcb + (2 - c_B - c_G) \nbdb] \big)
        \\
        & \ \ + (1 - g)^2 \big( g^2 [c_B \ngcg + (1  - c_B) \ngdg] + g (1-g) [c_B (\ngcb + \nbcg) 
        \\
        & \ \ \ + (1 - c_B) (\ngdb + \nbdg) ] + (1 - g)^2 [c_B \nbcb + (1 - c_B) \nbdb] \big)
        \\
        & = g^3 \big( c_G [\ngcg - \ngcb - \nbcg + \nbcb - \ngdg + \ngdb + \nbdg - \nbdb]
        \\
        & \ \ \ - c_B [\ngcg - \ngcb - \nbcg + \nbcb - \ngdg + \ngdb + \nbdg - \nbdb] \big)
        \\
        & \ \ + g^2 \big( c_G [\ngcb + \nbcg - 2\nbcb - \ngdb - \nbdg + 2\nbdb]
        \\
        & \ \ \ + c_B [\ngcg - 2\ngcb - 2\nbcg + 3\nbcb - \ngdg + 2\ngdb + 2\nbdg - 3\nbdb]
        \\
        & \ \ \ + \ngdg - \ngdb -\nbdg + \nbdb \big)
        \\
        & \ \ + g \big( c_G (\nbcb - \nbdb) + c_B (\ngcb + \nbcg - 3\nbcb - \ngdb - \nbdg + 3\nbdb)
        \\
        & \ \ \ + \ngdb + \nbdg - 2\nbdb \big)
        \\
        & \ \ + c_B (\nbcb - \nbdb) + \nbdb.
    \end{split}
\end{equation}
This is a rewritten version of equation 20 from the supplement of \citet{perret2021}, which considered third-order norms with private information but only one action rule (the classic ``discriminator'' strategy: cooperate with good and defect with bad, irrespective of one's own reputation).

\section{Multiple groups with disjoint strategic imitation}

In the preceding analyses, we have assumed that individuals freely copy strategies across groups (well-mixed copying), so that the only impact of population structure was to partition reputation information into distinct groups.
In this section we consider a model in which, in addition, strategic imitation occurs only within groups, disallowing imitation between groups (disjoint copying).
Even when strategic updates are constricted in this manner, game-play interactions between groups mean that the strategic composition of one group may shape the composition in another group. Here we consider the case of $K = 2$ groups, disallowing strategic imitation across groups. This model requires that we keep track of the strategy frequencies in each of the groups separately.

\subsection{Strategic type dynamics}

When strategies cannot be copied between groups, we must independently track the frequencies and fitnesses of types within each group.
We zero in on the case of two groups.
Fitnesses are given by
\begin{equation*}
    \begin{split}
        \Pi_X^1 & = (1 - \act) \Big[ b \big( \nu_1 [f_X^1 +  f_Z^1 g_X^{1,1}] + \nu_2 [f_X^2 + f_Z^2 g_X^{1,2}]) \big) - c \Big]
        \\
        \Pi_Y^1 & = (1 - \act) \Big[ b \big( \nu_1 [f_X^1 +  f_Z^1 g_Y^{1,1}] + \nu_2 [f_X^2 + f_Z^2 g_Y^{1,2}]) \big) \Big]
        \\
        \Pi_Z^1 & = (1 - \act) \Big[ b \big( \nu_1 [f_X^1 +  f_Z^1 g_Z^{1,1}] + \nu_2 [f_X^2 + f_Z^2 g_Z^{1,2}]) \big) - c g^{\bullet,1} \Big]
        \\
        \Pi_X^2 & = (1 - \act) \Big[ b \big( \nu_1 [f_X^1 +  f_Z^1 g_X^{2,1}] + \nu_2 [f_X^2 + f_Z^2 g_X^{2,2}]) \big) - c \Big]
        \\
        \Pi_Y^2 & = (1 - \act) \Big[ b \big( \nu_1 [f_X^1 +  f_Z^1 g_Y^{2,1}] + \nu_2 [f_X^2 + f_Z^2 g_Y^{2,2}]) \big) \Big]
        \\
        \Pi_Z^2 & = (1 - \act) \Big[ b \big( \nu_1 [f_X^1 +  f_Z^1 g_Z^{2,1}] + \nu_2 [f_X^2 + f_Z^2 g_Z^{2,2}]) \big) - c g^{\bullet,2} \Big].
    \end{split}
\end{equation*}

\begin{figure}[!ht]
    \begin{center}
    \includegraphics[width=0.8\textwidth]{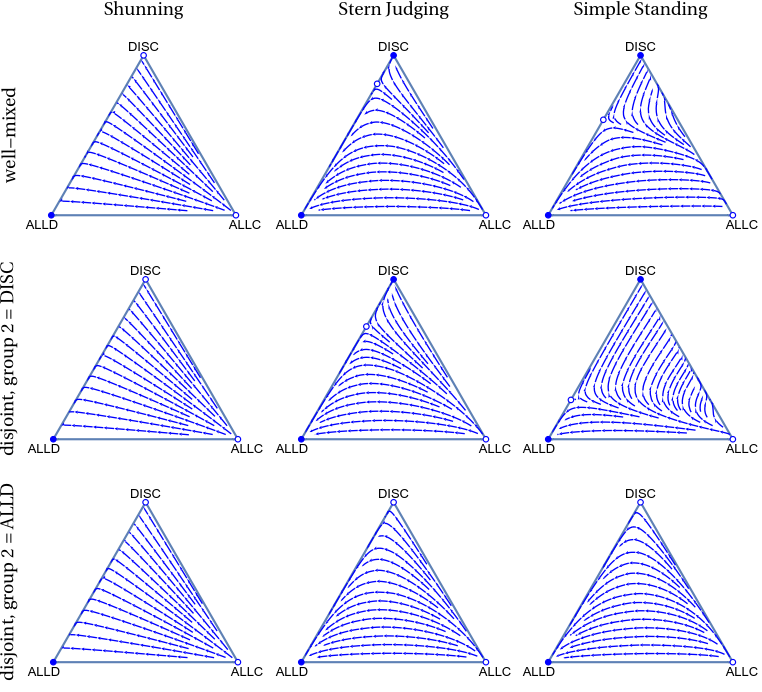}
    \end{center}
    \caption{\small{Strategy frequency dynamics for one of two equally-sized groups ($\nu_1 = \nu_2 = 1/2$) under \textit{Shunning}, \textit{Stern Judging}, and \textit{Simple Standing}.
    In all panels, the dynamics of strategy frequencies in group $1$ are shown.
    The top row corresponds to well-mixed copying; the bottom two rows correspond to disjoint copying, with group $2$ fixed for either DISC  or ALLD. Under \textit{Stern Judging} of \textit{Simple Standing}, fixing group $2$ for DISC increases the basin of attraction for cooperation in group $1$, whereas fixing group $2$ for ALLD reduces the cooperative basin in group $1$.
     In all panels, $b = 2$, $c  = 1$, $\ass = \act = 0.02$.}}
    \label{fig:disjoint_ternary}
\end{figure}

Armed with these fitness expressions, we can study how the strategic composition of one group affects the other.
We first consider the behavior of strategies in group $1$, when group $2$ is exogenously fixed for either DISC or ALLD.
Figure \ref{fig:disjoint_ternary} shows the dynamics that arise in these cases, for one choice of parameter values:  $\nu_1 = \nu_2 = 1/2$, $b = 2$, $c = 1$, and $\ass = \act = .02$.
When strategic types are copied in a well-mixed manner (top row), \textit{Shunning} cannot sustain cooperation, whereas \textit{Stern Judging} and \textit{Simple Standing} both maintain sizeable basins of attraction toward cooperation.
When group $2$ is exogenously fixed for DISC, this remains the case, as good behavior in group $1$ from the perspective of group $2$ is rewarded.
When group $2$ is exogenously fixed for ALLD, none of the three norms we consider can sustain cooperation, as discriminators waste fitness on cooperative acts with group $2$ that will never be repaid in kind.

In the subsequent section (\ref{sec:disjoint_equilibria}), we show that under disjoint copying, when group $2$ is fixed for defectors, the all-discriminator equilibrium in group $1$ is stable against invasion by defectors provided
\begin{equation}
    \frac{b}{c} > \frac{1}{\nu_1(\pgc - \pgd)} = \frac{1}{\nu_1(\epsilon - \ass)}.
\end{equation}
(This reduces to the one-group case in the limit $\nu_1 \to 1$.)
In the example shown in Figure \ref{fig:disjoint_ternary}, we have $b = 2$ and $c = 1$. The critical $b/c$ value is slightly greater than $2$, meaning that the all-discriminator equilibrium just barely fails to be stable (bottom row).
When groups are partitioned in this manner and copying is disjoint, discriminators in one group ``waste'' effort on defectors in the other group: the other group contains no discriminators, so they can only accrue a payoff due to discriminators in their own group.
This wasted effort manifests as a lower average payoff for discriminators, which increases the temptation to defect.

When group $2$ is fixed for discriminators, however, this can help group $1$ discriminators resist invasion by defectors (middle row of Figure \ref{fig:disjoint_ternary}).
How much help is provided by group 2 depends sensitively on the social norm, specifically how likely discriminators in one group are to look kindly upon discriminators with different reputational views.
Under \textit{Shunning}, any interaction with an individual with a bad reputation yields a bad reputation; under \textit{Simple Standing}, any such interaction yields a good reputation; and \textit{Stern Judging} is intermediate between the two.
This helps explain the behavior of equilibria along the $Y-Z$ edge of the simplex seen in Figure \ref{fig:disjoint_ternary}.
Under \textit{Shunning}, the all-$Z$ equilibrium is unstable; under \textit{Stern Judging} and \textit{Simple Standing}, it is stable, and there exists an unstable mixed $Y-Z$ equilibrium, corresponding to a slice of phase space that is drawn to the all-$Z$ stable equilibrium.
This slice of phase space is larger under \textit{Simple Standing} than under \textit{Stern Judging}, as expected.

When strategies are freely copied across groups (well-mixed copying), strategy frequencies equilibrate quickly, and their dynamics can be understood in terms of group-averaged reputations.
What we have shown here is that even in the polar opposite copying scenario (disjoint copying), the fact that individuals freely interact across group lines causes their dynamics to be linked.
The general tendency is that discriminators in one group make it easier for discriminators to proliferate in the other group, whether by making the other group's discriminators stable against invasion or even by making defectors vulnerable to invasion by discriminators.
Conversely, defectors in one group can render another group more vulnerable to invasion by defectors.
And so even without direct strategy copying, gameplay between disjoint groups can cause their strategy compositions to resemble each other.

\subsection{Behavior of equilibria}
\label{sec:disjoint_equilibria}

We now turn to an analytical treatment of the equilibria seen in Figure \ref{fig:disjoint_ternary}, specifically those along the DISC-ALLD edge; this will allow us to glean some insight into under what circumstances group $1$ can be invaded when group $2$'s strategic composition is fixed.
When DISC and ALLD are the only two strategic types present, their fitnesses are given by
\begin{equation*}
    \begin{split}
        \Pi_Y^1 & = (1 - \act) \Big[ b \big( \nu_1 f_Z^1 g_Y^{1,1} + \nu_2 f_Z^2 g_Y^{1,2} \big) \Big]
        \\
        \Pi_Z^1 & = (1 - \act) \Big[ b \big( \nu_1 f_Z^1 g_Z^{1,1} + \nu_2 f_Z^2 g_Z^{1,2} \big) - c g^{\bullet,1} \Big]
        \\
        \Pi_Y^2 & = (1 - \act) \Big[ b \big( \nu_1 f_Z^1 g_Y^{2,1} + \nu_2 f_Z^2 g_Y^{2,2} \big) \Big]
        \\
        \Pi_Z^2 & = (1 - \act) \Big[ b \big( \nu_1 f_Z^1 g_Z^{2,1} + \nu_2 f_Z^2 g_Z^{2,2} \big) - c g^{\bullet,2} \Big].
    \end{split}
\end{equation*}
\subsubsection{Both groups fixed for ALLD}
Let $f = f_Z^1$ be the frequency of $Z$ in group $1$.
For DISC to invade ALLD in group $1$, we would require
\begin{equation*}
    \begin{split}
        (\partial_f \dot{f})|_{f = 0} & > 0
        \\
        \Pi_Z^1|_{f_Y^1 = 1,f_Y^2 = 1} & > \Pi_Y^1|_{f_Y^1 = 1,f_Y^2 = 1}
        \\
        \therefore -c g^{\bullet,1} & > 0.
    \end{split}
\end{equation*}
Since $c$ and $g^{\bullet,2}$ are both positive numbers, $Z$ cannot invade.

\subsubsection{Both groups fixed for DISC}
Assume now that both groups are fixed for DISC.
For ALLD to invade DISC in group $1$, we would require
\begin{equation*}
    \begin{split}
        \Pi_Y^1|_{f_Z^1 = 1,f_Z^2 = 1} & > \Pi_Z^1|_{f_Z^1 = 1,f_Z^2 = 1}
        \\
        b \big( \nu_1 g_Y^{1,1} + \nu_2 g_Y^{1,2} \big)|_{f_Z^1 = 1,f_Z^2 = 1} & > b \big( \nu_1 g_Z^{1,1} + \nu_2 g_Z^{1,2} \big)|_{f_Z^1 = 1,f_Z^2 = 1} - c g^{\bullet,1}|_{f_Z^1 = 1,f_Z^2 = 1}
        \\
        \frac{b}{c} & < \frac{g^{\bullet,1}}{\nu_1 \big( g_Z^{1,1} - g_Y^{1,1} \big) + \nu_2 \big( g_Z^{1,2} - g_Y^{1,2} \big)}
        \\
        \frac{b}{c} & < \frac{g^{\bullet,1}}{\Gij{1}{2} \nu_2 (\pgc - \pgd - \pbc + \pbd) + g^{\bullet,1} \big[ \nu_1 (\pgc - \pgd) + \nu_2 (\pbc - \pbd)\big]}.
    \end{split}
\end{equation*}
This is less stringent than the standard condition $b/c < 1 / (\pgc - \pgd) = 1 / (\epsilon - \ass)$ for one group.
Discriminators in group $1$ contribute much more weakly to the fitness of discriminators in group $2$ and thus offer limited protection against invasion by defectors.

\subsubsection{One group fixed for ALLD, other for DISC}
Suppose that group $1$ is fixed for DISC and $2$ is fixed for ALLD.
We now investigate whether ALLD can invade $1$ and DISC can invade $2$, respectively.

In the first case, let $f$ be the frequency of ALLD in group 1 (that is, $f_Y^1$).
ALLD can invade $1$ provided
\begin{equation*}
    \begin{split}
        (\partial_f \dot{f})|_{f = 0} & > 0
        \\
        \therefore (\partial_f [f (\Pi_Y^1 - \bar{\Pi}^1)] )|_{f = 0} & > 0
        \\
        \therefore (\partial_f [f (\Pi_Y^1 - f \Pi_Y^1 - (1 - f) \Pi_Z^1 )] )|_{f = 0} & > 0
        \\
        \therefore (\partial_f [(f - f^2) \Pi_Y^1 - (f - f^2) \Pi_Z^1 )] )|_{f = 0} & > 0
        \\
        \therefore (\partial_f [f - f^2][\Pi_Y^1 - \Pi_Z^1 )] )|_{f = 0} & > 0
        \\
        \therefore ([1 - 2f][\Pi_Y^1 - \Pi_Z^1 )] )|_{f = 0}
        \\
        \therefore \Pi_Y^1|_{f = 0} & > \Pi_Z^1|_{f = 0}
        \\
        \therefore b \big( \nu_1 f_Z^1 g_Y^{1,1} + \nu_2 f_Z^2 g_Y^{1,2} \big)|_{f_Y^1 = 0, f_Y^2 = 1} & > \Big[ b \big( \nu_1 f_Z^1 g_Z^{1,1} + \nu_2 f_Z^2 g_Z^{1,2} \big) - c g^{\bullet,1} \Big] |_{f_Y^1 = 0, f_Y^2 = 1}
        \\
        \therefore b \nu_1 g_Y^{1,1} & > b \nu_1 g_Z^{1,1} - c g^{\bullet,1}|_{f_Y^1 = 0, f_Y^2 = 1}
        \\
        \therefore b \nu_1 (g_Z^{1,1} - g_Y^{1,1}) & < c g^{\bullet,1}
        \\
        \therefore \frac{b}{c} & < \frac{g^{\bullet,1}}{\nu_1 (g_Z^{1,1} - g_Y^{1,1})}
        \\
        \therefore \frac{b}{c} & < \frac{g^{\bullet,1}}{\nu_1 [g^{\bullet,1} \pgc + (1 - g^{\bullet,1}) \pbd - g^{\bullet,1} \pgd - (1 - g^{\bullet,1}) \pbd]}
        \\
        \therefore \frac{b}{c} & < \frac{g^{\bullet,1}}{\nu_1 g^{\bullet,1} (\pgc - \pgd)}
        \\
        \therefore \frac{b}{c} & < \frac{1}{\nu_1 (\pgc - \pbd)} = \frac{1}{\nu_1 (\epsilon - \ass)}.
    \end{split}
\end{equation*}
Letting $\nu_1 \to 1$ allows us to recover the one-group condition, $b/c < 1/(\pgc - \pgd) = 1/(\epsilon - \ass)$.
Since $\nu_1 < 1$, this is generally less strict than the one-group condition: the fact that the second group consists entirely of defectors makes it more difficult for the first group to resist invasion by defectors.

We now consider the second case, i.e., whether DISC can invade $2$, which is fixed for ALLD.
Let $f$ now be the frequency of DISC in group $2$ (that is, $f_Z^2$).
DISC being able to invade requires
\begin{equation*}
    \begin{split}
        (\partial_f \dot{f})|_{f = 0} & > 0
        \\
        \therefore \Pi_Z^2|_{f = 0} & > \Pi_Y^2|_{f = 0}
        \\
        \therefore b \Big[ \big( \nu_1 f_Z^1 g_Z^{2,1} + \nu_2 f_Z^2 g_Z^{2,2} - c g^{\bullet,2} \big) \Big]|_{f_Z^1 = 1, f_Z^2 = 0} & > b \big( \nu_1 f_Z^1 g_Y^{2,1} + \nu_2 f_Z^2 g_Z^{2,2} \big)|_{f_Z^1 = 1, f_Z^2 = 0}
        \\
        \therefore b \nu_1 (g_Z^{2,1} - g_Y^{2,1}) & > c g^{\bullet,1}|_{f_Z^1 = 1, f_Z^2 = 0}
        \\
        \therefore \frac{b}{c} & > \frac{g^{\bullet,1}}{\nu_1(g_Z^{2,1} - g_Y^{2,1})}
        \\
        \therefore \frac{b}{c} & > \frac{g^{\bullet,1}}{\nu_1 \big[ \Gij{2}{1} (\pgc - \pgd - \pbc + \pgd) + g^{\bullet,2}(\pbc - \pbd) \big]}.
    \end{split}
\end{equation*}
This is distinct from the single-group case, in which DISC ($Z$) can never invade ALLD ($Y$) (which corresponds to $\nu_1 \to 0$, blowing up the denominator).
In this scenario, discriminators in group $1$ can help discriminators in group $2$ rise in frequency, even though they are not guaranteed to have good opinions of discriminators in group $2$.

\section{Derivation of replicator equation under different copying models}\label{sec:replicator_derivation}

Here we explicitly derive the replicator dynamics for various group-wise strategy copying scenarios.
In this section we use $i$ and $j$ to denote strategic types, as opposed to $s$ and $s^\prime$.

\subsection{One group}
\label{sec:one_group}

Consider first the case of a single group.
We have the following events to take into account:
\begin{enumerate}
    \item \textbf{Increase}. A type $j (\neq i)$ individual is chosen to update with probability $f_j$.
    With probability $f_i$, the compared individual is type $i$.
    The update happens with probability $\phi(\Pi_j,\Pi_i) = 1/(1 + \exp [ \beta ( \Pi_j - \Pi_i)]$.
    The frequency of type $i$ increases by $1/N$.
    \item \textbf{Decrease}. A type $i$ individual is chosen to update with probability $f_i$.
    With probability $f_j$, the compared individual is type $j (\neq i)$.
    The update happens with probability $\phi(\Pi_i, \Pi_j)$.
    The frequency of type $i$ decreases by $1/N$.
\end{enumerate}
Thus,
\begin{equation*}
    \begin{split}
        \mathbb{E}[ \Delta f_i] & = \frac{1}{N} \mathbb{P} \Big( \Delta f_i^A = \frac{1}{N} \Big) - \frac{1}{N} \mathbb{P} \Big( \Delta f_i^A = -\frac{1}{N} \Big)
        \\
        & = \frac{1}{N} \Bigg( \sum_j f_j f_i \phi(\Pi_j,\Pi_i) - f_i \sum_j f_j \phi(\Pi_i,\Pi_j) \Bigg)
        \\
        & = \frac{1}{N} f_i \Bigg( \sum_j f_j \Big[ \phi(\Pi_j,\Pi_i) - \phi(\Pi_i,\Pi_j) \Big] \Bigg).
    \end{split}
\end{equation*}
Note that
\begin{equation*}
    \begin{split}
        \phi(\Pi_j, \Pi_i) = \frac{1}{1 + \exp[\beta(\Pi_j - \Pi_i)]} & \approx  \frac{1}{1 + \exp[\beta(\Pi_j - \Pi_i)]} \Big|_{\beta = 0} + \beta \Big( \frac{d}{d\beta}\Big[\frac{1}{1 + \exp[\beta(\Pi_j - \Pi_i)]} \Big] \Big)\Big|_{\beta = 0} + \mathcal{O}(\beta^2)
        \\
        & = \frac{1}{2} + \beta \Big[ \frac{(\Pi_i - \Pi_j)\exp[\beta(\Pi_j - \Pi_i)]}{(1 + \exp[\beta(\Pi_j - \Pi_i)])^2} \Big] \Big|_{\beta = 0} + \mathcal{O}(\beta^2)
        \\
        & = \frac{1}{2} + \beta \frac{\Pi_i - \Pi_j}{4} + \mathcal{O}(\beta^2).
    \end{split}
\end{equation*}
Hence
\begin{equation*}
    \phi(\Pi_j,\Pi_i) - \phi(\Pi_i,\Pi_j) \approx \beta \frac{\Pi_i- \Pi_j}{2} + \mathcal{O}(\beta^2).
\end{equation*}
We therefore have
\begin{equation*}
    \begin{split}
        \mathbb{E}[ \Delta f_i] & = \frac{1}{N} f_i \sum_j  \Big[ \frac{\beta}{2} f_j (\Pi_i - \Pi_j) + \mathcal{O}(\beta^2) \Big]
        \\
        & \approx \frac{\beta}{2N}f_i \sum_j f_j(\Pi_i - \Pi_j)
        \\
        & = \frac{\beta}{2N} f_i (\Pi_i \sum_j f_j - \sum_j f_j \Pi_j)
        \\
        & = \frac{\beta}{2N} f_i (\Pi_i - \bar{\Pi}).
    \end{split}
\end{equation*}
This is what ultimately justifies the use of the replicator equation under pairwise comparison. Rescaling time so that, on average, one update event occurs per time step yields
\begin{equation*}
    \dot{f}_i = f_i ( \Pi_i - \bar{\Pi}).
\end{equation*}

\subsection{Multiple groups, copying only group identity}

We now consider that there is more than one group ($K > 1$) and the entire population is fixed for the same strategy, but individuals can copy the group identity of others.
This turns out to be almost identical to the one-group case outlined in section \ref{sec:one_group}, except that the relevant transition probability is instead
\begin{equation}
    \phi( \Pi^J, \Pi^I ) = \frac{1}{1 + \exp [ \beta (\Pi^J - \Pi^I ) ]}.
\end{equation}
The remainder of the argument proceeds identically, except with $f_i$ and $f_j$ replaced with $\nu_I$ and $\nu_J$, and we obtain
\begin{equation}
    \dot{\nu}_I = \nu_I (\Pi^I - \bar{\Pi}).
\end{equation}

\subsection{Multiple groups, disjoint strategic imitation}

When there is more than one group ($K > 1$), the preceding analysis holds, except that we must specify that an individual with strategy $i$ in group $I$ can only copy from another individual in group $I$ (their in-group).
We thus obtain
\begin{equation}
    \begin{split}
        \dot{f}_i^I & = f_i ( \Pi_i^I - \bar{\Pi}^I), \text{with}
        \\
        \bar{\Pi}^I & = \sum_i f_i^I \Pi_i^I.
    \end{split}
    \label{eq:SI_disjoint}
\end{equation}

\subsection{Multiple groups, well-mixed strategic imitation}
\label{sec:well_mixed}

We now derive the analogous case for multiple groups ($K > 1$) with ``well-mixed copying'', i.e., individuals do not distinguish between their in-group and out-group when deciding whom to compare their fitness against and potentially imitate.
Let $\nu_I$ be the frequency of group $I$, and let $n_i^I = N \nu_I f_i^I$ be the \emph{absolute number} of individuals of type $I$ following strategy $i$.
The following events may occur.
\begin{enumerate}
    \item \textbf{Increase}. A type $j$ individual in group $I$ is chosen to update with probability $\nu_I f_j^I$.
    With probability $\nu_J f_i^J$, the compared individual is type $i,J$, with $J \in \{1 \ldots K \}$ (i.e., $J$ can take on the same value as $I$).
    The update happens with probability $\phi(\Pi_j^I,\Pi_i^J)$.
    $n_i^I$ increases by $1$.
    \item \textbf{Decrease}.
    A type $i$ individual in group $I$ is chosen to update with probability $\nu_I f_i^I$.
    With probability $\nu_J f_j^J$, the compared individual is type $j (\neq i),J$, with $J \in \{1 \ldots K \}$ (i.e., $J$ can take on the same value as $I$).
    The update happens with probability $\phi(\Pi_i^I,\Pi_j^J)$.
    $n_i^I$ decreases by $1$.
\end{enumerate}
Thus
\begin{equation*}
    \begin{split}
        \mathbb{E}\Big[ \Delta n_i^I \Big] & = \mathbb{P} \Big(\Delta n_i^I = 1 \Big) - \mathbb{P} \Big(\Delta n_i^I = -1 \Big)
        \\
        & = \nu_I \sum_j f_j^I \sum_J \nu_J f_i^J \phi(\Pi_j^I,\Pi_i^J) - \nu_I f_i^I \sum_j \sum_J \nu_J f_j^J \phi(\Pi_i^I,\Pi_j^J)
        \\
        & \approx  \nu_I \sum_j f_j^I \sum_J \nu_J f_i^J \Big( \frac{1}{2} + \beta \frac{\Pi_i^J - \Pi_j^I}{4} \Big) - \nu_I f_i^I \sum_j \sum_J \nu_J f_j^J \Big(\frac{1}{2} + \beta \frac{\Pi_j^J - \Pi_i^I}{4} \Big) \Big]
        \\
        & = \nu_I \frac{1}{2} \Big[ \sum_j f_j^I \sum_J \nu_J f_i^J - f_i^I \sum_j \sum_J \nu_J f_j^J \Big]
        \\
        & \ \ \ + \nu_I \frac{\beta}{4} \sum_j f_j^I \Big[ \sum_J \nu_J f_i^J(\Pi_i^J - \Pi_j^I) - f_i^I \sum_J \nu_J f_j^J (\Pi_j^J - \Pi_i^I) \Big]
        \\
        & = \nu_I \frac{1}{2} \Big[ \sum_J \nu_J \big( f_i^J \sum_j f_j^I - f_i^I \sum_j f_j^J \big) \Big]
        \\
        & \ \ \ + \nu_I \frac{\beta}{4} \Big[ \sum_J \nu_J \sum_j \Big( f_j^I f_i^J(\Pi_i^J - \Pi_j^I) - f_i^I f_j^J (\Pi_j^J - \Pi_i^I) \Big) \Big].
        \\
        & = \nu_I \frac{1}{2} \Big[ \sum_J \nu_J \big( f_i^J - f_i^I \big) \Big]
        \\
        & \ \ \ + \nu_I \frac{\beta}{4} \Big[ \sum_J \nu_J \big( f_i^J (\Pi_i^J - \sum_j f_j^I \Pi_j^I) + f_i^I (\Pi_i^I - \sum_j f_j^J \Pi_j^J) \big) \Big].
    \end{split}
\end{equation*}
Rescaling time allows us to recast this as an equation for $\dot{n}_i^I$.
Recalling that $n_i^I = N \nu_I f_i^I$, and dropping the $1/2$ prefactor, we have
\begin{equation}
    \begin{split}
        \dot{f}_i^I & \propto \sum_J \nu_J \Big[ f_i^J - f_i^I \Big] + \frac{\beta}{2}\Big[ \sum_J \nu_J \big( f_i^J (\Pi_i^J - \sum_j f_j^I \Pi_j^I) + f_i^I (\Pi_i^I - \sum_j f_j^J \Pi_j^J) \big) \Big] .
    \end{split}
    \label{eq:mixed_replicator}
\end{equation}
The proportionality constant will depend on how we rescale time.
Note that the first term \emph{does not have a $\beta$ prefactor} and roughly corresponds to ``neutral'' mixing between the two groups.
This means that \emph{that term will dominate}, and thus we expect $f_i^I$ to equilibrate rapidly to a value common to all $I$.
If we mandate this, the only dynamical quantity becomes $f_i = \sum_I \nu_I f_i^I$, so we have
\begin{equation*}
    \begin{split}
        \dot{f}_i & = \sum_I \nu_I \dot{f}_i^I
        \\
        & \propto \sum_I \nu_I \Big[ \sum_J \nu_J \big( f_i^J (\Pi_i^J - \sum_j f_j^I \Pi_j^I) + f_i^I (\Pi_i^I - \sum_j f_j^J \Pi_j^J) \big) \Big]
        \\
        & = \sum_J \nu_J f_i \Pi_i^J - f_i \sum_I \nu_I \sum_j f_j \Pi_j^I + \sum_I \nu_I f_i \Pi_i^I - f_i \sum_J \nu_J \sum_j f_j \Pi_j^J
        \\
        & = f_i \Big( \sum_J \nu_J \Pi_i^J + \sum_I \nu_I \Pi_i^I - \sum_j f_j \sum_I \nu_I \Pi_j^I  - \sum_j f_j \sum_J \nu_J  \Pi_j^J \Big)
        \\
        & \propto f_i \sum_J \nu_J \big(\Pi_i^J - \bar{\Pi}^J \big).
    \end{split}
\end{equation*}
Rescaling time allows us to write this as an equality:
\begin{equation}
    \begin{split}
        \dot{f}_i & = f_i \sum_J \nu_J \big( \Pi_i^J - \bar{\Pi}^J \big)
        \\
        & = f_i \Big( \sum_J \nu_J \Pi_i^J - \bar{\Pi} \Big) = f_i \big( \Pi_i - \bar{\Pi} \big), \text{with}
        \\
        \Pi_i & = \sum_L \nu_L \Pi_i^L,
        \\
        \bar{\Pi} & = \sum_L \nu_L \sum_i f_i \Pi_i^L = \sum_L \nu_L \bar{\Pi}^L = \sum_i f_i \Pi_i.
    \end{split}
    \label{eq:SI_well_mixed}
\end{equation}

\subsection{Multiple groups, in-group favored (``partially-mixed copying'')}\label{sec:groups_mixing}

We have seen that if individuals freely copy across group lines, strategy frequencies change much faster due to mixing than due to selection.
We now consider the possibility of partially, but not completely, restricting partner choice for strategy imitation.
Let $m$ (for ``imitation'' or, equivalently, for ``mixing'') be the weight that an individual assigns to the opposite group when deciding whom to imitate, so that $m = 0$ corresponds to no mixing (disjoint imitation) and $m = 1$ corresponds to full mixing.
An individual in group $I$ thus chooses an individual in their own group with probability $1 - m$ and chooses an individual in a random group $J$ (which could be $I$) with probability $\nu_J m$.
For $n_i^I$, the following events are possible.
\begin{enumerate}
    \item \textbf{Increase}. A type $j$ individual in group $I$ is chosen to update with probability $\nu_I f_j^I$.
    With probability $(1 - m) f_i^I$, the compared individual is type $i,I$, and with probability $\nu_J m f_i^J$, the compared individual is type $i,J$ ($J$ can be $I$).
    The update happens with probability $\phi(\Pi_j^I,\Pi_i^I)$ (for $i,I$) or $\phi(\Pi_j^I,\Pi_i^J)$ (for type $i,J$).
    In either case, $n_i^I$ increases by $1$.
    \item \textbf{Decrease}.
    A type $i$ individual in group $I$ is chosen to update with probability $\nu_I f_i^I$.
    With probability $(1 - m) f_j^I$, the compared individual is type $j (\neq i),I$, and with probability $\nu_J m f_j^J$, the compared individual is type $j (\neq i), J$ ($J$ can be $I$).
    The update happens with probability $\phi(\Pi_i^I,\Pi_j^I)$ (for $j,I$) or $\phi(\Pi_i^A,\Pi_j^J)$ (for $j,J$).
    In either case, $n_i^I$ decreases by $1$.
\end{enumerate}

We thus have
\begin{equation}
    \begin{split}
        \mathbb{E}\Big[ \Delta n_i^I \Big] & = \mathbb{P} \Big(\Delta n_i^I = 1 \Big) - \mathbb{P} \Big(\Delta n_i^I = -1 \Big)
        \\
        & = (1 - m) \Big[ \nu_I \sum_j f_j^I f_i^I \phi(\Pi_j^I,\Pi_i^I) - \nu_I f_i^I \sum_j f_j^I \phi(\Pi_i^I,\Pi_j^I) \Big]
        \\
        & \ \ \ + m \Big[  \nu_I \sum_j f_j^I \sum_J \nu_J f_i^J \phi(\Pi_j^I,\Pi_i^J) - \nu_I f_i^I \sum_j \sum_J \nu_J f_j^J \phi(\Pi_i^I,\Pi_j^J) \Big]
        \\
        & \approx (1 - m) \Big[ \nu_I \sum_j f_j^I f_i^I \big( \frac{1}{2} + w\frac{\Pi_i^I - \Pi_j^I}{4} \big) - \nu_I f_i^I \sum_j f_j^I \big(\frac{1}{2} + w\frac{\Pi_j^I - \Pi_i^I}{4} \big) \Big]
        \\
        & \ \ \ + m \Big[ \nu_I \sum_j f_j^I \sum_J \nu_J f_i^J \big( \frac{1}{2} + w\frac{\Pi_i^J - \Pi_j^I}{4} \big) - \nu_I f_i^I \sum_j \sum_J \nu_J f_j^J \big(\frac{1}{2} + w\frac{\Pi_j^J - \Pi_i^I}{4} \big) \Big]
        \\
        & = m \nu_I \frac{1}{2} \Big[ \sum_j f_j^I \sum_J \nu_J f_i^J - f_i^I \sum_j \sum_J \nu_J f_j^J \Big]
        \\
        & \ \ \ + (1 - m) \nu_I \frac{w}{2} \sum_j f_j^I f_i^I(\Pi_i^I - \Pi_j^I)
        \\
        & \ \ \ + m \nu_I \frac{w}{4} \sum_j f_j^I \Big[ \sum_J \nu_J f_i^J(\Pi_i^J - \Pi_j^I) - f_i^I \sum_J \nu_J f_j^J (\Pi_j^J - \Pi_i^I) \Big]
        \\
        & = m \nu_I \frac{1}{2} \Big[ \sum_J \nu_J \big( f_i^J \sum_j f_j^I - f_i^I \sum_j f_j^J \big) \Big]
        \\
        & \ \ \ + (1 - m) \nu_I \frac{w}{2} f_i^I \big( \Pi_i^I - \sum_j f_j^I \Pi_j^I \big)
        \\
        & \ \ \ + m \nu_I \frac{w}{4} \Big[ \sum_J \nu_J \sum_j \big( f_j^I f_i^J(\Pi_i^J - \Pi_j^I) - f_i^I f_j^J (\Pi_j^J - \Pi_i^I) \big) \Big].
        \\
        & = m \nu_I \frac{1}{2} \Big[ \sum_J \nu_J \big( f_i^J - f_i^I \big) \Big]
        \\
        & \ \ \ + (1 - m) \nu_I \frac{w}{2} f_i^I \big( \Pi_i^I - \bar{\Pi}^I \big)
        \\
        & \ \ \ + m \nu_I \frac{w}{4} \Big[ \sum_J \nu_J \big( f_i^J (\Pi_i^J - \bar{\Pi}^I) + f_i^I (\Pi_i^I - \bar{\Pi}^J) \big) \Big].
    \end{split}
    \label{eq:niA_mixing}
\end{equation}
Recalling that $n_i^I = N \nu_I f_i^I$, the replicator dynamics are given by
\begin{equation*}
    \begin{split}
        \dot{f}_i^I & \propto m \frac{1}{2} \Big[ \sum_J \nu_J \big( f_i^J - f_i^I \big) \Big]
        \\
        & \ \ \ + (1 - m) \frac{w}{2} f_i^I \big( \Pi_i^I - \bar{\Pi}^I \big)
        \\
        & \ \ \ + m \frac{w}{4} \Big[ \sum_J \nu_J \big( f_i^J (\Pi_i^J - \bar{\Pi}^I) + f_i^I (\Pi_i^I - \bar{\Pi}^J) \big) \Big]
        \\
        & \propto \frac{m}{w} \Big[ \sum_J \nu_J \big( f_i^J - f_i^I \big) \Big]
        \\
        & \ \ \ + (1 - m) f_i^I \big( \Pi_i^I - \bar{\Pi}^I \big)
        \\
        & \ \ \ + \frac{m}{2} \Big[ \sum_J \nu_J \big( f_i^J (\Pi_i^J - \bar{\Pi}^I) + f_i^I (\Pi_i^I - \bar{\Pi}^J \big) \Big]
    \end{split}
\end{equation*}
As usual, the $\propto$ can be converted into $=$ by rescaling time.
In each equation, the first term (proportional to $m/w$) sets the rate of between-group ``neutral'' mixing, the second corresponds to within-group selection, and the third corresponds to between-group selection.
Note that setting $m = 0$ yields equation \eqref{eq:SI_disjoint} and setting $m = 1$ yields equation \eqref{eq:SI_well_mixed}, subject to rescaling.

\subsection{Multiple groups, copying both strategy and group identity}

We now assume that individuals engage in both well-mixed strategic copying \emph{and} copying of group identity.
With probability $\tau$, an individual resolves to update their group identity; with probability $1 - \tau$, they resolve to update their behavioral strategy.
We consider here the possible change in $n_i^I$.
\begin{enumerate}
    \item \textbf{Increase...}
    \begin{enumerate}
        \item \textbf{...by changing group identity}.
        A type $i$ individual in group $J \in \{1 \ldots K \}$ is chosen to update with probability $\nu_J f_i^J$.
        With probability $\tau$, they choose to update their group identity.
        With probability $\nu_I f_j^I$, the comparison partner is type $j$ (any strategy) and in group $I$.
        The update happens with probability $\phi( \Pi_i^J, \Pi_j^I)$.
        \item \textbf{...by changing behavioral strategy}.
        A type $j$ (any strategy) individual in group $I$ is chosen to update with probability $\nu_I f_j^I$.
        With probability $1 - \tau$, they choose to update their behavioral strategy.
        With probability $\nu_J f_i^J$, the comparison partner is type $i$ and in group $J \in \{1 \ldots K \}$.
        The update happens with probability $\phi( \Pi_j^I, \Pi_i^J )$.
    \end{enumerate}
    \item \textbf{Decrease...}
    \begin{enumerate}
        \item \textbf{...by changing group identity}.
        A type $i$ individual in group $I$ is chosen to update with probability $\nu_I f_i^I$.
        With probability $\tau$, they choose to update their group identity.
        With probability $\nu_J f_j^J$, the comparison partner is type $j$ (any strategy) and in group $J \in \{1 \ldots K \}$.
        The update happens with probability $\phi( \Pi_i^I, \Pi_j^J)$.
        \item \textbf{...by changing behavioral strategy}.
        A type $i$ individual in group $I$ is chosen to update with probability $\nu_I f_i^I$.
        With probability $1 - \tau$, they choose to update their behavioral strategy.
        With probability $\nu_J f_j^J$, the comparison partner is type $j$ (any strategy) and in group $J \in \{1 \ldots K \}$.
        The update happens with probability $\phi( \Pi_i^I, \Pi_j^J )$.
    \end{enumerate}
\end{enumerate}
We have
\begin{equation*}
    \begin{split}
        \mathbb{E}\Big[ \Delta n_i^I \Big] & = \mathbb{P} \Big(\Delta n_i^I = 1 \Big) - \mathbb{P} \Big(\Delta n_i^I = -1 \Big)
        \\
        & = \tau \Big( \sum_J \nu_J f_i^J \nu_I \sum_j f_j^I \phi(\Pi_i^J,\Pi_j^I) - \nu_I f_i^I \sum_J \sum_j \nu_J f_j^J \phi( \Pi_i^I, \Pi_j^J) \Big)
        \\
        & \ \ \ + (1 - \tau) \Big( \nu_I \sum_j f_j^I \sum_J \nu_J f_i^J \phi( \Pi_j^I, \Pi_i^J) - \nu_i f_i^I \sum_J \sum_j \nu_J f_j^J \phi( \Pi_i^I, \Pi_j^J) \big)
        \\
        & = \tau \Big( \nu_I \sum_J \nu_J f_i^J \sum_j f_j^I \phi(\Pi_i^J,\Pi_j^I) - \nu_I f_i^I \sum_J \sum_j \nu_J f_j^J \phi( \Pi_i^I, \Pi_j^J) \Big)
        \\
        & \ \ \ + (1 - \tau) \Big( \nu_I \sum_j f_j^I \sum_J \nu_J f_i^J \phi( \Pi_j^I, \Pi_i^J) - \nu_i f_i^I \sum_J \sum_j \nu_J f_j^J \phi( \Pi_i^I, \Pi_j^J) \big)
        \\
        & \approx \tau \nu_I \sum_J \nu_J f_i^J \sum_j f_j^I \Big( \frac{1}{2} + \beta \frac{\Pi_j^I - \Pi_i^J}{4} \Big) + (1 - \tau) \nu_I \sum_j f_j^I \sum_J \nu_J f_i^J \Big( \frac{1}{2} + \beta \frac{\Pi_i^J - \Pi_j^I}{4} \Big)
        \\
        & \ \ \ - \nu_I f_i^I \sum_J \sum_j \nu_J f_j^J \Big( \frac{1}{2} + \beta \frac{\Pi_j^J - \Pi_i^I}{4} \Big)
        \\
        & = \nu_I \frac{1}{2} (f_i - f_i^I) +  \nu_I \frac{\beta}{4} \sum_J \nu_J f_i^J \sum_j f_j^I \big[ \tau (\Pi_j^I - \Pi_i^J) + (1 - \tau) (\Pi_i^J - \Pi_j^I) \big]
        \\
        & \ \ \ - \nu_I f_i^I \frac{\beta}{4} \sum_J \sum_j \nu_j f_j^J (\Pi_j^J - \Pi_i^I)
        \\
        & = \nu_I \frac{1}{2} (f_i - f_i^I) +  \nu_I \frac{\beta}{4} \sum_J \nu_J f_i^J \sum_j f_j^I (1 - 2\tau) (\Pi_i^J - \Pi_j^I)
        \\
        & \ \ \ - \nu_I f_i^I \frac{\beta}{4} \sum_J \sum_j \nu_j f_j^J (\Pi_j^J - \Pi_i^I)
        \\
        & = \nu_I \frac{1}{2} (f_i - f_i^I) +  \nu_I (1 - 2\tau) \frac{\beta}{4} \Big( \sum_J \nu_J f_i^J \sum_j f_j^I \Pi_i^J - \sum_J \nu_J f_i^J \sum_j f_j^I \Pi_j^I \Big)
        \\
        & \ \ \ - \nu_I f_i^I \frac{\beta}{4} \Big( \sum_J \sum_j \nu_j f_j^J \Pi_j^J - \sum_J \sum_j \nu_j f_j^J \Pi_i^I \Big)
        \\
        & = \nu_I \frac{1}{2} (f_i - f_i^I) +  \nu_I (1 - 2\tau) \frac{\beta}{4} \Big( \sum_J \nu_J f_i^J \Pi_i^J \sum_j f_j^I - \sum_j f_j^I \Pi_j^I \sum_J \nu_J f_i^J \Big)
        \\
        & \ \ \ - \nu_I f_i^I \frac{\beta}{4} \Big( \sum_J \sum_j \nu_J f_j^J \Pi_j^J - \sum_J \sum_j \nu_J f_j^J \Pi_i^I \Big)
        \\
        & = \nu_I \frac{1}{2} (f_i - f_i^I) +  \nu_I (1 - 2\tau) \frac{\beta}{4} \Big( \sum_J \nu_J f_i^J \Pi_i^J - f_i \sum_j f_j^I \Pi_j^I \Big)
        \\
        & \ \ \ - \nu_I f_i^I \frac{\beta}{4} \Big( \sum_J \sum_j \nu_J f_j^J \Pi_j^J - \Pi_i^I \Big)
        \\
        & = \nu_I \frac{1}{2} (f_i - f_i^I) +  \nu_I (1 - 2\tau) \frac{\beta}{4} \Big( \sum_J \nu_J f_i^J \Pi_i^J - f_i \bar{\Pi}^I \Big) - \nu_I f_i^I \frac{\beta}{4} \Big( \bar{\Pi} - \Pi_i^I \Big)
        \\
        & \propto \nu_I \big( f_i - f_i^I + \frac{\beta}{2} \Big[ (1 - 2\tau)  \Big( \sum_J \nu_J f_i^J \Pi_i^J - f_i \bar{\Pi}^I \Big) + f_i^I \Big( \Pi_i^I - \bar{\Pi} \Big) \Big]).
    \end{split}
\end{equation*}
Observe that the $f_i - f_i^I$ term lacks a prefactor and therefore will dominate, so we expect that all $f_i^I$ will rapidly equilibrate to a common value $f_i$.
The result is actually a system of equations in both $f_i^I$ and $\nu_I$, since $\nu_I = \sum_j n_j^I/N$ and $f_i = n_i/(N \nu_I)$.
Thus
\begin{equation}
    \begin{split}
        \nu_I & = \frac{1}{N} \sum_j n_j^I
        \\
        \therefore \dot{\nu}_I & = \frac{1}{N} \sum_j \dot{n}_j^I
        \\
        & \propto \frac{1}{N} \sum_j \nu_I \big( f_j - f_j^I + \frac{\beta}{2} \Big[ (1 - 2\tau)  \Big( \sum_J \nu_J f_j^I \Pi_j^I - f_j \bar{\Pi}^I \Big) + f_j^I \Big( \Pi_j^I - \bar{\Pi} \Big) \Big])
        \\
        & = \frac{1}{N} \nu_I \frac{\beta}{2} \Big[ (1 - 2 \tau) \sum_j \Big( \sum_J \nu_J f_j^I \Pi_j^I - f_j \bar{\Pi}^I \Big) + \sum_j f_j^I (\Pi_j^I - \bar{\Pi}) \Big]
        \\
        & = \frac{1}{N} \nu_I \frac{\beta}{2} \Big[ (1 - 2 \tau) \big( \bar{\Pi} -  \bar{\Pi}^I \big) + \bar{\Pi}^I - \bar{\Pi} \Big]
        \\
        & = \frac{1}{N} \nu_I \beta \tau (\bar{\Pi}^I - \bar{\Pi}).
    \end{split}
\end{equation}
As expected, when $\tau \to 0$ (i.e., individuals never update their group identity), this vanishes.
For positive $\tau$, $\nu_I$ changes at a rate that depends on the difference between $\nu_I$'s fitness and the population average.
We can take advantage of the fact that the $f_i^I$ equilibrate rapidly to a common value $f_i$ and average out the fact that fitnesses $\Pi_i^I$ may differ by group, by considering only
\begin{equation}
    \begin{split}
        f_i & = \sum_I \nu_I f_i^I
        \\
        \therefore \dot{f}_i & = \frac{d}{dt} \Big( \sum_I \nu_I f_i^I )
        \\
        & = \sum_I ( \dot{\nu}_I f_i^I + \nu_I \dot{f}_i^I )
        \\
        & = \sum_I \nu_I \dot{f}_i^I
        \\
        & = \sum_I \nu_I \frac{1}{N} \frac{d}{dt} \frac{n_i^I}{\nu_I}
        \\
        & = \frac{1}{N} \sum_I \nu_I \frac{ \dot{n}_i^I \nu_I - n_i^I \dot{\nu}_I}{(\nu_I)^2}
        \\
        & = \frac{1}{N} \sum_I \nu_I \frac{ \dot{n}_i^I \nu_I - N \nu_I f_i^I \dot{\nu}_I}{(\nu_I)^2}
        \\
        & = \frac{1}{N} \sum_I \Big( \dot{n}_i^I - N f_i^I \dot{\nu}_I \Big)
        \\
        & = \frac{1}{N} \sum_I \nu_I \big( f_i - f_i^I + \frac{\beta}{2} \Big[ (1 - 2\tau)  \Big( \sum_J \nu_J f_i^J \Pi_i^J - f_i^I \bar{\Pi}^I \Big) + f_i^I \Big( \Pi_i^I - \bar{\Pi} \Big) \Big] - f_i^I \beta \tau (\bar{\Pi}^I - \bar{\Pi}) \big)
        \\
        & = \frac{1}{N} \sum_I \nu_I \big( \frac{\beta}{2} \Big[ (1 - 2\tau)  \Big( \sum_J \nu_J f_i^J \Pi_i^J - f_i^I \bar{\Pi}^I \Big) + f_i^I \Big( \Pi_i^I - \bar{\Pi} \Big) \Big] - f_i^I \beta \tau (\bar{\Pi}^I - \bar{\Pi}) \big)
        \\
        & = \frac{1}{N} \sum_I \nu_I \big( \frac{\beta}{2} \Big[ (1 - 2\tau)  \Big( \sum_J \nu_J f_i \Pi_i^J - f_i \bar{\Pi}^I \Big) + f_i \Big( \Pi_i^I - \bar{\Pi} \Big) \Big] - f_i \beta \tau (\bar{\Pi}^I - \bar{\Pi}) \big)
        \\
        & = \frac{1}{N} \big( \frac{\beta}{2} \Big[ (1 - 2\tau)  \Big( f_i \sum_I \nu_I \sum_J \nu_J \Pi_i^J - f_i \sum_I \nu_I \bar{\Pi}^I \Big) + f_i \Big( \sum_I \nu_I \Pi_i^I - 
        \sum_I \nu_I \bar{\Pi} \Big) \Big]
        \\
        & \ \ \ - f_i \beta \tau (\sum_I \nu_I \bar{\Pi}^I - \sum_I \nu_I \bar{\Pi}) \big)
        \\
        & = \frac{1}{N} f_i \big( \frac{\beta}{2} \Big[ (1 - 2\tau)  \Big( \sum_I \nu_I \bar{\Pi}_i - \bar{\Pi} \Big) + \Big( \bar{\Pi}_i - 
        \bar{\Pi} \Big) \Big] - \beta \tau (\bar{\Pi} - \bar{\Pi}) \big)
        \\
        & = \frac{1}{N} f_i \beta \Big[ (1 - \tau) ( \bar{\Pi}_i - \bar{\Pi} ) \Big].
    \end{split}
\end{equation}
Sending $\tau \to 1$ (i.e., individuals only ever update their group identity, not their behavioral strategy) yields $\dot{f}_i = 0$ due to selection.
(The leading term in the expression for $\dot{n}_i^I$ still has no $\beta$ prefactor and, thus, corresponds to the $f_i^I$ equilibrating as a result of random group identity switching, even in the absence of behavioral strategy updating, so the necessary assumption that the $f_i^I$ equilibrate rapidly is not violated.)
Rescaling time again so as to drop the $\beta/N$ prefactor yields the system of equations
\begin{equation}
    \begin{split}
        \dot{\nu}_I & = \nu_I \tau (\bar{\Pi}^I - \bar{\Pi}),
        \\
        \dot{f}_i & = f_i (1 - \tau) (\bar{\Pi}_i - \bar{\Pi}).
    \end{split}
\end{equation}

\clearpage
\bibliographystyle{apalike}
\bibliography{supplement}